\documentclass{aa}
\usepackage[varg]{txfonts}
\usepackage{graphicx}
\usepackage{natbib}
\usepackage{booktabs}
\usepackage{morefloats}
\bibpunct{(}{)}{;}{a}{}{,}
\usepackage{xcolor}
\usepackage{hyperref}
\begin{document}

\newcommand{\orcid}[1]{\href{https://orcid.org/#1}{\includegraphics[width=10pt]{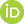}}}

\title{Lightcurve and spectral modelling of the Type IIb SN 2020acat}
\subtitle{Evidence for a strong Ni bubble effect on the diffusion time}

\author{Mattias~Ergon\inst{\ref{inst1}}\orcid{0000-0002-6209-6015},
Peter~Lundqvist\inst{\ref{inst1}}, Claes~Fransson\inst{\ref{inst1}}, Hanindyo~Kuncarayakti\inst{\ref{inst2},\ref{inst21}},
Kaustav~K.~Das\inst{\ref{inst4}}\orcid{0000-0001-8372-997X},
Kishalay~De\inst{\ref{inst5}},
Lucia~Ferrari\inst{\ref{inst9},\ref{inst91}},
Christoffer~Fremling\inst{\ref{inst3}},
Kyle~Medler\inst{\ref{inst6}},
Keiichi~Maeda\inst{\ref{inst7}}\orcid{0000-0003-2611-7269},
Andrea~Pastorello\inst{\ref{inst10}},
Jesper~Sollerman\inst{\ref{inst1}}\orcid{0000-0003-1546-6615},
Maximilian~D.~Stritzinger\inst{\ref{inst8}}\orcid{0000-0002-5571-1833}}

\institute{The Oskar Klein Centre, Department of Astronomy, AlbaNova, Stockholm University, 106 91 Stockholm, Sweden 
\label{inst1}
\and
Tuorla Observatory, Department of Physics and Astronomy, FI-20014 University of Turku, Finland
\label{inst2}
\and
Finnish Centre for Astronomy with ESO (FINCA), FI-20014 University of Turku, Finland
\label{inst21}
\and
Cahill Center for Astrophysics, California Institute of Technology, MC 249-17, 1200 E California Boulevard, Pasadena, CA, 91125, USA
\label{inst4}
\and
MIT-Kavli Institute for Astrophysics and Space Research, 77 Massachusetts Ave., Cambridge, MA 02139, USA
\label{inst5}
\and
Facultad de Ciencias Astronómicas y Geofísicas, Universidad Nacional de La Plata, Paseo del Bosque S/N, B1900FWA La Plata, Argentina
\label{inst9}
\and
Instituto de Astrofísica de La Plata (IALP), CONICET, Argentina
\label{inst91}
\and
Division of Physics, Mathematics, and Astronomy, California Institute of Technology, Pasadena, CA 91125, USA
\label{inst3}
\and
Astrophysical Research Institute Liverpool John Moores University, Liverpool L3 5RF, UK
\label{inst6}
\and
Department of Astronomy, Kyoto University, Kitashirakawa-Oiwake-cho, Sakyo-ku, Kyoto, 606-8502. Japan
\label{inst7}
\and
INAF – Osservatorio Astronomico di Padova, Vicolo
dell’Osservatorio 5, 35122 Padova, Italy
\label{inst10}
\and
Department of Physics and Astronomy, Aarhus University, Ny Munkegade 120, DK-8000 Aarhus C, Denmark
\label{inst8}}

\date{Accepted for publication by Astronomy \& Astrophysics.}

\abstract{We use the light curve and spectral synthesis code JEKYLL to calculate a set of macroscopically mixed Type IIb supernova (SN) models, which are compared to both previously published and new late-phase observations of SN 2020acat. The models differ in the initial mass, the radial mixing and expansion of the radioactive material, and the properties of the hydrogen envelope. The best match to the photospheric and nebular spectra and lightcurves of SN 2020acat is found for a model with an initial mass of 17 M$_\odot$, strong radial mixing and expansion of the radioactive material, and a 0.1 M$_\odot$ hydrogen envelope with a low hydrogen mass-fraction of 0.27. The most interesting result is that strong expansion of the clumps containing radioactive material seems to be required to fit the observations of SN 2020acat both in the diffusion phase and the nebular phase. These "Ni bubbles" are expected to expand due to heating from radioactive decays, but the degree of expansion is poorly constrained. Without strong expansion there is a tension between the diffusion phase and the subsequent evolution, and models that fit the nebular phase produce a diffusion peak that is too broad. The diffusion phase lightcurve is sensitive to the expansion of the "Ni bubbles", as the resulting Swiss-cheese-like geometry decreases the effective opacity and therefore the diffusion time. This effect has not been taken into account in previous lightcurve modelling of stripped-envelope SNe, which may lead to a systematic underestimate of their ejecta masses. In addition to strong expansion, strong mixing of the radioactive material also seems to be required to fit the diffusion peak. It should be emphasized, though, that JEKYLL is limited to a geometry that is spherically symmetric on average, and large-scale asymmetries may also play a role. The relatively high initial mass found for the progenitor of SN 2020acat places it at the upper end of the mass distribution of Type IIb SN progenitors, and a single star origin can not be excluded.}


\titlerunning{Lightcurve and spectral modelling of the Type IIb SN 2020acat.}
\authorrunning{M. Ergon et al.}
\maketitle

\defcitealias{Erg18}{E18}
\defcitealias{Erg22}{E22}
\defcitealias{Med22}{M22}
\defcitealias{Med23}{M23}

\section{Introduction}
\label{s_intruduction}

\citet[][hereafter \citetalias{Erg18}]{Erg18} and \citet[][herafter \citetalias{Erg22}]{Erg22} presented and tested the light curve and spectral synthesis code JEKYLL, and demonstrated its capability to model both the photospheric and nebular phase of Supernovae (SNe). In particular we demonstrated that both NLTE and the macroscopic mixing of the ejecta that occurs in the explosion need to be taken into account for the models to be realistic. As discussed in \citetalias{Erg22}, the macroscopic mixing of the ejecta influences the SN in several ways. Both by preventing compositional mixing of the nuclear burning zones, which affects the strength of important lines in the nebular phase, and by expansion of clumps containing radioactive material, 
which tends to decrease the effective opacity and therefore the diffusion time in the photospheric phase. The latter effect, which can be dramatic, has also been discussed by \citet{Des19} with respect to Type IIP SNe, although in their case the clumping was not directly linked to the expansion of the radioactive material. The magnitude of the effect depends on uncertain properties of the small-scale 3D ejecta structure, like the typical scale at which the fragmentation occurs in the explosion, and to what extent clumps containing radioactive material subsequently expand due to heating from radioactive decays. It is therefore of great interest to further constrain these properties. Note, however, that JEKYLL do not simulate the hydrodynamics giving rise to the macroscopically mixed ejecta, but use a parametrized representation of such ejecta consisting of clumps of different composition, density, filling factor and size (see \citealt{Jer11} and \citetalias{Erg22}).

\citetalias{Erg22} applied JEKYLL to the Type IIb SN 2011dh, and showed that a macroscopically mixed SN model based on a progenitor with an initial mass of $\sim$12 M$_\odot$ well reproduces the observed spectra and lightcurves of SN 2011dh, both in the photospheric and nebular phase. This is in line with previous work on this SN \cite[see][]{Mau11,Ber11,Erg15,Jer15} and underpins the emerging consensus that Type IIb SNe mainly originate from relatively low-mass progenitors, in turn suggesting a binary origin. However, this conclusion is mainly based on approximate modelling, although for a few Type IIb SNe (like SNe 2011dh and 1993J) there are constraints from both detailed NLTE modelling in the nebular phase \citep{Jer15} and progenitor detections \citep{Ald94,Mau11}. It is therefore interesting to apply JEKYLL to another nearby, well-observed Type IIb SN, to explore which constraints can be obtained on the SN and progenitor parameters.

SN 2020acat was discovered on December 9 2020 \citep{Srv20}, and classified as a Type IIb by \citet{Pes20}. \citet[][hereafter \citetalias{Med22}]{Med22} presented an extensive photometric and spectroscopic dataset, observational analysis and approximate modelling of the SN. A complementing set of NIR spectra was presented by \citet[][hereafter \citetalias{Med23}]{Med23}. Here we present further late-time optical spectroscopy and photometry, and all-together the data-set for SN 2020acat is one of the best obtained for Type IIb SNe so far. Using the highly approximate (but classical) \citet{Arn82} model for the diffusion phase lightcurve, \citetalias{Med22} find SN 2020acat to have an ejecta mass similar to that of SN 2011dh, which is well constrained to originate from a relatively low mass progenitor. On the other hand, using a one-zone NLTE model for the nebular spectra they find an oxygen mass of $\sim$1 M$_\odot$, indicating a progenitor of considerably higher initial mass than that of SN 2011dh. This tension motivates more detailed modelling, and it is interesting to see if it can be resolved by using JEKYLL, which self-consistently models both the photospheric and nebular phase using more elaborate physics.

The paper is outlined as follows. In Sect.~\ref{s_observations} we describe the observations of SN 2020acat and compare them to the observations of SN 2011dh, which provides a starting point for the modelling with JEKYLL. In Sect.~\ref{s_methods_and_models} we briefly summarise the methods used by JEKYLL and describe our grid of Type IIb SN models, and in Sect.~\ref{s_comparison_to_observations} we compare these models to the observations of SN 2020acat in order to constrain the SN and progenitor parameters. Finally, in Sect.~\ref{s_conclusions} we summarize the paper.

\section{Observations}
\label{s_observations}

\subsection{Photometry}
\label{s_obs_phot}

The bulk of the photometry for SN 2020acat is adopted from \citetalias{Med22}, and was obtained in the $B$, $V$, $r$, $i$ and $z$ bands with the Nordic Optical Telescope (NOT), the Liverpool Telescope, the Asiago Copernico Telescope, the Palomar Samuel Oschin Telescope, the Mount Ekar Schmidt Telescope and several telescopes part of the Las Cumbres Observatory, in the $U$ and $\textit{UVM2}$ bands with the Swift Observatory, and in the $J$, $H$ and $K$ bands with the NOT and the New Technology Telescope (NTT). The reduction and calibration of these data are described in \citetalias{Med22}. In addition to this we present new late-time optical photometry obtained at $\sim$400 days with the Gemini South Telescope and the Faulkes 
Telescope North, part of the Las Cumbres Observatory Global Telescope (LCOGT; \citealt{Bro13}) network. Complementary $K$-band photometry was also performed on the acquisition images for the NIR spectra obtained with the Keck Telescope (see Sect.~\ref{s_obs_spec}). These additional photometric observations of SN 2020acat are listed in Table~\ref{t_phot_obs}.

The Gemini South observations were obtained in the $g$ and $r$ bands at an epoch of 387 days using the Gemini Multi-Object Spectrograph (GMOS-S) instrument. The observations were reduced using the Gemini package included in IRAF, and PSF photometry performed with the DAOPHOT package. Instrumental magnitudes were calibrated to the standard AB system using 12 stars in the SN field to compute zero-point corrections relative to the Panoramic Survey Telescope and Rapid Response System (PanSTARRS) catalog. 

The Faulkes Telescope North observations were obtained in the $r$ band at an epoch of 404 days using the Spectral Camera, as part of program NOAO2020B-012 (PI: De). The individual reduced images were retrieved from the online LCOGT archive, followed by stacking and photometric calibration against the PS1 catalog. We performed subtraction of the host galaxy light using archival PS1 images as templates, following the method described in \citet{De20}. The flux of the source was estimated by performing forced PSF photometry on the difference images.

The Keck $K$-band observations were obtained at epochs of 16, 77, 135 and 167 days using the NIRES instrument as part of the spectroscopic observations. The images were reduced and the photometry performed and calibrated to the 2MASS system using the IRAF based SNE pipeline \citep{Erg14}. For the calibration a two-step procedure was used, where the magnitudes of the stars visible in the Keck images were first measured and calibrated to the 2MASS system using NIR images with a wider field-of-view obtained with the NOT.

Finally, in addition to what was done in \citetalias{Med22}, we also applied spectral corrections (S-corrections, see \citealt{Str02}) to the photometry. In the nebular phase these corrections can be substantial, and for a discussion about this with respect SN 2011dh and details about the procedure see \citet{Erg18} and references therein. Instrumental filter response functions were constructed from filter and CCD data provided by the observatory or the manufacturer and extinction data for the site. S-corrections were then calculated based on the these instrumental response functions, the filter response functions for the Johnson-Cousins (JC), Sloan Digital Sky Survey (SDSS) and 2 Micron All Sky Survey (2MASS) standard systems and the spectral evolution of SN 2020acat.

\subsection{Spectra}
\label{s_obs_spec}

The bulk of the spectra for SN 2020acat are adopted from \citetalias{Med22} and \citetalias{Med23}, and were obtained with the NOT using the ALFOSC instrument, the Asiago Copernico Telescope, and the Keck Telescope using the NIRES instruments. In addition to this we present new late-time spectra obtained with the VLT using the FORS2 and X-Shooter instruments, the Gemini South Telescope using the GMOS-S instrument, and the Keck Telescope using the LRIS instrument. These additional spectral observations of SN 2020acat are listed in Table \ref{t_spec_obs}.

The VLT observations were obtained at an epoch of 106 days using the X-shooter instrument, and at an epoch of 204 days using the FORS2 instrument with Grism 300V. The X-shooter and FORS2 spectra were reduced and calibrated using ESOReflex \citep{Freu13} following standard procedures, which include bias subtraction, flatfielding, wavelength calibration, and flux calibration with a spectrophotometric standard star. 

The Gemini South observations were obtained at an epoch of 387 days using the GMOS-S instrument with the R400 grating. The wavelength calibration was done using Cu-Ar lamps, and the flux calibration was done with a spectrophotometric standard star. The VLT FORS2 observations were obtained as part of the FORS+ Survey of Supernovae in Late Times program (FOSSIL, Kuncarayakti et al. in prep; see \citealt{Kun22}).

The Keck/LRIS observations were obtained at an epoch of 422 days. The data were reduced using the fully automated data reduction pipeline LPipe \citep{Per19}. An observation of G191-B2B taken on the same night was used for flux calibration.

Unfortunately, simultaneous NIR photometry to flux-calibrate the Keck NIR spectra was not originally obtained. Therefore, as mentioned in the previous section, we have measured additional $K$-band photometry from the acquisition images, and otherwise rely on interpolations from the $J$ and $H$-band photometry obtained with NOT and NTT. However, after 115 days, there is no $J$ and $H$-band photometry available, so in this case we decided to extrapolate the $J$-band evolution using our optimal model for SN 2020acat and linearly interpolate between this and the measured $K$-band magnitudes. This should be kept in mind while examining the $J$ and $H$-band regions of the NIR spectra obtained after 115 days.

\begin{table*}[tb]
\caption{List of photometric observations of SN 2020acat presented in this paper.}
\begin{center}
\begin{tabular}{lllllll}
\toprule
MJD (days) & Phase (days) & $g$ & $r$ & $K$ & Telescope & Instrument\\
\midrule
59207.65 & 16 & - & - & 14.40 $\pm$ 0.07 & Keck & NIRES\\
59268.32 & 77 & - & - & 15.56 $\pm$ 0.06 & Keck & NIRES\\
59326.33 & 135 & - & - & 16.95 $\pm$ 0.17 & Keck & NIRES\\
59358.30 & 167 & - & - & 17.58 $\pm$ 0.07 & Keck & NIRES\\
59578.27 & 387 & 22.7 $\pm$ 0.2 & 21.4 $\pm$ 0.1 & - & Gemini South & GMOS-S\\
59595.70 & 404 & - & 21.76 $\pm$ 0.22 & - & Faulkes Telescope North & Spectral Camera\\
\bottomrule
\end{tabular}
\end{center}
\label{t_phot_obs}
\end{table*}

\begin{table*}[tb]
\caption{List of spectral observations of SN 2020acat presented in this paper.}
\begin{center}
\begin{tabular}{llllll}
\toprule
MJD (days) & Phase (days) & Range (\AA) & Resolution & Telescope & Instrument\\
\midrule
59297.61 & 106 & 3100 - 24400 & 6000 & VLT & X-Shooter\\
59395.61 & 204 & 3500 - 9500 & 300& VLT & FORS2\\
59578.27 & 387 & 4500 - 9000 & 600 & Gemini South & GMOS-S\\
59613.50 & 422 & 3100 - 10000 & 1000 & Keck & LRIS\\
\bottomrule
\end{tabular}
\end{center}
\label{t_spec_obs}
\end{table*}

\subsection{Distance, extinction and explosion epoch}
\label{s_dist_ext}

According to the NASA/IPAC Extragalactic Databse (NED), the host galaxy PGC037027 has a redshift of $z=0.00793$, which using a cosmology with H$_0$=$73.0 \pm 5$ km s$^{-1}$ Mpc$^{-1}$, $\Omega_\mathrm{m}$=0.27, and $\Omega_\Lambda$=0.73 corresponds to a Hubble flow distance of $35.3 \pm 4$ Mpc (see \citetalias{Med22} for a discussion on the error bar), corrected for the influence of the Virgo Cluster, the Great Attractor, and the Shapley Supercluster. This, in turn, corresponds to a distance modulus $m -M = 32.74 \pm 0.27$ mag. 

As in \citetalias{Med22}, we assume that the extinction within the host galaxy is negligible, which is supported by the absence of host galaxy \ion{Na}{i} D lines and the position of SN within the host galaxy. Given this assumption, the total extinction is the same as the extinction within the Milky Way along the line of sight, which is $E(B-V) = 0.021$ mag according to NED.

The constraints on the explosion epoch are good and there are only two days between the last non-detection at MJD = 59190.61 and the first detection at MJD = 59192.65. Contrary to \citetalias{Med22}, who used a fit to the pseudo-bolometric lightcurve to determine the explosion epoch, we simply adopt the midpoint between the last non-detection and the first detection (MJD = 59191.63) as the explosion epoch.

\subsection{Comparison to SN 2011dh} 
\label{s_comparison_to_11dh}

As SN 2011dh was modelled by JEKYLL in \citetalias{Erg22}, and has both excellent data and well-constrained SN and progenitor parameters, it is of particular interest to compare the observations of SN 2020acat to this SN. The main purpose is to provide a starting point for the modelling of SN 2020acat with JEKYLL, but we also discuss some other topics.

\subsubsection{Lightcurves}

\begin{figure}[tbp!]
\includegraphics[width=0.49\textwidth,angle=0]{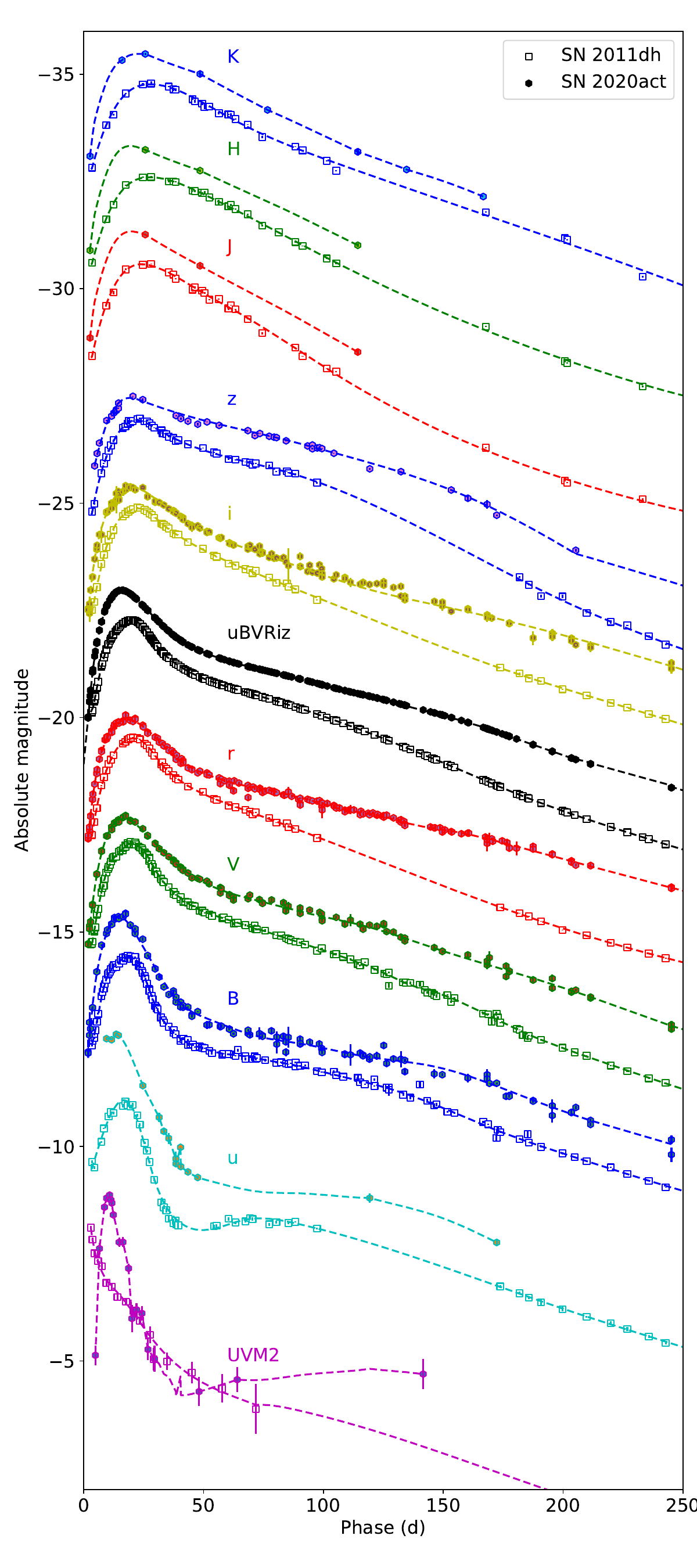}
\caption{Broadband and bolometric light curves until 250 days for SN 2020acat (filled circles) compared to SN 2011dh (unfilled squares). From the bottom to top we show the \textit{UVM2} (magenta), u (cyan), B (blue), V (green), r (red), ugBVriz pseudo-bolometric (black), i (yellow), z (blue), J (red), H (green) and K (blue) light curves, which, for clarity, have been shifted by 6.0, 4.3, 2.0, 0.0, -2.3, -7.7, -10.0, -13.0, -15.0 and -17.0 mags, respectively.}
\label{f_acat_dh_lightcurve_comp}
\end{figure}

Figure \ref{f_acat_dh_lightcurve_comp} shows the optical, NIR and pseudo-bolometric \textit{uBVriz} lightcurves of SN 2020acat compared to SN 2011dh. In the figure we also show cubic spline fits to the data, and for sparsely sampled bands, interpolations in colour. In general, the lightcurves are quite similar, and show a rise to a bell-shaped maximum followed by a tail with a roughly linear decline, characteristic for Type IIb and other stripped-envelope (SE) SNe. The maximum is less pronounced and occurs later for redder bands, and the decline rate on the tail is initially lower for bluer bands but subsequently increase. The maximum is shaped by diffusion of the energy deposited by the radioactive $^{56}$Ni synthesised in the SN explosion, and the tail, where the SN becomes optically thin, by the instant release of this energy. For SN 2011dh there was also an initial decline phase observed by PTF \citep{Arc11}, seen in many Type IIb SNe, and caused by the cooling of their low-mass hydrogen envelopes. This phase is not observed in SN 2020acat and given that the first observation is from $\sim$1 day it has to be short. For a more detailed discussion of the lightcurves of SN 2011dh and Type IIb SNe in general see \citet{Erg14,Erg15} and \citetalias{Erg22}.

However, there are also differences. SN 2020acat is more luminous than SN 2011dh, peaks earlier and decline more slowly on the tail. This is further illustrated by Table \ref{t_lc_char_comp}, where we list the times and magnitudes of the peak as well as the tail decline rates for the pseudo-bolometric \textit{uBVriz} lightcurves. Note that, the tail decline rate is roughly similar in the beginning, then increases for SN 2011dh at $\sim$100 days and subsequently for SN 2020acat at $\sim$150 days, after which it becomes roughly similar again. This is consistent with SN 2020acat becoming optically thin to the $\gamma$-rays later than SN 2011dh. In addition, in Fig.~\ref{f_acat_dh_lightcurve_comp} we see that on the rise to peak luminosity, SN 2020acat is significantly bluer than SN 2011dh, at least if we focus on the optical bands. The evolution in the UV is also quite different, where we see a continues decline in SN 2011dh, but a pronounced diffusion peak in SN 2020acat. This could be related to the much shorter cooling phase for SN 2020acat. Another clear difference is the evolution of the r-band in the nebular phase, which is a directly related to the evolution of the [\ion{O}{i}] 6300,6364 \AA~lines. 

\begin{table*}[tb]
\caption{Lightcurve characteristics for the pseudo-bolometric \textit{uBVriz} lightcurves of SN 2020acat and SN 2011dh measured from cubic spline fits.}
\begin{center}
\begin{tabular}{llllll}
\hline\hline \\ [-1.5ex]
SN & Maximum & Bolometric magnitude & Decline rate (75 d) & Decline rate (125 d ) & Decline rate (200 d)\\ [0.5ex]
& (days) & (mag) & (mag day$^{-1}$) & (mag day$^{-1}$) & (mag day$^{-1}$)\\
\hline \\ [-1.5ex]
2020acat & 16.00 & -17.27 & 0.014 & 0.015 & 0.018\\
2011dh & 20.01 & -16.57 & 0.016 & 0.022 & 0.020\\ [0.5ex]
\hline
\end{tabular}
\end{center}
\label{t_lc_char_comp}
\end{table*}

\subsubsection{Spectra}

\begin{figure*}[tbp!]
\includegraphics[width=1.0\textwidth,angle=0]{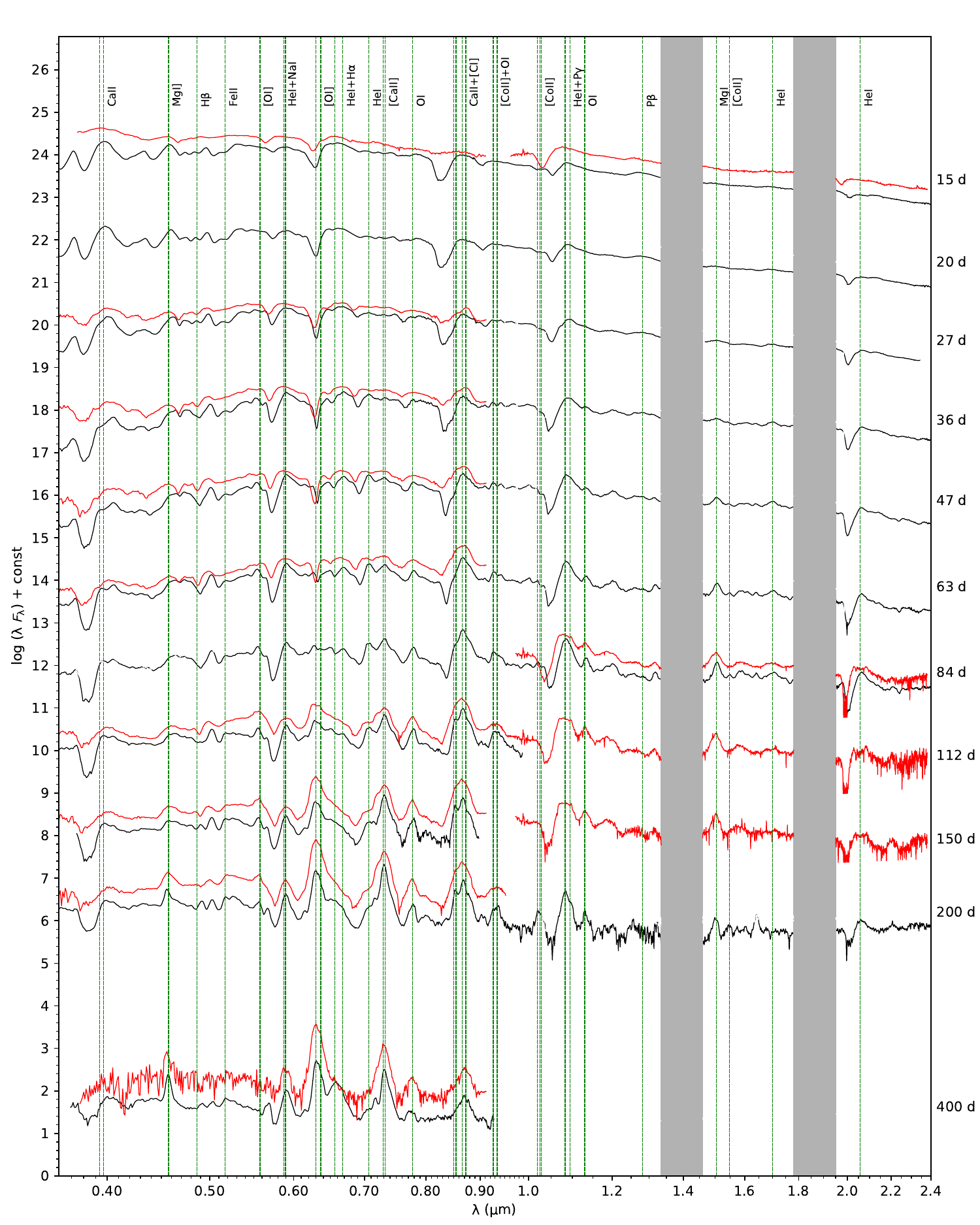}
\caption{Spectral evolution of SN 2020acat (red) compared to SN 2011dh (black). Spectra from ten logarithmically spaced epochs between 15 and 200 days and a single epoch at 400 days are shown. In addition, the rest wavelengths of the most important lines are shown as dashed green lines.}
\label{f_acat_dh_spec_evo_comp}
\end{figure*}

Figure \ref{f_acat_dh_spec_evo_comp} shows the optical and NIR spectral evolution of SN 2020acat compared to SN 2011dh. In this and all following figures, the spectra have been time-interpolated as described in \citet{Erg14}. If not otherwise stated, we only show interpolated spectra that have observed counterparts close in time. In general, the spectra are quite similar, showing the transition from a hydrogen dominated to a helium dominated spectrum, characteristic of Type IIb SNe. H$\alpha$ is initially the strongest line but gradually disappears on the rise to the peak, whereas absorption in H$\alpha$ and H$\beta$ lines remains for a longer time. The helium lines appear on the rise to the peak, and during the decline to the tail they grow strong. The spectra also show lines from heavier elements, in particular the \ion{Ca}{ii} 3934,3968 \AA~and \ion{Ca}{ii} 8498,8542,8662 \AA~lines (hereafter \ion{Ca}{ii} HK and \ion{Ca}{ii} NIR triplet), which are present throughout most of the evolution, and the forbidden [\ion{Ca}{ii}] 7291,7323 \AA~and [\ion{O}{i}] 6300,6364 \AA~lines, which become the strongest lines during the nebular phase. For a more detailed discussion of the spectra of SN 2011dh and Type IIb SNe in general see \citet{Erg14,Erg15}, \citet{Jer15} and \citetalias{Erg22}.

\begin{figure}[tb]
\includegraphics[width=0.5\textwidth,angle=0]{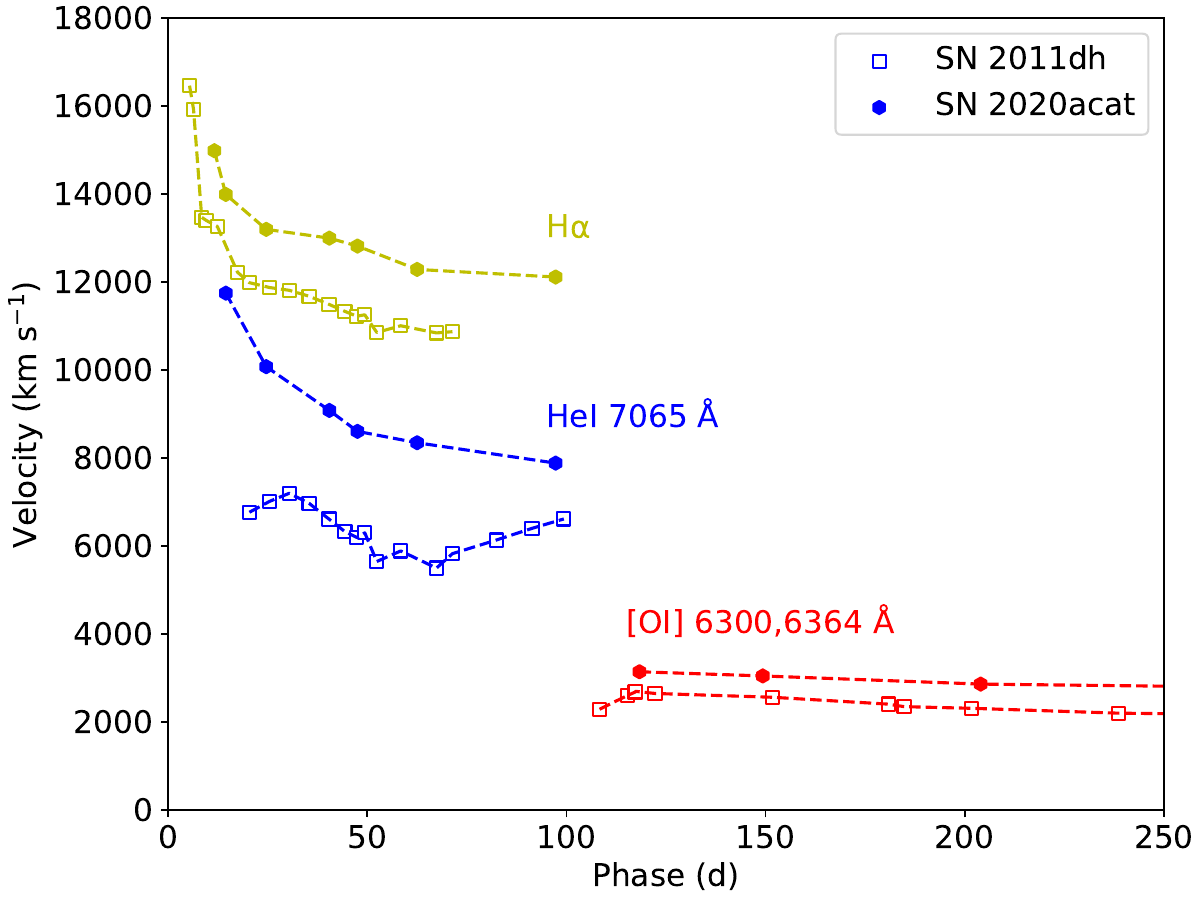}
\caption{Comparison of the evolution of the absorption minimum for the H$\alpha$ (yellow) and \ion{He}{i} 7065 \AA~(blue) lines and the HWHM of the [\ion{O}{i}] 6300,6364 \AA~doublet for SN 2020acat (filled circles) and SN 2011dh (empty squares).}
\label{f_acat_dh_vel_comp}
\end{figure}

However, there are also differences, and the lines of SN 2020acat are broader and the velocities higher. This is further illustrated by Fig.~\ref{f_acat_dh_vel_comp} where we show the velocity evolution of the absorption minimum for the H$\alpha$ and \ion{He}{i} 7065 \AA~lines and the half-width half-maximum (HWHM) velocity for the [\ion{O}{i}] 6300,6364 \AA~lines for SNe 2020acat and 2011dh. The asymptotic H$\alpha$ velocity, which likely corresponds to the interface between the helium core and hydrogen envelope \citep[see][]{Erg14,Erg18} is $\sim$12000 km s$^{-1}$ for SN 2020acat compared to $\sim$11000 km s$^{-1}$ for SN 2011dh. The \ion{He}{i} 7065 \AA~velocity, which may be thought of as a representative for the helium envelope, is 38 percent higher (on average) for SN 2020acat, and the HWHM velocity of the [\ion{O}{i}] 6300,6364 \AA~line, which may be thought of as a representative for the carbon-oxygen core, is 20 percent higher (on average) for SN 2020acat. We also measured the velocity of the absorption minimum for the \ion{O}{i} 7774 \AA~line and the HWHM velocity of the [\ion{Ca}{ii}] 7291,7323 \AA~line, which are 30 and 33 percent higher (on average) for SN 2020acat, respectively.

\begin{figure}[tb]
\includegraphics[width=0.5\textwidth,angle=0]{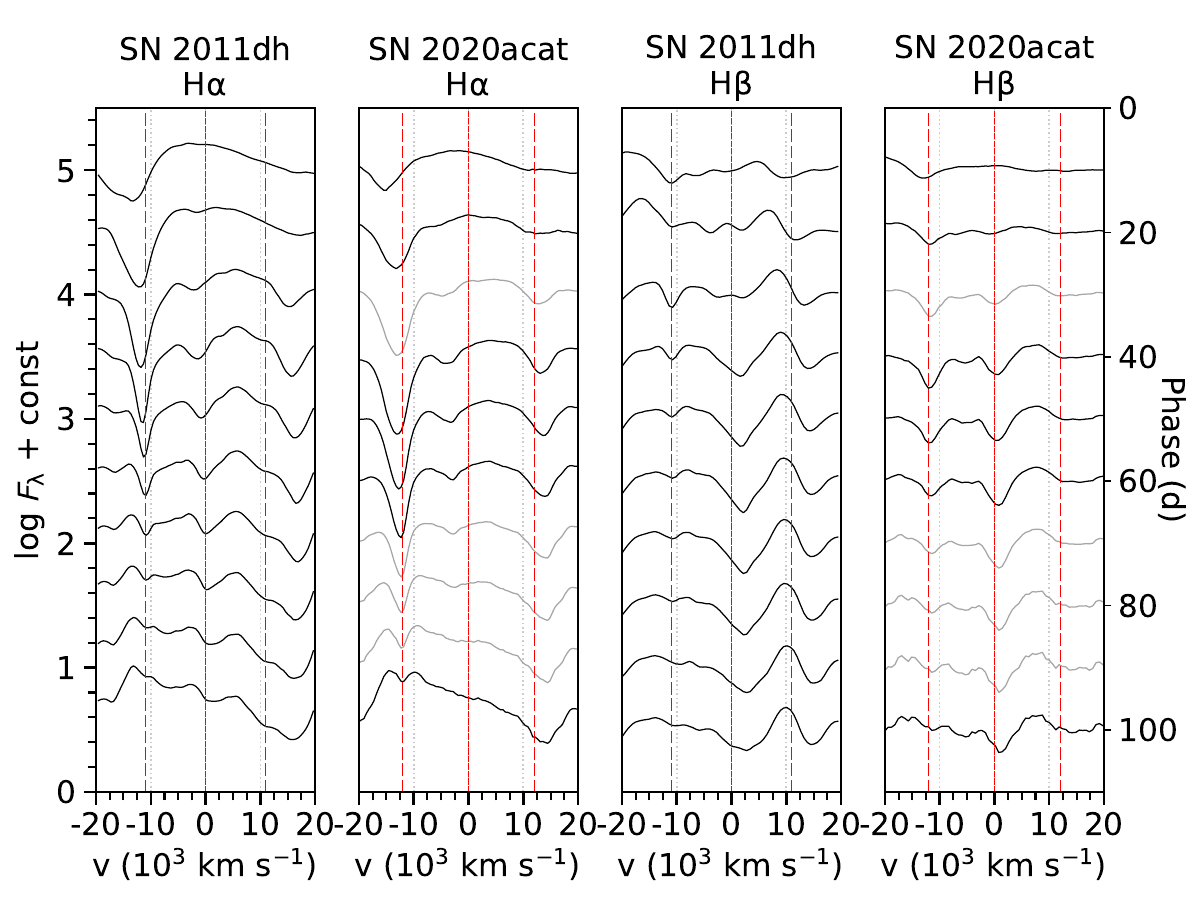}
\caption{Comparison of the evolution of the H$\alpha$ and H$\beta$ lines in SN 2020acat and SN 2011dh. Spectra from 10 equally spaced epochs between 10 and 100 days are shown. 
Interpolated spectra that have no observed counterpart close in time are shown in gray.
The inferred helium-hydrogen interface velocities of SNe 2011dh (11000 km s$^{-1}$) and 2020acat (12000 km s$^{-1}$) are shown as dashed red lines.}
\label{f_acat_dh_spec_comp_H}
\end{figure}

In Fig.~\ref{f_acat_dh_spec_comp_H} we provide a closeup of the H$\alpha$ and H$\beta$ lines for SN 2020acat and SN 2011dh. Similar to the H$\alpha$ line, the velocity of the H$\beta$ line is higher in SN 2020acat, and the asymptotic velocity of the absorption minimum approaches 12000 km s$^{-1}$ for both lines. It also appears that the hydrogen signature is a bit stronger. In particular, these lines remain in absorption for a longer time in SN 2020acat. Whereas the H$\alpha$ line disappears in absorption at $\sim$80 days in SN2011dh, it remains in absorption at 100 days in SN 2020acat.

In Fig.~\ref{f_acat_dh_spec_comp_H} we provide a closeup of the \ion{He}{i} 5876 \AA~and \ion{He}{i} 1.083 $\mu m$ lines between 10 and 150 days for SN 2020acat and SN 2011dh. Similar to the \ion{He}{i} 7065 \AA~line, the velocities of these lines are higher in SN 2020acat, in particular at early times, and in particular for the \ion{He}{i} 1.083 $\mu m$ line.  This line is also much stronger in SN 2020acat at early times. As discussed by \citetalias{Med23}, at late times the \ion{He}{i} 1.083 $\mu m$ line (as well as the \ion{He}{i} 2.058 $\mu m$ line) attain a very distinct flat-topped shape for SN 2020acat, which is not seen in SN 2011dh.

\begin{figure}[tb]
\includegraphics[width=0.5\textwidth,angle=0]{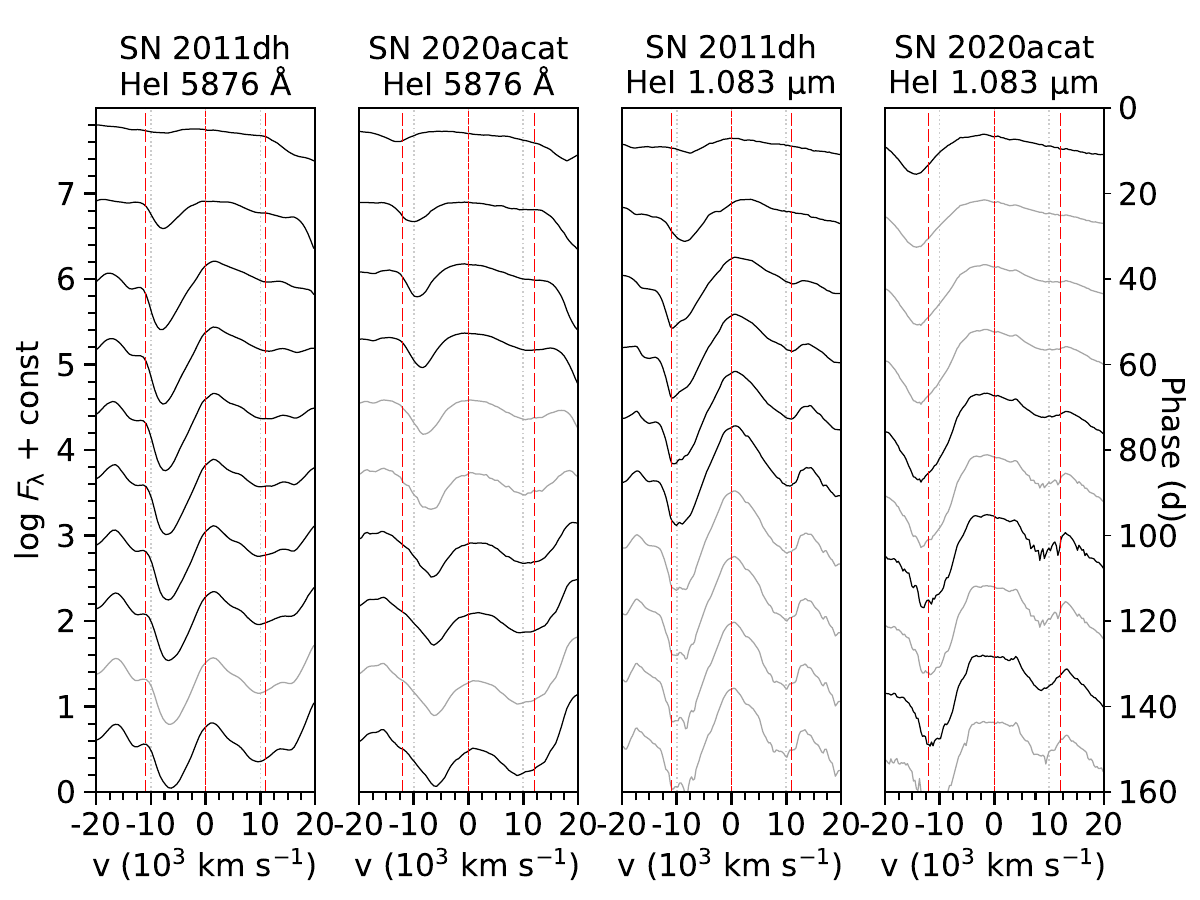}
\caption{Comparison of the evolution of the \ion{He}{i} 5876 \AA~and \ion{He}{i} 10830 \AA~lines in SN 2020acat and SN 2011dh. Spectra from 10 equally spaced epochs between 10 and 150 days are shown.
Otherwise the same as in Fig.~\ref{f_acat_dh_spec_comp_H}.}
\label{f_acat_dh_spec_comp_He}
\end{figure}

\begin{figure}[tb]
\includegraphics[width=0.5\textwidth,angle=0]{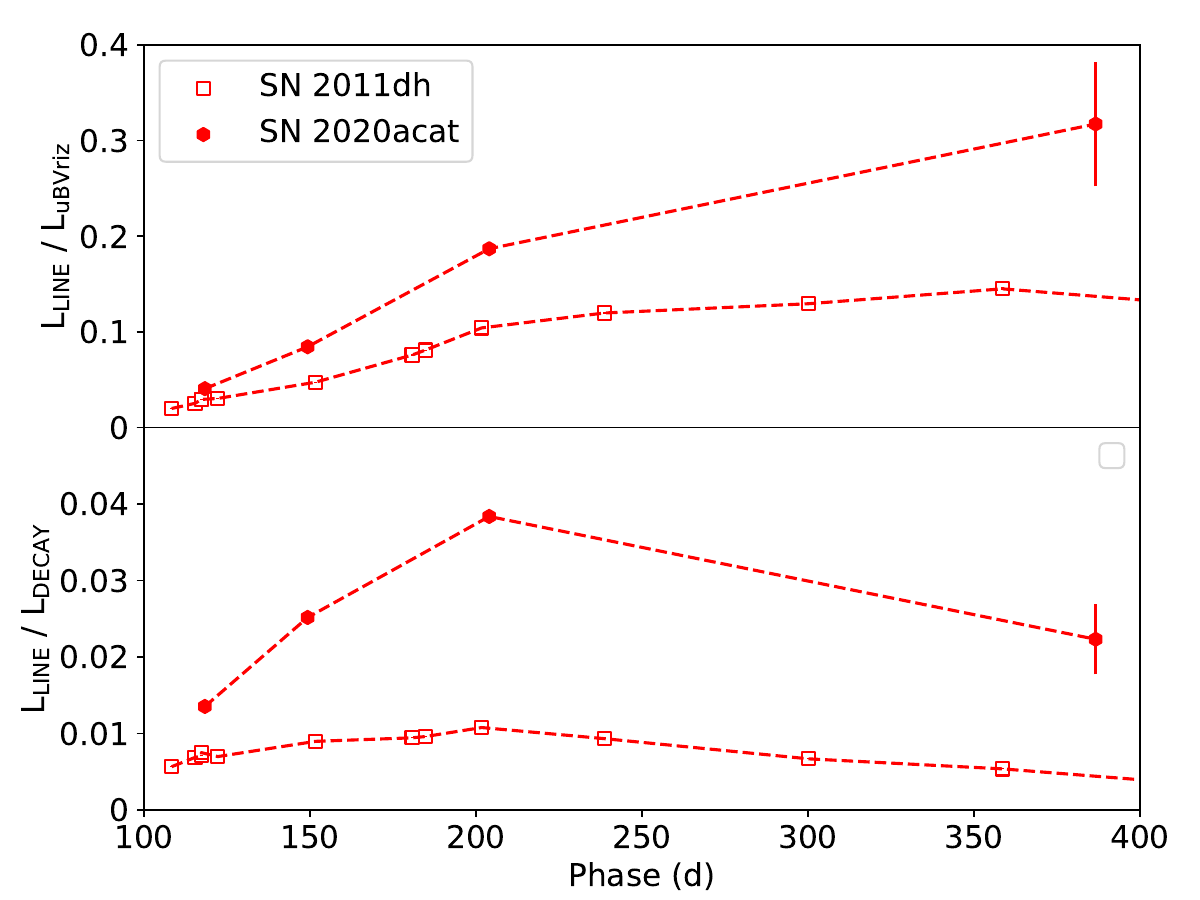}
\caption{Comparison of the evolution of the [\ion{O}{i}] 6300,6364 \AA~luminosity normalized with the pseudo-bolometric \textit{uBVriz} luminosity (upper panel) and the $^{56}$Ni decay chain luminosity (lower panel) for SN 2020acat (filled circles) and SN 2011dh (empty squares).}
\label{f_acat_dh_flux_comp}
\end{figure}

It is also clear that the [\ion{O}{i}] 6300,6364 \AA~line, which is a tracer of the initial mass of the progenitor \citep[see e.g][]{Jer15}, is stronger in SN 2020acat. This is further illustrated in Fig~\ref{f_acat_dh_flux_comp}, where we show the [\ion{O}{i}] 6300,6364 \AA~luminosity normalized with the pseudo-bolometric \textit{uBVriz} luminosity and the luminosity of the $^{56}$Ni decay chain for SN 2020acat and SN 2011dh. The line luminosity was measured with the same method as in \citet{Jer15} to allow for a comparison with fig.~15 in that paper. Relative to the pseudo-bolometric \textit{uBVriz} luminosity, the [\ion{O}{i}] 6300,6364 \AA~luminosity is 2.0 times higher (on average after 150 days) for SN 2020acat, and relative to the $^{56}$Ni decay chain luminosity it is 3.8 times higher (on average after 150 days) for SN 2020acat.

Finally, there are also some other differences worth mentioning. First, the evolution of the \ion{Ca}{ii} HK and NIR triplet lines  differ. Early on these lines are absent in SN 2020acat, and later on they are much weaker in absorption in SN 2020acat. Second, the quite strong [\ion{N}{ii}] 6548,6583 \AA~lines emerging on the red shoulder of the [\ion{O}{i}] 6300,6364 \AA~lines towards $\sim$300 days in SN 2011dh \citep[see][]{Jer15} seem to be absent or at least much weaker in SN 2020acat.

\subsection{Estimates of the SN parameters}
\label{s_educated_guess}

We may attempt to use the comparison for an educated guess of how the SN parameters scale between SN 2011dh and SN 2020acat. In \citet{Erg15b} we fitted scaling relations for the SN parameters as a function of the observed quantities to a large grid of hydrodynamical SN models \citep[see also][]{Erg15}. These were as follows:

\begin{equation}
\mathrm{log~M_\mathrm{ej} = -3.42 + 1.81~log~t_\mathrm{m} - 0.18~log~L_\mathrm{m} + 1.47~log~v_\mathrm{m}}
\label{eq_dep_1}
\end{equation}

\begin{equation}
\mathrm{log~E_\mathrm{ej} = -3.95 + 0.75~log~t_\mathrm{m} - 0.07~log~L_\mathrm{m} + 2.90~log~v_\mathrm{m}}
\label{eq_dep_2}
\end{equation}

\begin{equation}
\mathrm{log~M_\mathrm{Ni} = -4.96 + 2.08~log~t_\mathrm{m} + 0.93~log~L_\mathrm{m} + 1.19~log~v_\mathrm{m}}
\label{eq_dep_3}
\end{equation}
where $\mathrm{t_{m}}$, $\mathrm{L_{m}}$ and $\mathrm{v_{m}}$ are the time, luminosity and photospheric velocity at the maximum. Measurements of $\mathrm{t_{m}}$ and $\mathrm{L_{m}}$ for SN 2020acat and SN 2011dh are listed in Table~\ref{t_lc_char_comp}. Measuring $\mathrm{v_m}$ from Fig.~\ref{f_acat_dh_vel_comp} using the \ion{He}{i} 7065 \AA~line as a proxy for the photosphere, Eqs.~\ref{eq_dep_1}-\ref{eq_dep_3} give scale factors of 1.0, 2.1 and 1.7 for $\mathrm{M_{ej}}$, $\mathrm{E_{ej}}$ and $\mathrm{M_{Ni}}$, respectively, compared to SN 2011dh. Applying these scaling factors to the optimal model for SN 2011dh from \citetalias{Erg22}, we get M$_\mathrm{ej}$=1.7 M$_\odot$, E$_\mathrm{ej}$=1.4 B and M$_\mathrm{Ni}$=0.13 M$_\odot$. This is qualitatively similar to the results in \citetalias{Med22} using the Arnett model; as compared to SN 2011dh, the ejecta mass of SN 2020acat seems to be similar, whereas the kinetic energy of the ejecta and the mass of $^{56}$Ni seem to be much higher. 

However, as is evident from Fig.~\ref{f_acat_dh_flux_comp}, the strength of the [\ion{O}{i}] 6300,6364 \AA~lines points in another direction, suggesting a considerably higher oxygen mass in SN 2020acat, corresponding to a considerably larger ejecta mass (given that we assume the progenitor to be an almost bare helium core). Assuming everything else equal, the [\ion{O}{i}] 6300,6364 \AA~luminosity normalized with the pseudo-bolometric luminosity would just scale with the fractional oxygen mass, which would then be 2.0 times higher. This corresponds to an $\sim$2 times higher ejecta mass, and an initial mass of $\sim$17 M$_\odot$ in the \citet{Woo07} models. This is again qualitatively in line with the results by \citetalias{Med22}, who used one-zone NLTE modelling to find an oxygen mass of $\sim$1~M$_\odot$ corresponding to an initial mass of 16-17 M$_\odot$.

Our educated guess for the SN parameters provides a starting point for the modelling with JEKYLL and a guideline for the SN models. Given the inconclusive results for the initial mass, we treat that as a free parameter, whereas the velocities of the interfaces between the carbon-oxygen core, helium envelope and hydrogen envelope and the mass of $^{56}$Ni are kept fixed based on the comparison. Instead, we take the opportunity to explore the parameters of the macroscopic mixing, which are not well constrained, as well as the properties of the hydrogen envelope.

\section{Methods and models}
\label{s_methods_and_models}

The SN models presented in this work were calculated with the JEKYLL code, which is described in detail in \citetalias{Erg18} and \citetalias{Erg22}. Here, we briefly repeat the general methods used in JEKYLL. The configuration of JEKYLL and the atomic data used are described in Appendix \ref{a_configuration} and \ref{a_atomic_data}, respectively.

Like in \citetalias{Erg22}, the ejecta models are phenomenological models based on results from hydrodynamical modelling and the observed velocities of the ejecta. For the comparison with SN 2020acat we present a set of models differing in initial mass, radial mixing and expansion of the radioactive material, and the mass and mass-fraction of hydrogen in the hydrogen envelope.

\subsection{JEKYLL}
\label{s_method_general}

JEKYLL is a light curve and spectral-synthesis code based on a Monte-Carlo (MC) method for the time-dependent 3D radiative transfer developed by \citet{Luc02,Luc03,Luc05}, and extended as described in \citetalias{Erg18}. To calculate the radiation field and the state of matter\footnote{With `state of matter' we refer to the temperature and the populations of ionised and excited states.} an iterative procedure is used, which is similar to an accelerated $\Lambda$-iteration (see discussion in \citetalias{Erg18}). The statistical and thermal equilibrium equations are solved taking into account all relevant processes. In particular, this includes heating, excitation and ionisation by non-thermal electrons calculated using the method by \citet{Koz92}. In the inner region, where the matter and radiation field are assumed to be coupled, we use a diffusion solver to calculate the temperature. 

JEKYLL also takes into account the macroscopic mixing of the ejecta by use of the virtual grid method \citep{Jer11}, in which the fragmentation of the ejecta due to hydrodynamical instabilities is represented by spherical clumps characterized by their composition, density, size and filling factor. The clumps are drawn based on their filling factor and geometrical cross-section as the MC packets propagate through the ejecta, and are virtual in the sense that they only exist as long as a MC packet propagates through them.

The main limitations in JEKYLL are the assumptions of homologous expansion, thermal and statistical equilibrium, and a spherically symmetric distribution of the matter. The latter is, however, only assumed on large scales and on average, and small-scale asymmetries are taken into account through the virtual grid method. Another important limitation is the lack of a treatment of the ejecta chemistry (i.e. molecules and dust).

\subsection{Ejecta models}
\label{s_ejecta_model}

The ejecta models are based on SN models by \citet{Woo07} for non-rotating single stars at solar metallicity with initial masses of 13, 15, 17, 19 and 21 M$_\odot$, from which the helium core has been carved out and the masses and abundances for the compositional layers adopted. The stellar models were evolved to the verge of core-collapse and exploded with an energy of 1.2 B using the 1-D code KEPLER. Note, that the evolution depends on the assumed stellar parameters (no rotation and solar metallicity) as well as on the assumed progenitor system (single star). Note also, that the explosive nucleosynthesis and the amount of fallback onto the remnant depends on the assumed explosion energy (1.2 B) as well as the simplified 1-D explosion treatment in KEPLER.

In our ejecta models, the carbon-oxygen core is assumed to have a constant (average) density, and the helium envelope to have the same (average) density profile as the best-fit model for SN 2011dh by \citet{Ber12}. In addition, a low-mass hydrogen envelope based on models by \citet{Woo94} is attached. Based on the comparison with SN 2011dh (Sect.~\ref{s_comparison_to_11dh}) and our previous successful model of this SN, the velocities of the interfaces between the carbon-oxygen core, the helium envelope and the hydrogen envelope are set to 4200 and 12000 km s$^{-1}$, respectively, and the explosive nucleosynthesis adjusted to match a $^{56}$Ni mass of 0.13 M$_\odot$ (see below). It should be emphasised that the models are not self-consistent hydrodynamical models, but rather phenomenological models based on results from hydrodynamical simulations and the observed velocities and luminosity of SN 2020acat.

Based on the original onion-like compositional structure, we identify five compositional zones (O/C, O/Ne/Mg, O/Si/S, Si/S, and Ni/He) in the carbon-oxygen core and two compositional zones (He/N and He/C) in the helium envelope. The explosive nucleosynthesis is adjusted by scaling the mass of the Ni/He zone, whereas the Si/S and O/Si/S zones (which are also affected by the explosive nucleosynthesis) are left untouched. To mimic the mixing of the compositional zones in the explosion, three scenarios with different degrees of mixing of the radioactive material (weak, medium, and strong) are explored. In the weak mixing scenario, the core is homogeneously mixed, but no core material is mixed into the envelope. In the medium mixing scenario, 50 percent of the radioactive Ni/He material is mixed into the inner helium envelope, and in the strong mixing scenario, 20 percent of this is mixed further into the outer helium envelope. The other material in the core is \emph{not} mixed into the helium envelope in any of these scenarios, which is a simplification.

In our parametrization, given the mass-fractions of the compositional zones, the clumping geometry is determined by the sizes (or masses) of the clumps and their filling factors \citepalias[see][]{Erg22}. As discussed in Sect.~\ref{s_macro_type_IIb}, the constraints on the clumping geometry in Type IIb SNe are rather weak, in particular with respect to the helium envelope. In this work we assume a clump mass of $2.8 \times 10^{-5}$ M$_\odot$ and explore three scenarios with different amounts of expansion (none, medium and strong) of the radioactive material. In the medium expansion scenario, we assume a density contrast factor between the expanded and compressed material of 10 in the core and 5 in the helium envelope, and in the strong expansion scenario, we assume a density contrast factor of 60 in the core and 30 in the helium envelope. The main reason to keep the clump mass fixed is to limit the computational cost (which is considerable). Note, however, that the clump mass mainly affects the effective opacity, as the decrease of that in a clumpy medium disappears when the clumps become optically thin \citepalias[see][]{Erg22}. It is therefore somewhat degenerate with the expansion of the radioactive material, which further motivates our choice to keep one of these parameters fixed. 

We also investigate the effect of the mass and mass-fraction of hydrogen in the  hydrogen envelope (which together determine the total mass of the hydrogen envelope), and explore three different masses (low, medium and high), and two different mass-fractions (low and medium). The medium scenario corresponds to an hydrogen mass of 0.027 M$_\odot$ and X$_\mathrm{H}$=0.54. The low and high hydrogen mass scenarios corresponds to 0.0135 and 0.054 M$_\odot$, and the low mass-fraction scenario to X$_\mathrm{H}$=0.27. Our set of models thus differ in initial mass, radial mixing and expansion of the radioactive material, and the mass and mass-fraction of the hydrogen in the hydrogen envelope. All models are listed in Table \ref{t_ejecta_models} and a detailed description of each model is given in Appendix \ref{a_ejecta_models}.

In addition to this set of models, which is used to constrain the model parameters, we have calculated a few variants on the M17-s-m model varying in the metallicity and the mass of $^{56}$Ni as well as a model with very strong expansion in the core (a contrast factor of 210). These models are listed in Table \ref{t_ejecta_models_additional}, and are referred to in Sect~\ref{s_comparison_to_11dh}, where we make a detailed comparison of our optimal model with observations of SN 2020acat.

\begin{table*}[tb]
\caption{Main set of ejecta models. For each model we list the initial mass, the ejecta mass and (kinetic) energy, the radial mixing and the expansion of the radioactive material, and the mass and mass-fraction of hydrogen in the hydrogen envelope.}
\begin{center}
\begin{tabular}{llllllll}
\toprule
Model & M$_\mathrm{ZAMS}$ (M$_\odot$) & M$_\mathrm{ej}$ (M$_\odot$) & E$_\mathrm{ej}$ (B) & Radial mixing & Expansion & M$_\mathrm{H}$ (M$_\odot$) & X$_\mathrm{H}$\\
\midrule
M13-m-s & 13 & 2.1 & 0.82 & medium & strong & 0.027 & 0.54 \\
M15-m-s & 15 & 2.6 & 0.92 & medium & strong & 0.027 & 0.54 \\
M17-w-n & 17 & 3.5 & 1.0 & weak & none & 0.027 & 0.54 \\
M17-m-n & 17 & 3.5 & 1.0 & medium & none & 0.027 & 0.54 \\
M17-s-n & 17 & 3.5 & 1.0 & strong & none & 0.027 & 0.54 \\
M17-m-m & 17 & 3.5 & 1.0 & medium & medium & 0.027 & 0.54 \\
M17-w-s & 17 & 3.5 & 1.0 & weak & strong & 0.027 & 0.54 \\
M17-m-s & 17 & 3.5 & 1.0 & medium & strong & 0.027 & 0.54 \\
M17-s-s & 17 & 3.5 & 1.0 & strong & strong & 0.027 & 0.54 \\
M17-s-s-H-l & 17 & 3.5 & 0.95 & medium & strong & 0.0135 & 0.54  \\
M17-s-s-H-h & 17 & 3.6 & 1.1 & medium & strong & 0.054 & 0.54 \\
M17-s-s-XH-l & 17 & 3.6 & 1.1 & medium & strong & 0.027 & 0.27\\
M19-m-s & 19 & 4.5 & 1.3 & medium & strong & 0.027 & 0.54 \\
M21-m-s & 21 & 5.4 & 1.4 & medium & strong & 0.027 & 0.54 \\
\bottomrule
\end{tabular}
\end{center}
\label{t_ejecta_models}
\end{table*}

\begin{table*}[tb]
\caption{Additional set of ejecta models All are based on the M17-m-s model and has an initial mass of 17 M$_\odot$, medium mixing, strong expansion if not otherwise stated, M$_\mathrm{H}$=0.027 and X$_\mathrm{H}$=0.54.}
\begin{center}
\begin{tabular}{ll}
\toprule
Model & Description\\
\midrule
M17-m-s-z-smc & Same as M17-m-s but with SMC metallicity\\
M17-m-s-z-lmc & Same as M17-m-s but with LMC metallicity\\
M17-m-s-Ni-l & Same as M17-m-s but with a $^{56}$Ni mass of 0.1 M$_\odot$\\
M17-m-s-Ni-h & Same as M17-m-s but with a $^{56}$Ni mass of 0.15 M$_\odot$\\
M17-m-vs & Same as M17-m-s but with a contrast factor of 210 in the core \\
\bottomrule
\end{tabular}
\end{center}
\label{t_ejecta_models_additional}
\end{table*}

\subsection{Macroscopic mixing in Type IIb SNe}
\label{s_macro_type_IIb}

Our knowledge of the macroscopic mixing in Type IIb SNe is limited, but there are some constraints, although they are generally weak. Some insights might also be gained from other types of SNe, not the least from SN 1987A.

For SN 1987A a filling factor of 0.2 was estimated for the Ni/He material in the core by \citet{Koz98b} using MIR fine-structure Fe lines, and a filling factor of 0.1 was estimated for the oxygen-rich material in the core by \citet{Spy91} using the optical depth of the [\ion{O}{i}] 6300,6364 \AA~lines. Given the core model for SN 1987A by \citet{Jer11}, this corresponds to an expansion factor of $\sim$10 for the Ni/He material, a compression factor of $\sim$5 for the oxygen-rich material, and a contrast factor of $\sim$50. Based on a similar line of arguments \citet{Jer12} found a density contrast of $\sim$30 between the Ni/He material and the oxygen-rich material in the core of the Type IIP SN 2004et.

Due to differences in the progenitor structure, this does not necessarily apply to Type IIb SNe. In particular, the hydrodynamical instabilities near the interface between the helium and hydrogen envelope are expected to be weaker in a Type IIb SN. However, a high density contrast in the core is consistent with constraints on the filling factor of the oxygen-rich material ($0.02<\Phi<0.07$) derived for SN 2011dh from small-scale variations in the [\ion{O}{i}] 6300, 6364 \AA~and \ion{Mg}{i}] 4571 \AA~line profiles \citep{Erg15} and the optical depth of the [\ion{O}{i}] 6300, 6364 \AA~lines \citep{Jer15}. The cavities observed in the Type IIb SN remnant Cas A also seem to indicate a considerable expansion of the radioactive material, even at high velocities \citep{Mil13b,Mil15}. Overall, however, the constraints on the expansion of the radioactive material in Type IIb SNe are weak, in particular with respect to the helium envelope.

For SN 1987A the number of clumps in the oxygen-rich zones in the core was estimated to $\sim$2000 by \citet{Chu94}, who used a statistical model to analyse small-scale variations in the [\ion{O}{i}] 6300, 6364 \AA~line profiles. Given the core-model for SN 1987A by \citet{Jer15}, this corresponds to a clumps mass of $\sim$10$^{-3}$ M$_\odot$. However, as for the contrast factor, this does not necessarily apply to Type IIb SNe. Applying the \citet{Chu94} model to SN 2011dh, \citet{Erg15} found a lower limit on the number of clumps in the O/Ne/Mg zone in the core of $\sim$900 from small-scale variations in the [\ion{O}{i}] 6300, 6364 \AA~and \ion{Mg}{i}] 4571 \AA~line profiles. A similar limit was derived for SN 1993J by \citet{Mat00} using the same statistical model. Given the core-model of SN 2011dh from \citet{Jer15}, the former limit corresponds to an upper limit on the clump mass of $\sim$$1.5 \times 10^{-4}$ M$_\odot$. To our best knowledge there are no constraints on the sizes of the clumps in the helium envelope, and overall, the constraints on the sizes of the clumps in Type IIb SNe are weak.

The extent of the mixing in Type IIb SNe is better constrained, and most lightcurve modelling require mixing of the He/Ni material far out in the helium envelope to reproduce the rise to peak luminosity \citep[e.g.][]{Erg15,Tad17}. Note, however, that such modelling typically ignore the opacity increase such mixing give rise to in the envelope, a limitation not present in our JEKYLL simulations. Extensive mixing is also supported by explosion modelling \citep[e.g.][]{Wong17}, and by the distribution of O- and Si-burning products in Cas A \citep[e.g.][]{Wil02}. In this works we assume that the mixing is macroscopic, which is supported by both theoretical arguments \citep[e.g.][]{Fry91}, and observations of SNe \citep[e.g.][]{Fra89}, and SNRs \citep[e.g.][]{Enn06}. In \citetalias{Erg22}, we discussed this issue in more detail, and showed that microscopically mixed models of SN 2011dh give a very poor match to the [\ion{Ca}{ii}] 7291, 7323 \AA~and [\ion{O}{i}] 6300, 6364 \AA~lines in the nebular phase.

\subsection{SN models} 
\label{s_sn_models}

The ejecta models described in Sect.~\ref{s_ejecta_model} were first (homologously) rescaled to one day. Based on an initial temperature profile, the SN models were then evolved with JEKYLL using 135 logarithmically spaced time steps to 501 days. The SN models were calculated using a frequency grid of 5000 logarithmically spaced intervals between 10 \AA~and 20 $\mu$m, and each model required $\sim$9000 CPU (Central Processing Unit) hours, which, using 128 CPUs, resulted in a computing time of $\sim$3 days. The initial temperature profile was taken from a HYDE \citep{Erg15} SN model for a 5 M$_\odot$ bare helium core exploded with an energy of 1.1 B and ejecting 0.13 M$_\odot$ of $^{56}$Ni. As this SN model was based on a bare helium core, the cooling of the thermal explosion energy, lasting for a few days in a model with a hydrogen envelope, is ignored. The subsequent evolution is powered by the continuous injection of radioactive decay energy, and the choice of initial temperature profile is not critical, although it may have some effect on the early evolution.

Note, that there is a switch in the JEKYLL setup at 100 days, when charge-transfer and a more extended scheme for non-thermal excitation is turned on (See Appendix~\ref{a_configuration}). This can be visible as a slight shift in some of the model lightcurves. Note also, that some MC noise is present in the models, and in order to reduce that gentle smoothing has been applied in some figures.

\section{Comparisons to observations}
\label{s_comparison_to_observations}

We now proceed by comparing our JEKYLL models to the observations of SN 2020acat. First, in Sect. \ref{s_constrain} we use the comparison to constrain the parameters of the model; the initial mass, the mixing and expansion of the radioactive material, and the mass and mass-fraction of hydrogen in the hydrogen-envelope. Then, in Sect. \ref{s_optimal_model} we compare the spectra and lightcurves of SN 2020acat in more detail to our optimal model, and discuss remaining differences and their possible origin.

\subsection{Constraining the model parameters}
\label{s_constrain}

It is important to point out that as a full scan of parameter space is not computationally feasible, and as several limitations exist even in advanced SN models, we can not hope for a perfect match. We should rather use a set of well motivated key measures to search for a model that best fits the observations. To constrain the parameters of our models we therefore apply five criteria, three for the properties of the helium core, and two for the properties of the hydrogen envelope.

First, to constrain the properties of the helium core, i.e. the initial mass and the mixing and expansion of the radioactive material, the optimal model should show the best overall match to the flux in the [\ion{O}{i}] 6300, 6364 \AA~lines in the nebular phase and the pseudo-bolometric lightcurve in both the diffusion and tail phase. These are all well established criteria that have been used in a wide range of cases, and are also well motivated from a physical point of view. In the nebular phase, the flux of the [\ion{O}{i}] 6300, 6364 \AA~lines provides a measure of the oxygen mass, which is related to the helium core mass in our models. In the diffusion phase, the bolometric lightcurve provides a measure of the diffusion time for thermal radiation, whereas in the tail phase, it provides a measure of the optical depth to the $\gamma$-rays. Both of these are related to the ejecta mass, which in turn is related to the helium core mass in our models. In addition, the diffusion time is related to the expansion of the radioactive material \citepalias[see][]{Erg22}, whereas the optical depth to the $\gamma$-rays is related to the mixing of this material. The capabilities of the JEKYLL code to model both the photospheric and nebular phase allows us to apply these three criteria in a self-consistent way based on highly sophisticated physics. 

Second, to constrain the properties of the hydrogen envelope, the optimal model should show the best match to the hydrogen and helium lines in the photospheric phase. The strength and shape of these lines are related to the optical depths of these lines in the hydrogen envelope, which in turn are related to the mass of hydrogen and helium in the  hydrogen envelope. As the hydrogen envelope in a Type IIb has a relatively low mass and soon gets more or less transparent, it does not have a significant impact on the other key quantities, and can be constrained separately.

\subsubsection{The helium core}

To explore the properties of the helium core we use the diffusion phase pseudo-bolometric lightcurve, the tail phase pseudo-bolometric lightcurve and the nebular phase [\ion{O}{i}] 6300, 6364 \AA~flux. First, the latter two are used two constrain the initial mass of the progenitor, and then the former is used to constrain the mixing and expansion of the radioactive material.

\begin{figure}[tbp!]
\includegraphics[width=0.49\textwidth,angle=0]{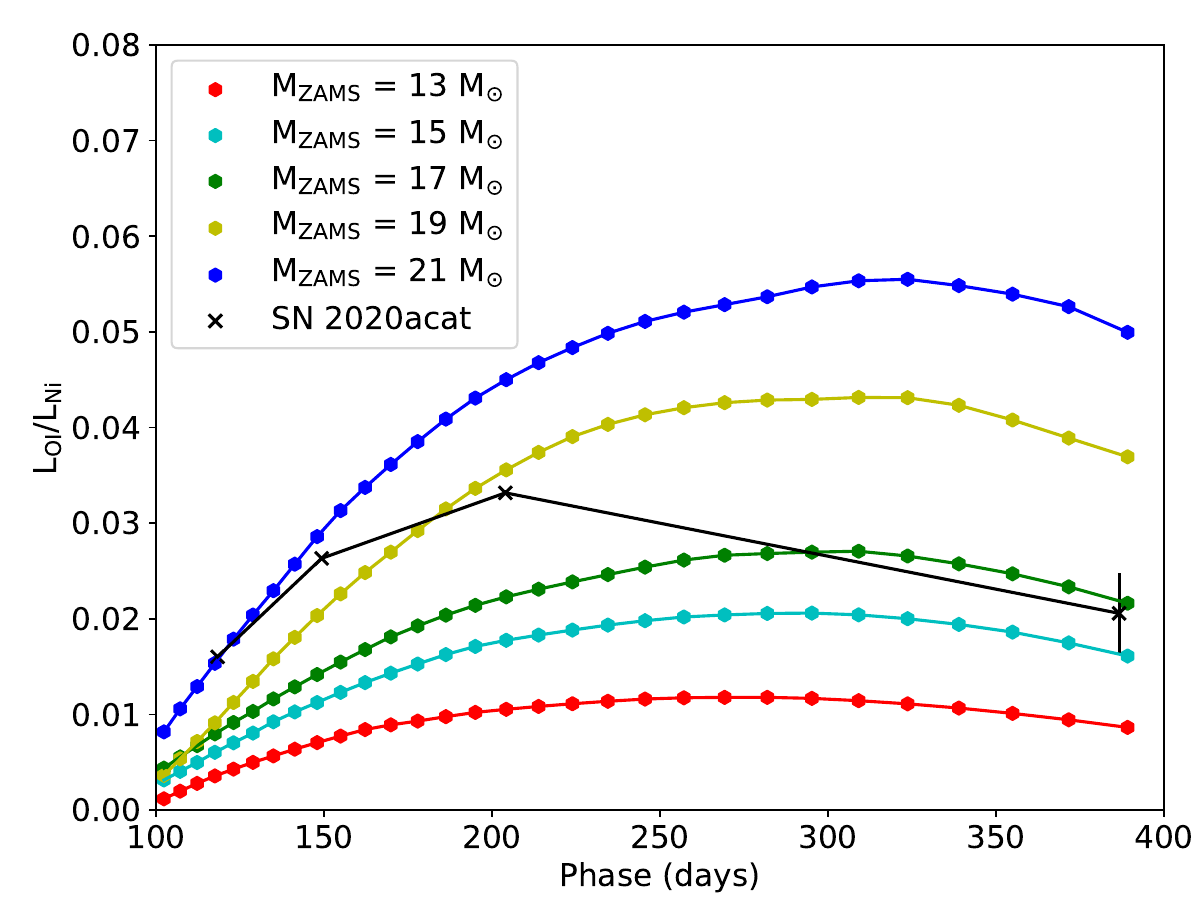}
\caption{Evolution of luminosity in the [\ion{O}{i}] 6300,6364 \AA~lines normalized with the $^{56}$Ni decay luminosity for SN 2020acat (black crosses) and the JEKYLL models with strong expansion and medium mixing of the radioactive material and initial masses of 13 M$_\odot$ (red), 15 M$_\odot$ (cyan), 17 M$_\odot$ (green), 19 M$_\odot$ (yellow) and 21 M$_\odot$ (blue).}
\label{f_acat_spec_line_flux_comp_OI}
\end{figure}

Figure \ref{f_acat_spec_line_flux_comp_OI} shows the evolution of the luminosity in the [\ion{O}{i}] 6300,6364 \AA~lines for SN 2020acat compared to models M13-m-s, M15-m-s, M17-m-s, M19-m-s and M21-m-s, which all have medium mixing and strong expansion of the radioactive material, and only differ in the initial mass. Clearly, the 13 M$_\odot$ model has far too low flux in the [\ion{O}{i}] 6300,6364 \AA~lines at all epochs, whereas the 21 M$_\odot$ model has too high flux from $\sim$150 days and onwards, and far to high flux at $\sim$400 days. It could be argued that later epochs are more reliable as the SN has then become more nebular and optical depth effects play less of a role. In that case both the 13 and 21 M$_\odot$ model seems to be excluded, and the 17-19 M$_\odot$ models gives the best match to the observations. Note, that the JEKYLL models overall evolve slower than observed for SN 2020acat, an issue we will return to in Sect.~\ref{s_optimal_model}.

\begin{figure}[tbp!]
\includegraphics[width=0.49\textwidth,angle=0]{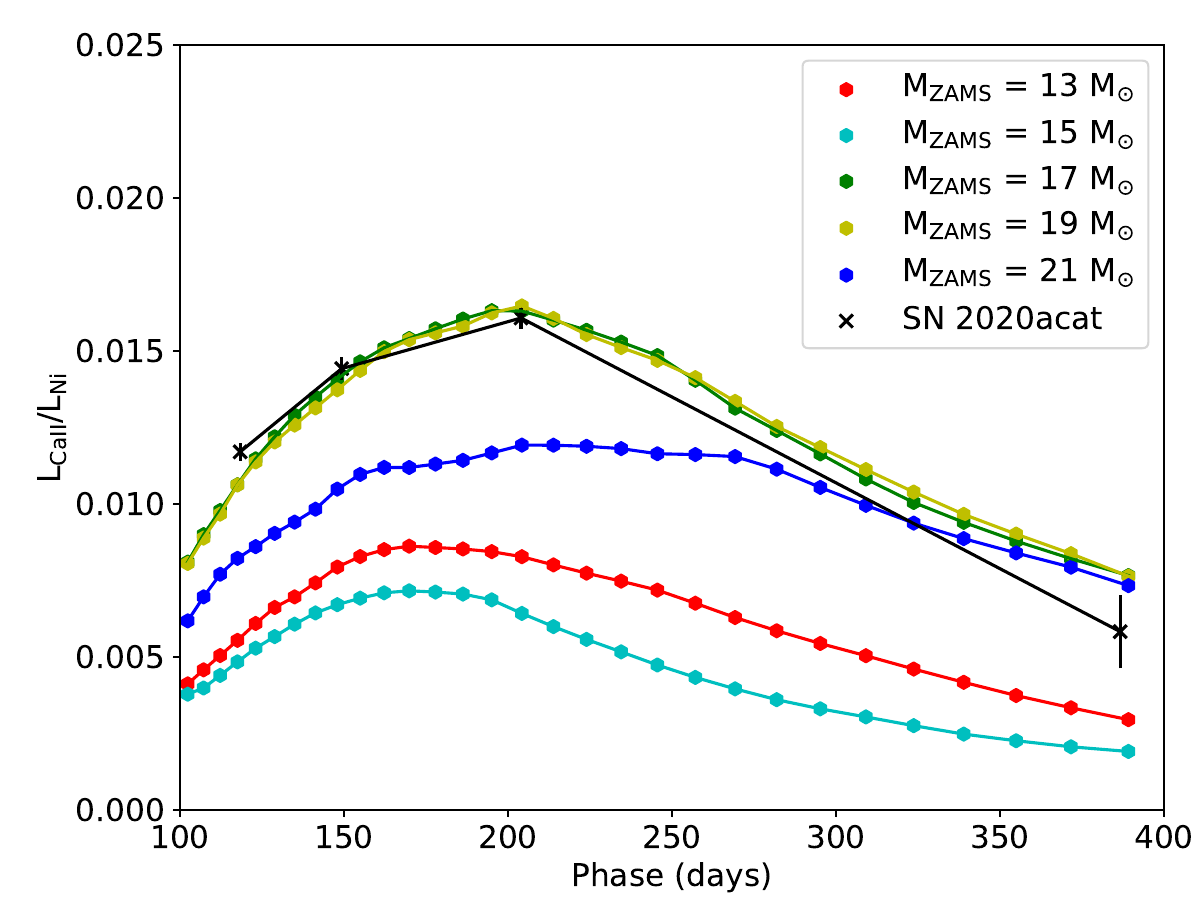}
\caption{Evolution of luminosity in the [\ion{Ca}{ii}] 7291,7323 \AA~lines normalized with $^{56}$Ni decay luminosity for SN 2020acat (black crosses) and the JEKYLL models with strong expansion and medium mixing of the radioactive material and initial masses of 13 M$_\odot$ (red), 15 M$_\odot$ (cyan), 17 M$_\odot$ (green), 19 M$_\odot$ (yellow) and 21 M$_\odot$ (blue).} 
\label{f_acat_spec_line_flux_comp_CaII}
\end{figure}

In Figure \ref{f_acat_spec_line_flux_comp_CaII} we show the evolution of the luminosity in the [\ion{Ca}{ii}] 7291,7323 \AA~lines for SN 2020acat compared to the same models. As these lines may overtake the cooling from the [\ion{O}{i}] 6300,6364 \AA~lines if calcium-rich material is somehow mixed with the oxygen-rich material \citepalias[see][]{Erg22}, it is important to examine these lines as well. Note, however, that the [\ion{Ca}{ii}] 7291,7323 \AA~lines mainly originate from the Si/S and O/Si/S zones, which are not adjusted to comply with the adopted $^{56}$Ni mass (see Sect.~\ref{s_ejecta_model}), so this comparison has to be taken with a grain of salt. Nevertheless, the 17-19 M$_\odot$ models, which matched the evolution of the flux in the [\ion{O}{i}] 6300,6364 \AA~line best, also give a reasonable match to the evolution of the flux in the [\ion{Ca}{ii}] 7291,7323 \AA~lines, which is assuring. 

\begin{figure}[tbp!]
\includegraphics[width=0.49\textwidth,angle=0]{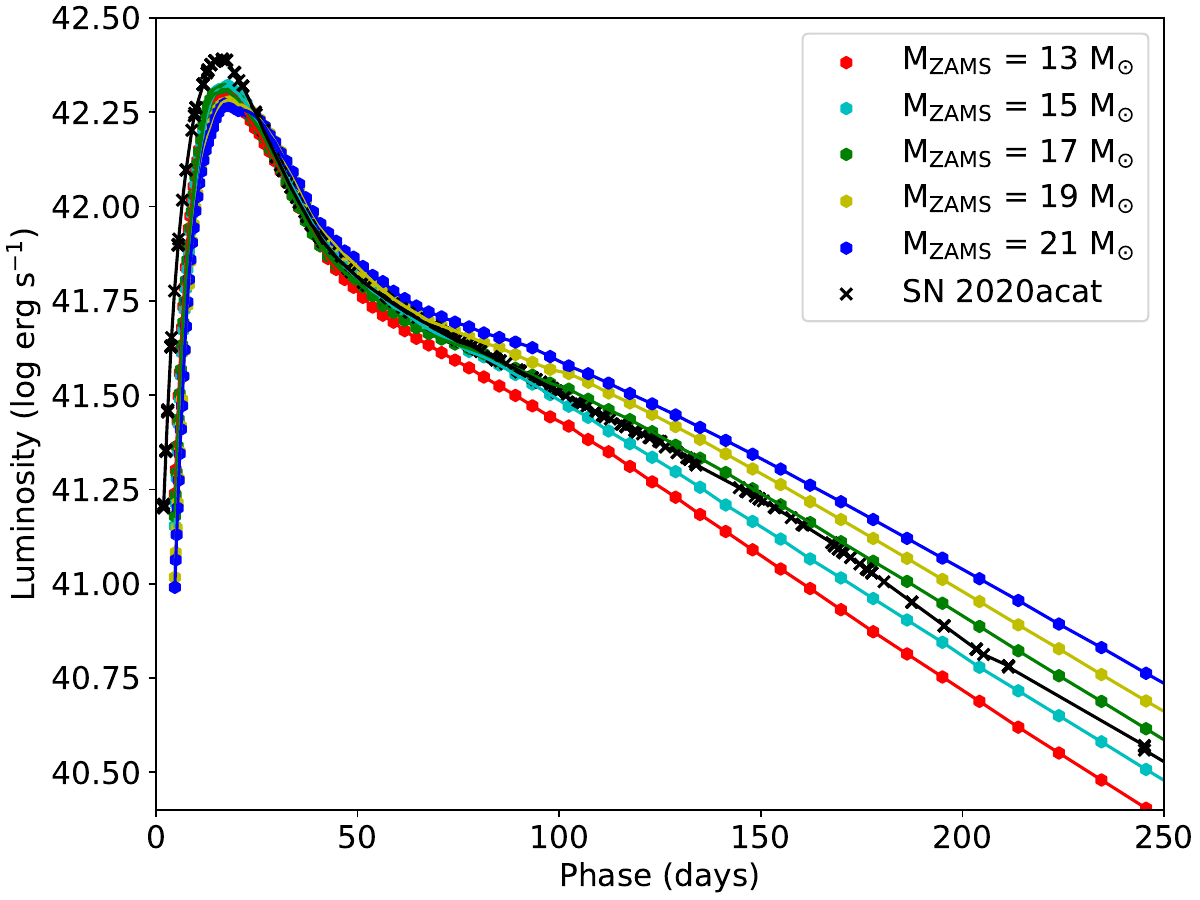}
\caption{Pseudo-bolometric \textit{uBVriz} lightcurves until 250 days for SN 2020acat and JEKYLL models with strong expansion and medium mixing of the radioactive material and initial masses of 13 M$_\odot$ (red), 15 M$_\odot$ (cyan), 17 M$_\odot$ (green), 19 M$_\odot$ (yellow) and 21 M$_\odot$ (blue).}
\label{f_acat_lightcurve_bol_comp_M}
\end{figure}

Figure \ref{f_acat_lightcurve_bol_comp_M} shows the pseudo-bolometric \textit{uBVriz} lightcurve for SN 2020acat compared to models M13-m-s, M15-m-s, M17-m-s, M19-m-s and M21-m-s, which all have medium mixing and strong expansion of the radioactive material, and only differ in the initial mass. During the diffusion phase the model lightcurves are fairly similar, but during the tail phase they progressively diverge. Compared to SN 2020acat, the 13 M$_\odot$ model has too low luminosity and declines too fast whereas the 21 M$_\odot$ model has too high luminosity and declines too slowly, indicating that the optical depth to the $\gamma$-rays in the 13 and 21 M$_\odot$ models are too low and too high, respectively. The best agreement with SN 2020acat in the tail phase is shown by the 15-17 M$_\odot$ models. The similarity of the models in the diffusion phase may look a bit surprising given the quite large difference in ejecta mass, but it should be kept in mind that the early evolution is largely determined by the helium envelope, which is not that different in the models. The mass of the helium envelope is only slowly increasing with initial mass and the interface velocities are fixed by observations. Note, that the tail luminosity and decline rate also depend on the mixing of the radioactive material, and high-mass models with extreme mixing and low-mass models with weak mixing may fit the tail better than the medium mixing models shown here. The tail phase comparison is therefore not conclusive in itself. However, in combination with the evolution of the flux in the [\ion{O}{i}] 6300,6364 \AA~lines the 13 and 21 M$_\odot$ models seem excluded, and we find the 17 M$_\odot$ model to give the best overall match. The diffusion phase does not provide much constraints on the initial mass, but instead we use it to constrain the mixing and expansion of the radioactive material.

\begin{figure}[tbp!]
\includegraphics[width=0.49\textwidth,angle=0]{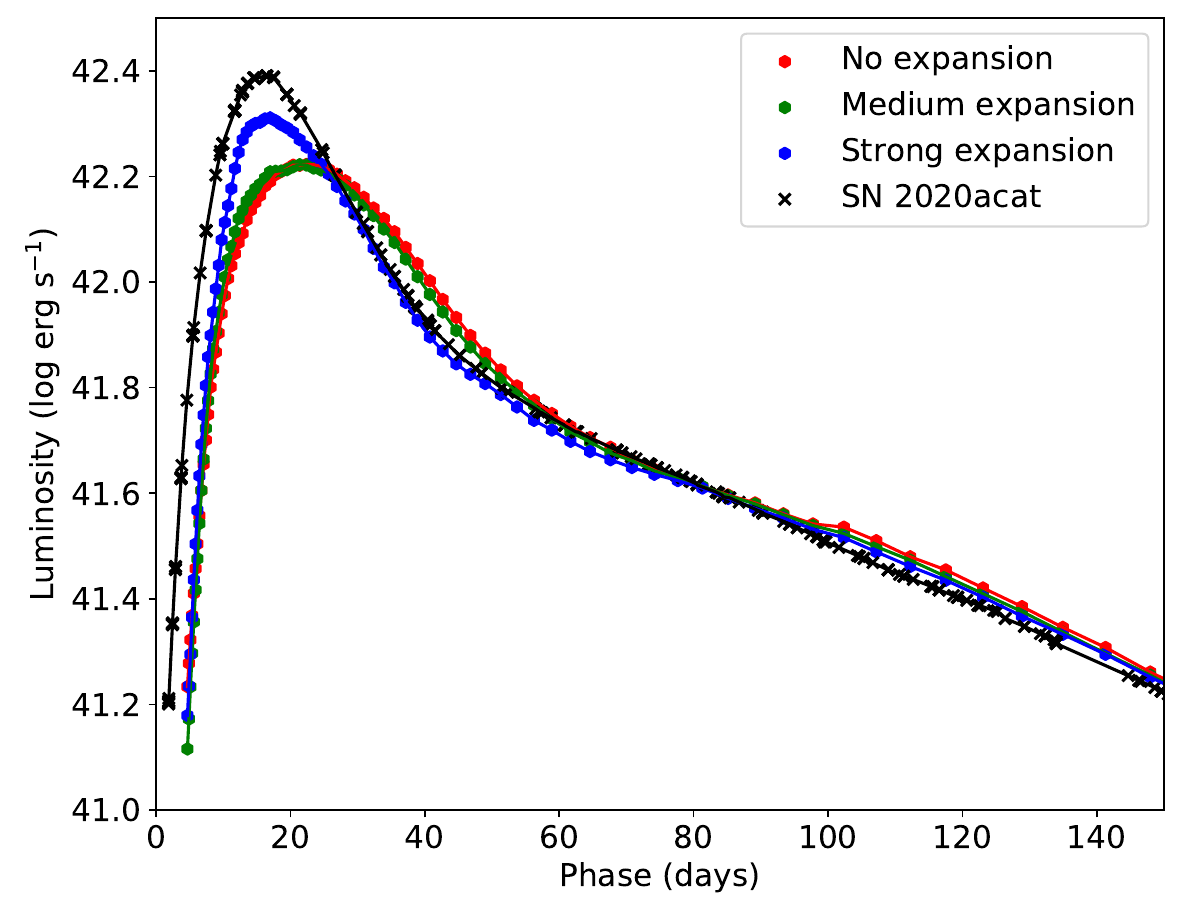}
\caption{Pseudo-bolometric \textit{uBVriz} lightcurves until 150 days for SN 2020acat and JEKYLL models with an initial mass of 17 M$_\odot$, medium mixing and no, medium and strong expansion of the radioactive material.}
\label{f_acat_lightcurve_bol_comp_exp}
\end{figure}

Figure \ref{f_acat_lightcurve_bol_comp_exp} shows the pseudo-bolometric \textit{uBVriz} lightcurve for SN 2020acat compared to models M17-m-n, M17-m-m and M17-m-s, which all have an initial mass of 17 M$_\odot$ and only differ in the expansion of the radioactive material. Contrary to the previous case, the tail phase lightcurves are similar, whereas the diffusion phase lightcurves differ. The diffusion peak is clearly too broad for the models without or with only mild expansion of the radioactive material, whereas the model with strong expansion of the radioactive material gives a much better fit. The reason for the differences in the diffusion phase lightcurve is that the expansion of the radioactive material decreases the effective opacity of the ejecta, a small-scale 3-D effect discussed in detail in \citetalias{Erg22} (see also \citealt{Des19} for a discussion of this effect on Type IIP SN lightcurves).

\begin{figure}[tbp!]
\includegraphics[width=0.49\textwidth,angle=0]{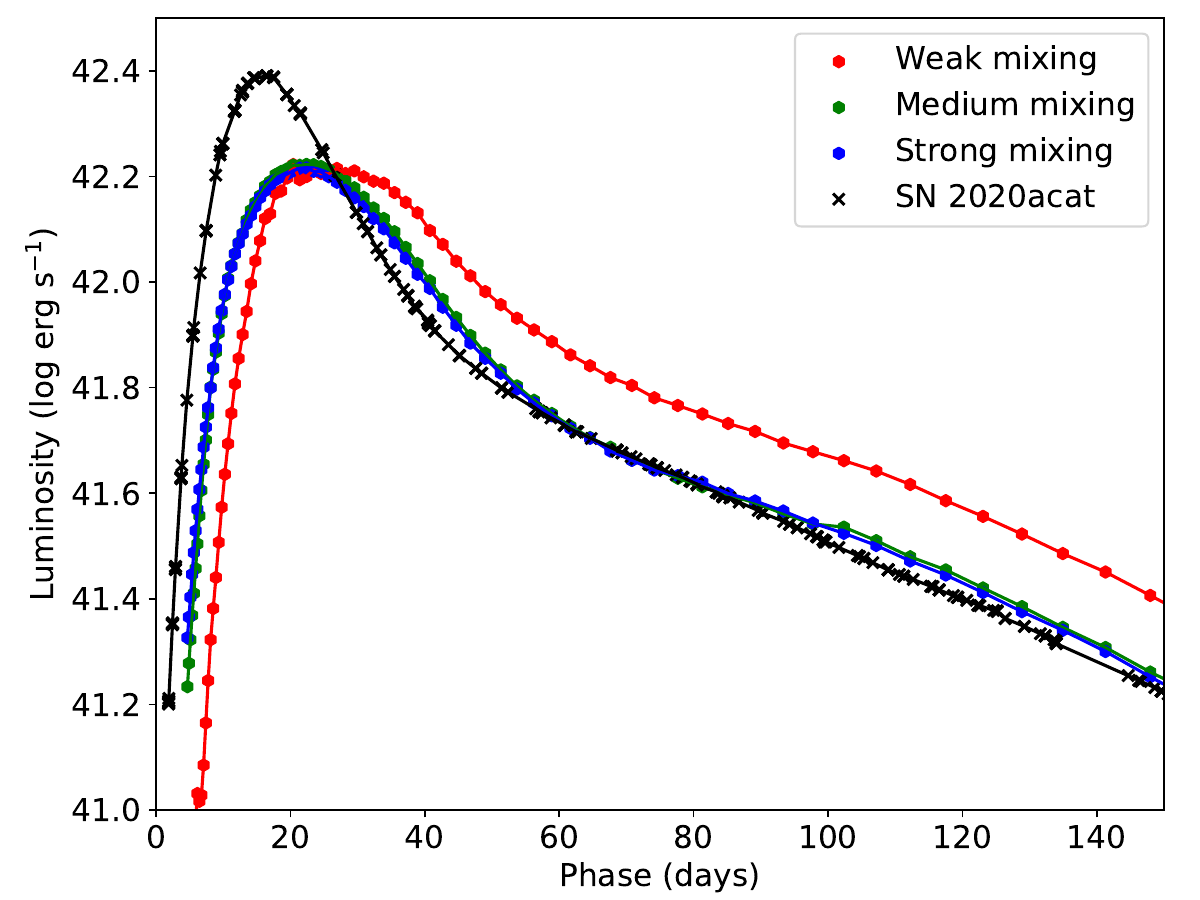}
\caption{Pseudo-bolometric \textit{uBVriz} lightcurves until 150 days for SN 2020acat and JEKYLL models with an initial mass of 17 M$_\odot$, no expansion and weak, medium and strong mixing of the radioactive material.}
\label{f_acat_lightcurve_bol_exp_n_comp_mix}
\end{figure}

\begin{figure}[tbp!]
\includegraphics[width=0.49\textwidth,angle=0]{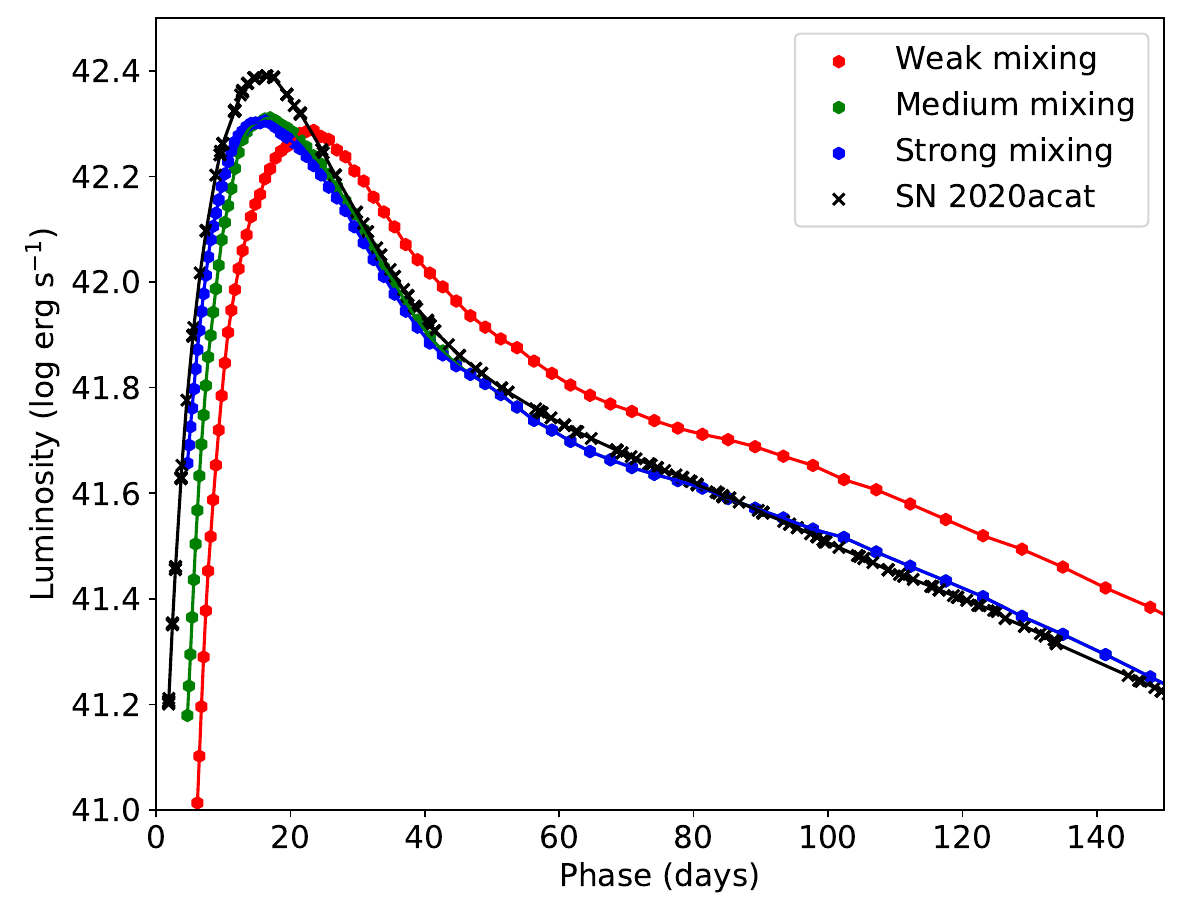}
\caption{Pseudo-bolometric \textit{uBVriz} lightcurves until 150 days for SN 2020acat and JEKYLL models with an initial mass of 17 M$_\odot$, strong expansion and weak, medium and strong mixing of the radioactive material.}
\label{f_acat_lightcurve_bol_exp_s_comp_mix}
\end{figure}

Figures \ref{f_acat_lightcurve_bol_exp_n_comp_mix} and \ref{f_acat_lightcurve_bol_exp_s_comp_mix} show the pseudo-bolometric \textit{uBVriz} lightcurve for SN 2020acat compared to models with an initial mass of 17 M$_\odot$ which only differ in the mixing of the radioactive material. In Fig.~\ref{f_acat_lightcurve_bol_exp_n_comp_mix} we show models M17-w-n, M17-m-n and M17-s-n which have no expansion of the radioactive material, and in Fig.~\ref{f_acat_lightcurve_bol_exp_s_comp_mix} we show models M17-w-s, M17-m-s and M17-s-s, which have strong expansion of the radioactive material. Clearly, the models with no expansion of the radioactive material give too broad diffusion peaks regardless of the mixing of this material. If the radioactive material is strongly expanded the width of the diffusion peak becomes narrower and agrees better with the observations, and the best agreement is achieved with strong mixing of this material. With only weak mixing of the material, the peak becomes far too broad, so both strong mixing and strong expansion of the radioactive material seem to be required to fit the diffusion peak of SN 2020acat. Note, that the peak luminosity is not fully reproduced by any of the models, and is $\sim$20 percent fainter than for SN 2020acat. We return to this issue in Sect.~\ref{s_optimal_model}. Note also, that the tail luminosity is too bright for the weakly mixed models. Although the match to the tail could be improved for these models by lowering the mass of $^{56}$Ni, this would give an even worse fit to the diffusion phase, as it roughly corresponds to a scaling of the lightcurve.

\subsubsection{The hydrogen envelope}

To explore the properties of the hydrogen envelope we use the hydrogen and the helium lines, where we first use the former to constrain the mass of hydrogen, and then the latter to constrain the mass-fraction of hydrogen, which for a given mass of hydrogen is inversely proportional to the mass of helium (as well as the total mass of the hydrogen envelope).

\begin{figure}[tbp!]
\includegraphics[width=0.49\textwidth,angle=0]{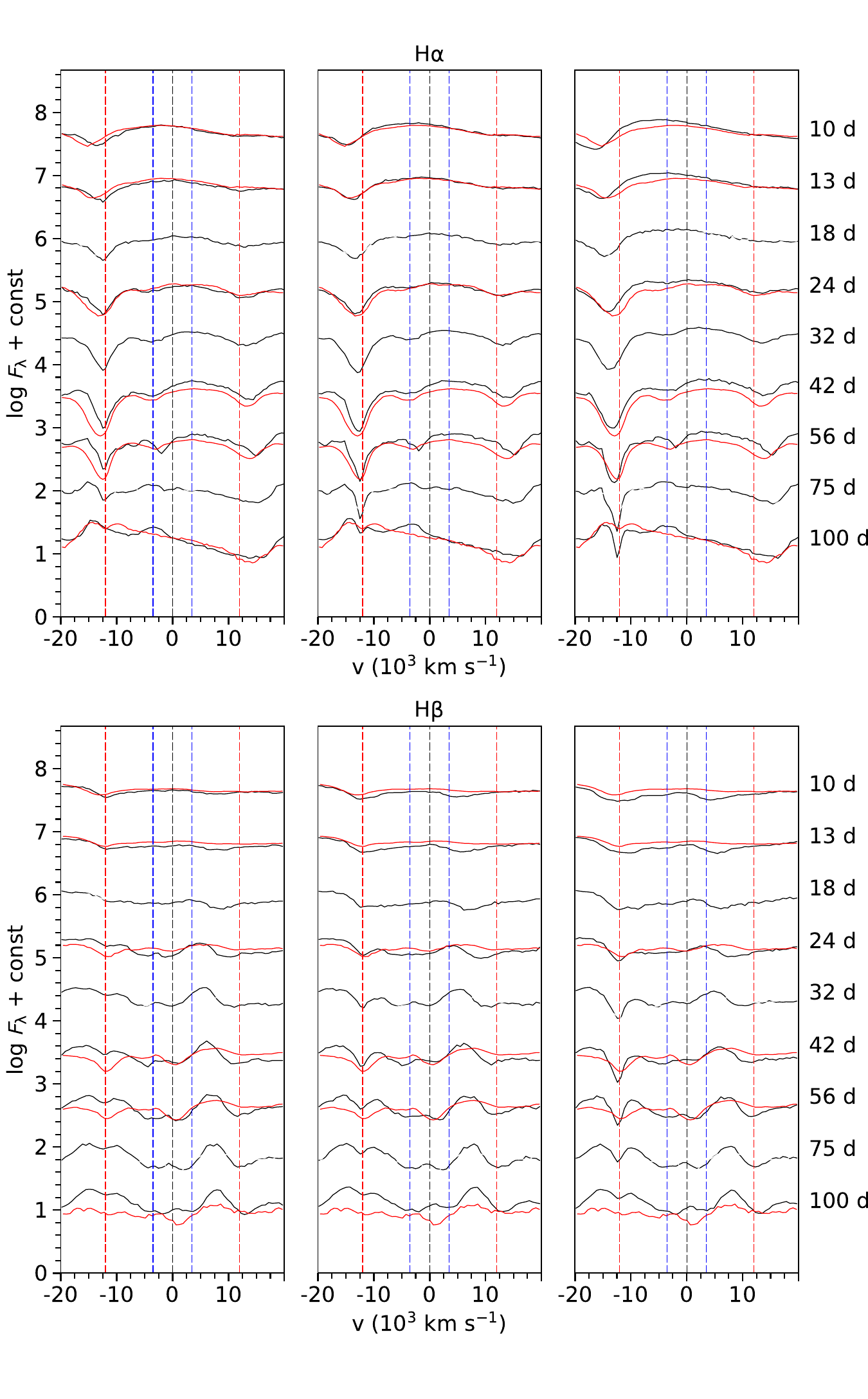}
\caption{Evolution of the H$\alpha$ (upper panel) and H$\beta$ (lower panel) lines for SN 2020acat (red) and the JEKYLL models (black) with an initial mass of 17 M$_\odot$, strong mixing and strong expansion of the radioactive material, and a mass of hydrogen in the envelope of 0.0135 (left), 0.027 (middle) and 0.054 (right) M$_\odot$. Spectra from nine logarithmically spaced epochs are shown, and the model C/O-He (blue) and He-H (red) interface velocities are indicated with dashed lines.}
\label{f_acat_spec_evo_H}
\end{figure}

Figure \ref{f_acat_spec_evo_H} shows the evolution of the H$\alpha$ and H$\beta$ lines for SN 2020acat compared to models M17-s-s-H-low, M17-s-s and M17-s-s-H-high, which only differ in the mass of hydrogen in the envelope. The H$\alpha$ and H$\beta$ absorption becomes too strong when approaching 100 days in the model with a high hydrogen mass (M$_{\mathrm{H}}$=0.054 M$_\odot$), and is overall too weak and appears at too low velocities in the model with a low hydrogen mass (M$_{\mathrm{H}}$=0.0135 M$_\odot$). On the other hand, both the H$\alpha$ and H$\beta$ lines are reasonably well reproduced in the model with a medium hydrogen mass (M$_{\mathrm{H}}$=0.027 M$_\odot$), and this model gives the best overall fit to the evolution of these lines.

\begin{figure}[tbp!]
\includegraphics[width=0.49\textwidth,angle=0]{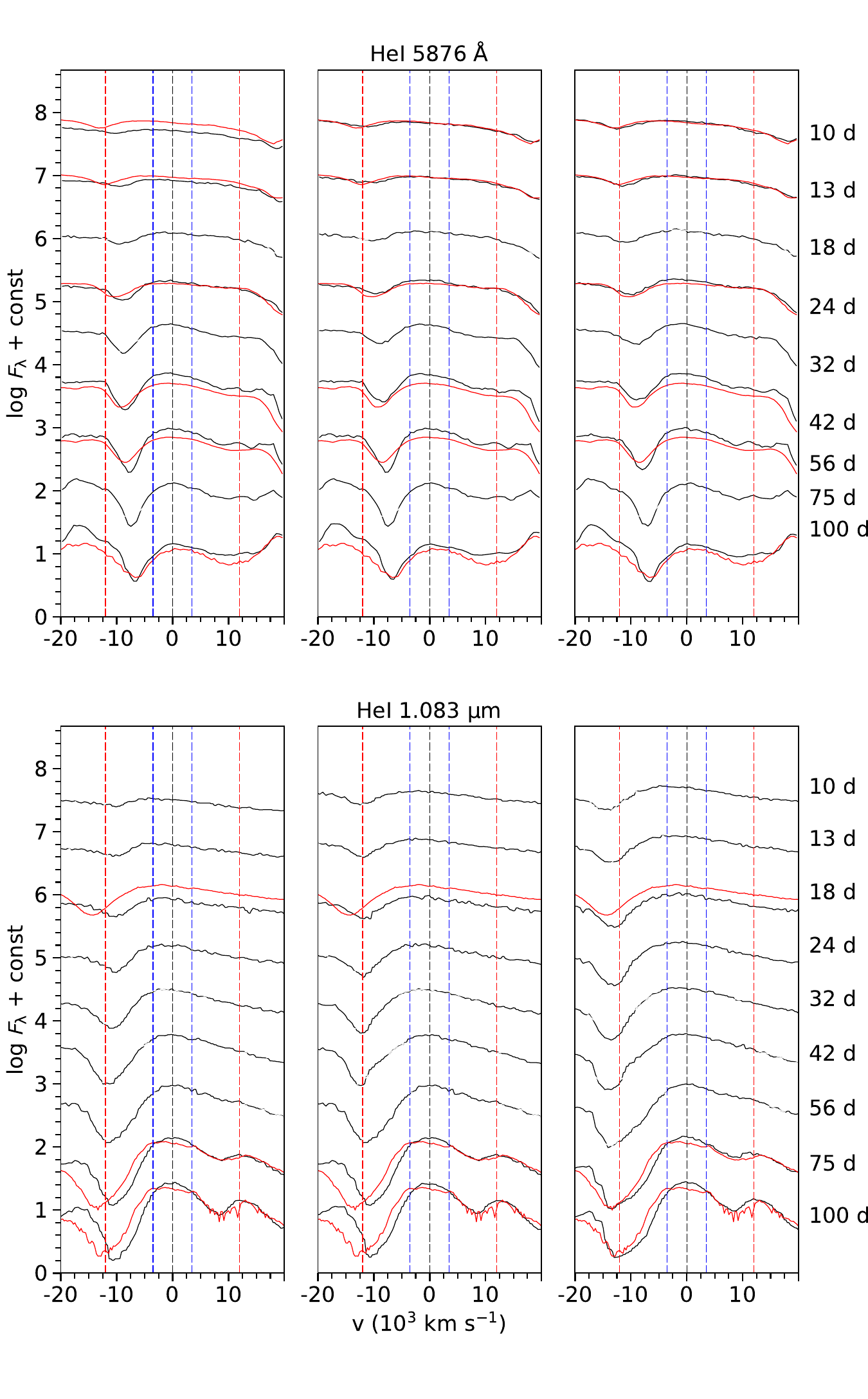}
\caption{Evolution of the \ion{He}{i} 5876 \AA~(upper panel) and \ion{He}{i} 1.083 $\mu m$ (lower panel) lines for SN 2020acat (red) and the JEKYLL models (black) with an initial mass of 17 M$_\odot$, strong expansion of the radioactive material, and medium mixing plus X$_\mathrm{H}$=0.54 (left), strong mixing plus X$_\mathrm{H}$=0.54 (middle) and strong mixing plus X$_\mathrm{H}$=0.27 (right). Otherwise the same as in Fig~\ref{f_acat_spec_evo_H}.}
\label{f_acat_spec_evo_He}
\end{figure}

Figure \ref{f_acat_spec_evo_He} shows the evolution of the \ion{He}{i} 5876 \AA~and \ion{He}{i} 1.083 $\mu m$ lines for SN 2020acat compared to models M17-m-s, M17-s-s and M17-s-s-XH-low, which all have a medium hydrogen mass, and differ in the mass-fraction of hydrogen and the mixing of the radioactive material. Clearly, the match is worse for the models with a high hydrogen mass-fraction (X$_{\mathrm{H}}$=0.54) than for the model with a low hydrogen mass-fraction (X$_{\mathrm{H}}$=0.27), although a stronger mixing of the radioactive material improves the match somewhat due to non-thermal excitation and ionisation. In particular, there is too weak absorption in the \ion{He}{i} 1.083 $\mu m$ line at velocities above the interface between the helium ad hydrogen envelope. Overall, the model with strong mixing and a low hydrogen mass-fraction gives the best match to the evolution of the \ion{He}{i} 5876 \AA~and \ion{He}{i} 1.083 $\mu m$ lines. 

\subsubsection{Summary and discussion}

In summary we arrive at an optimal model with an initial mass of 17 M$_\odot$, strong mixing and strong expansion of the radioactive material, and an 0.1 M$_\odot$ hydrogen envelope with X$_{\mathrm{H}}$=0.27. In Sect.~\ref{s_optimal_model} we have a more detailed look at this model, compare it with the observations of SN 2020acat, and discuss similarities as well as remaining differences and their possible origin. First, however, we provide a discussion about the main properties derived for our optimal model, and what can be learned from that.

The most interesting result is perhaps the strong expansion of the radioactive material that seems to be required, and the accompanying strong effect of that on the diffusion phase lightcurve. As discussed in detail in \citetalias{Erg22}, the expansion of the Ni/He clumps creates a density contrast between these and other clumps, which affects the radiative transfer and leads to a decrease of the effective opacity. This can be imagined as a Swiss-cheese-like geometry where the photons diffuse faster through the low-density "Ni bubbles". The effect is strongest in the limit of optically thick clumps and disappears in the limit of optically thin clumps. As derived in \citetalias{Erg22}, the decrease of the effective opacity in the limit of optically thick clumps is roughly given by the product of the (volume) expansion and filling factors for the Ni/He clumps. 

\begin{figure}[tbp!]
\includegraphics[width=0.49\textwidth,angle=0]{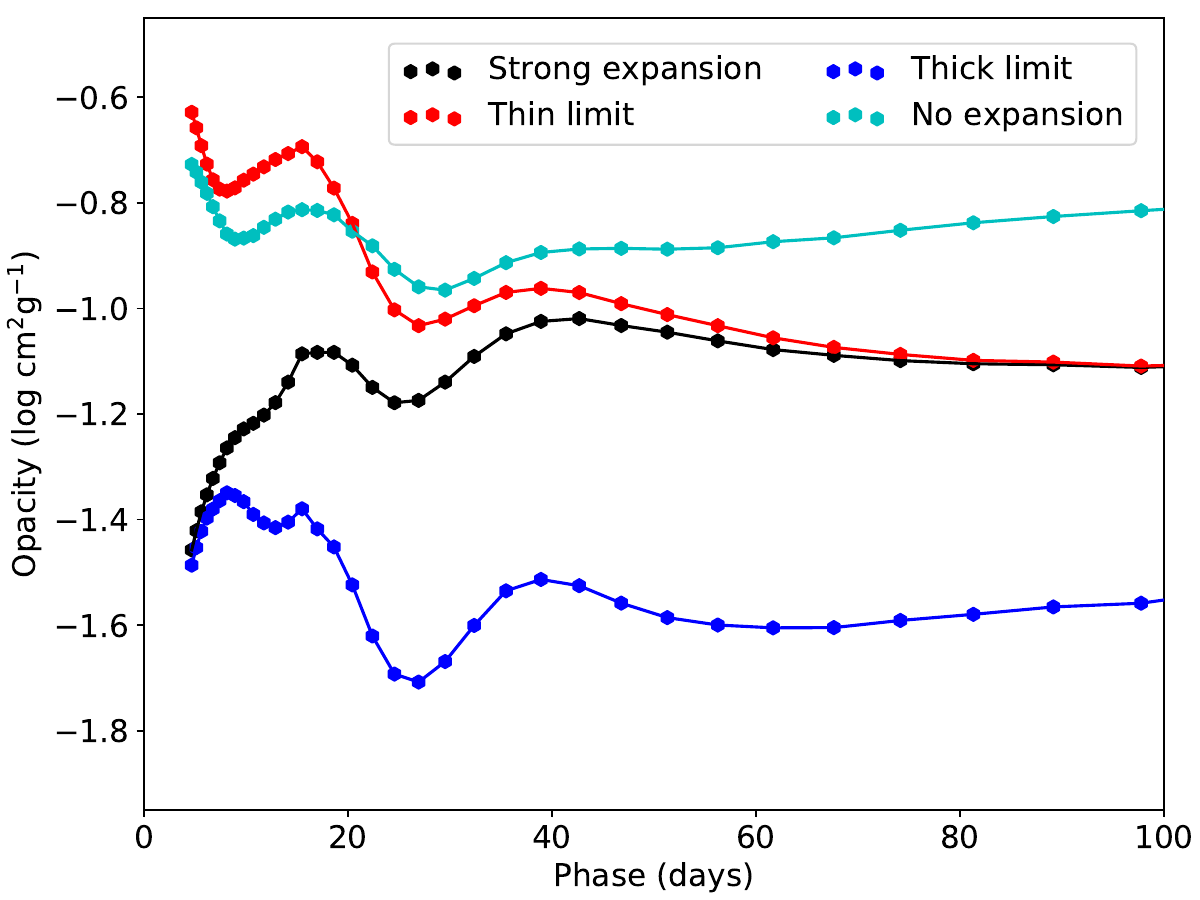}
\caption{Evolution of the (mass-averaged) effective Rosseland mean opacity in the inner helium envelope for the model with strong expansion (black) and the corresponding effective Rosseland mean opacity in the limits of optically thick (blue) and thin (red) clumps. In addition, we show the evolution of the effective Rosseland mean opacity for the model without expansion (cyan).}
\label{f_acat_kappa_evo}
\end{figure}

This is illustrated by Fig.~\ref{f_acat_kappa_evo}, where we show the (average) effective Rosseland mean opacity in the inner helium envelope for the model with strong expansion (M-17-s-m) and the corresponding limits for optically thick and thin clumps. Initially, the effective opacity follows the thick limit, which is a factor of $\sim$5 below the thin limit, and then gradually approaches the thin limit towards $\sim$60 days, where the radiative transfer effect disappears. However, in Fig.~\ref{f_acat_kappa_evo} we also show the model without expansion (M17-n-m), and compared to that model, the effective opacity remains lower even after $\sim$60 days. This is due to a density driven recombination effect, discussed in more detail in \citetalias{Erg22} and by \citet{Des18}. The radiative transfer and recombination effects are complementary, but the radiative transfer effect is stronger and dominates during the diffusion phase.

It should be pointed out, though, that we can not rule out that physics outside the limitations of JEKYLL (see Sect.~\ref{s_method_general}) may cause a similar effect on the diffusion phase lightcurve as the expansion of the "Ni bubbles". Early-time CSM interaction seems unlikely as there are no signs of this in the spectra, and as the mass-loss rate estimated from radio observations (Poonam et al. In prep.) is more similar to SN 2011dh than interacting Type IIb SNe as 1993J. Large-scale asymmetries are harder to rule out, and may or may not give rise to a similar effect on the diffusion phase lightcurve. We discuss this issue further below. Note also, that the degree of expansion of the radioactive material is somewhat degenerate with the assumed clumps size (see Sect.~\ref{s_macro_type_IIb}), and larger clumps would require less expansion of this material to achieve the same effect on the lightcurve.

If the magnitude of the effect in our model of SN 2020acat would be typical for Type IIb and other SE SNe, it has important implications for the entire literature of 1-D lightcurve modelling of such SNe. This applies to both simple \citep[e.g][]{Can13,Lym16,Pre16} or more advanced \citep[e.g][]{Erg15,Tad17} 1-D models, as none of these take the effect of the Ni bubbles on the effective opacity into account. Depending somewhat on which weight is given to the diffusion phase lightcurve, the ejecta masses derived from such modelling could be systematically and quite strongly underestimated. 

Ignoring the effect may give rise to a tension between quantities derived from the diffusion and tail phases, similar to what we find for SN 2020acat. Interestingly, such a tension has been reported by \citet{Whe15} for a literature sample of SE SNe, although this tension may at least partly arise from other simplifications in their methods \citep[see][]{Nag22}. Such a tension has also been reported for several Type Ic Broad-Lined (BL) SNe. For SN 1998bw, \citet{Des17} found that the tail phase required much more massive ejecta than the diffusion phase. \citet{Mae03} proposed that this could be explained by large-scale asymmetries in a jet-driven explosion and introduced a simple two-component ejecta model (see also \citealt{Val08} for a similar model). In this model the low density jet-component, supposed to contain most of the Ni/He material, gives rise to a fast and luminous diffusion peak, whereas the high density disk-component, supposed to contain most of the oxygen-rich material, gives rise to the tail. While such an ejecta geometry is not entirely far-fetched in the case of a Typ Ic-BL SN, supposedly originating from a fast-rotating progenitor star, it makes less sense in the case of a Type IIb SN.

To explain the early lightcurve of SN 1998bw, \citet{Hof99} proposed an oblate ejecta geometry. Oblate or prolate ejecta geometries are viewing angle dependent, and may boost or suppress the luminosity during the diffusion phase due to the projected area of the photosphere (see also \citealt{Kro09} for a prolate Type Ia toy model). Such an effect seems more plausible in the case of Type IIb SNe, and would give rise to another form of tension between the diffusion and the tail phases, more related to the mass of $^{56}$Ni than the ejecta mass. Direct observational evidence for large-scale asymmetries in the ejecta of SNe can be searched for using polarimetry. Although there was no polarimetry obtained for SN 2020acat, such observations have been obtained for several other Type IIb SNe. The results show a continuum polarisation of $\sim$0.5 percent during the helium dominated phase for most objects \citep{mau15}, which might be interpreted as moderately aspherical ejecta. This interpretation is not clear, however, as clumpy ejecta may also contribute to the continuum polarisation. Explosion models \citep{Wong17} indicates that asymmetries arise on a wide range of scales in Type IIb SNe, and although our results does not disprove that large-scale asymmetries affect their lightcurves, it does prove that small- and medium-scale asymmetries may also have an important effect.

The strong mixing of the radioactive material required to fit the early lightcurve is in line with results from hydrodynamical modelling of Type IIb SNe \citep[e.g.][]{Ber12,Erg15,Tad17}. Note, however, that in our models strong expansion of this material is also required to reduce the effective opacity in the layers into which the material is mixed. This is likely related to the fact that in our models the mixing of the radioactive material has an opposite effect, and increases the opacity, both through the higher line-opacity of this material and through non-thermal ionisation. Strong mixing of the radioactive material is also in line with results from Type IIb explosion models \citep{Wong17} and observations of the Type IIb SN remnant Cas A \citep[e.g.][]{Wil02}.

The relatively high initial mass of $\sim$17 M$_\odot$ derived places SN 2020acat at the upper end of the mass-distribution for Type IIb SNe. \citet{Jer15} estimated initial masses well below 17 M$_\odot$ for the progenitors of SNe 2008ax, 2011dh and 1993J using modelling of their nebular spectra, which for the latter two is supported by stellar-evolutionary analysis of pre-explosion imaging of the progenitors \citep{Ald94,Mau11}. \citet{Erg15b} found 56 percent of the Type IIb progenitors to have an initial mass less than 15 M$_\odot$ and 75 percent to have an initial mass below 20 M$_\odot$ using hydrodynamical lightcurve modelling. The simplified treatment of the opacity and the 1-D limitation (preventing the effect of the Ni bubbles on the diffusion time) makes this result uncertain though. The relatively high initial mass found for SN 2020acat also makes a single star origin more plausible. This is in contrast to SN 2011dh, for which a single star origin seems to be excluded.

The low mass-fraction of hydrogen in the envelope derived is more in line with a binary origin though, as such a low mass-fraction may naturally arise in a binary system during mass-transfer to the companion star \citep{Yoo10}. The low mass-fraction of hydrogen in the envelope is also in line with the short cooling phase that SN 2020acat seems to have experienced (<1 day), as this tends to result in smaller progenitor radii. The extent of the cooling phase depends on several factors though, and hydrodynamical modelling is needed to shed more light on this issue.

\subsection{Detailed comparison to SN 2020acat.}
\label{s_optimal_model}

In Sect.~\ref{s_constrain} we constrained the model parameters by comparing some key observables for SN 2020acat to our model grid, and arrived at an optimal model (M17-s-s-XH-low) with an initial mass of 17 M$_\odot$, strong mixing and expansion of the radioactive material, and a 0.1 M$_\odot$ hydrogen envelope with X$_{\mathrm{H}}$=0.27. Here we take a more in-depth look at this model and compare the spectra and lightcurves in more detail to SN 2020acat. In addition, it is interesting to compare to the optimal model for SN 2011dh, first presented in \citet{Jer15}, and then refined for the photospheric phase and discussed in detail in \citetalias{Erg22}. This model has an initial mass of 12 M$_\odot$, somewhat weaker mixing and expansion of the radioactive material, and an 0.05 M$_\odot$ hydrogen envelope with X$_{\mathrm{H}}$=0.54. The models also differ in the mass of $^{56}$Ni and the interface velocities, reflecting the lower luminosity and line velocities observed in SN 2011dh.

\begin{figure*}[tbp!]
\includegraphics[width=1.0\textwidth,angle=0]{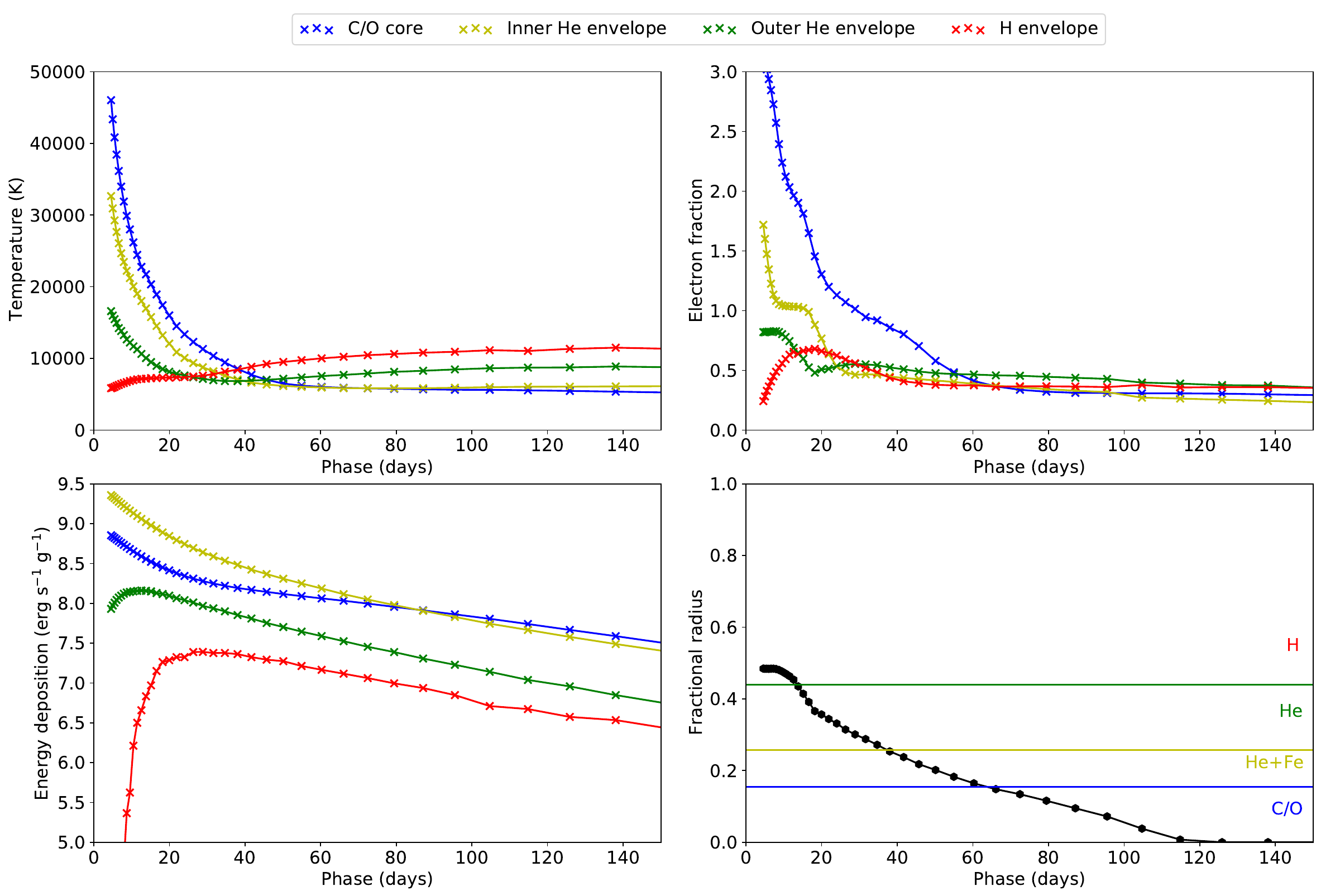}
\caption{Evolution of the temperature (upper-left panel), electron fraction (upper-right panel), and radioactive energy deposition (lower-left panel) in the oxygen core (blue), inner and outer (yellow and green) helium envelope and the hydrogen envelope (red) for the optimal model (M17-s-s-XH-low). In the lower-right panel we show the evolution of the (Rosseland mean) continuum photosphere (black) as well as the outer borders of the carbon-oxygen core (blue) and the inner and outer (green and yellow) helium envelope.}
\label{f_acat_matter_evo}
\end{figure*}

\begin{figure*}[tbp!]
\includegraphics[width=1.0\textwidth,angle=0]{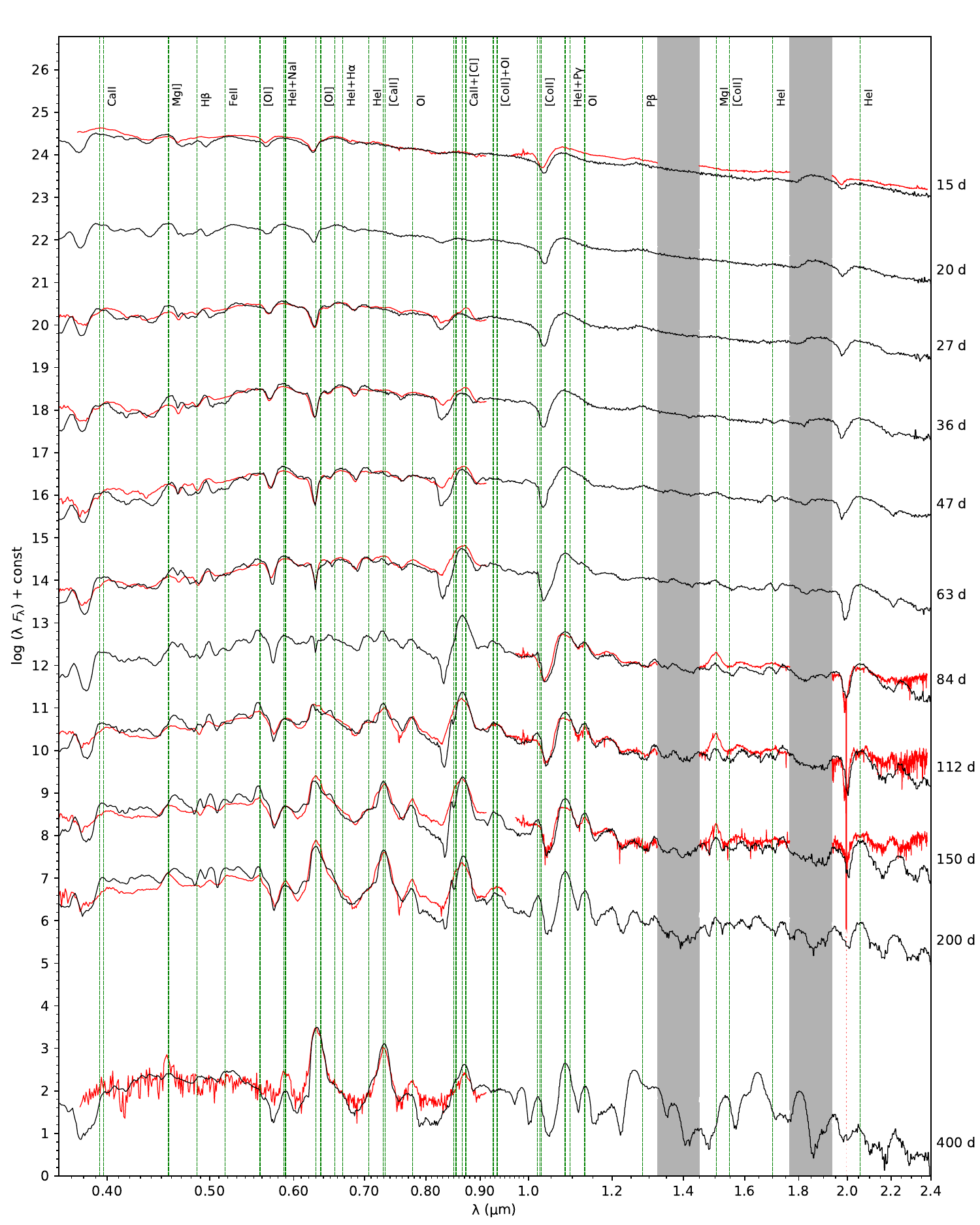}
\caption{Spectral evolution for the optimal model (M17-s-s-XH-low; black) compared to the observations of SN 2020acat (red). Spectra from ten logarithmically spaced epochs between 15 and 200 days and a single epoch at 400 days are shown. In addition, the rest wavelengths of the most important lines are shown as dashed green lines.} 
\label{f_acat_spec_evo_opt_nir}
\end{figure*}

\begin{figure}[tbp!]
\includegraphics[width=0.49\textwidth,angle=0]{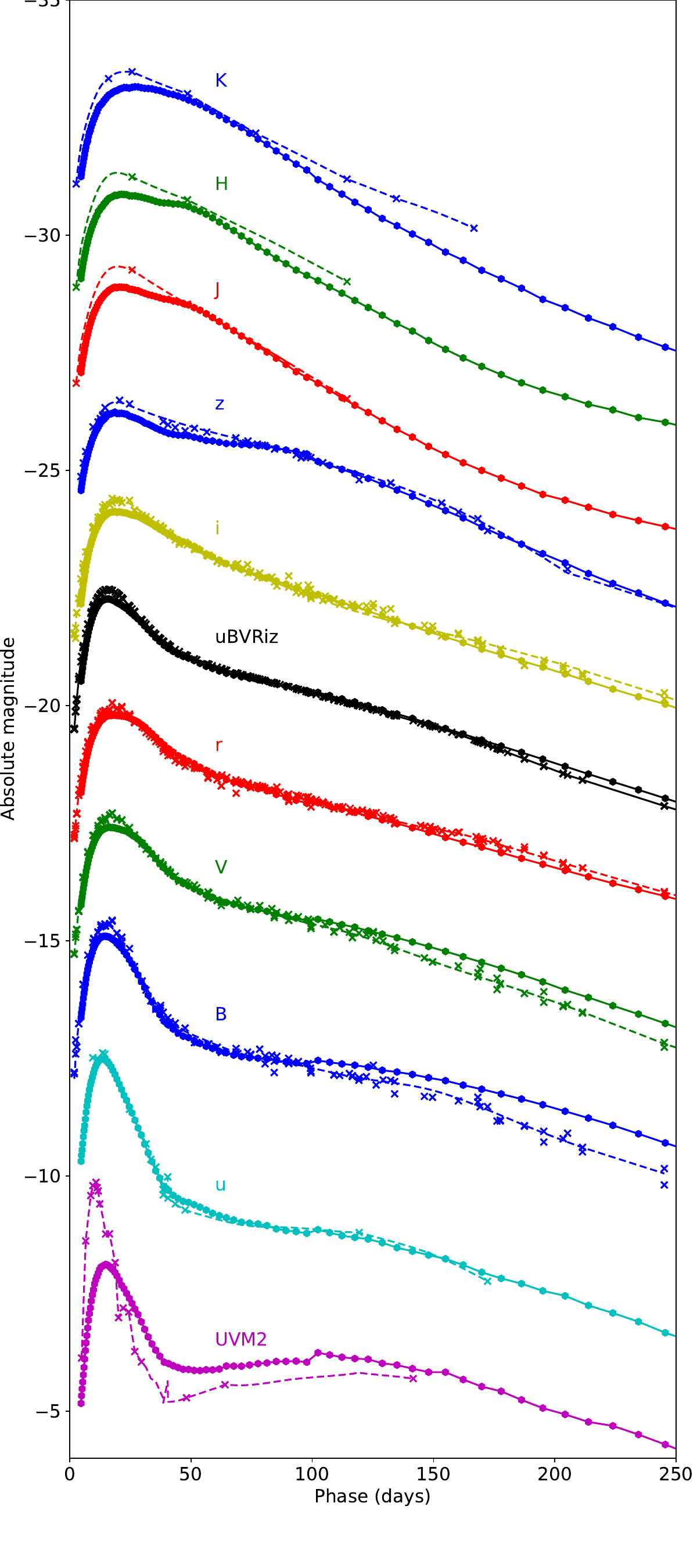}
\caption{Broadband and bolometric light curves until 250 days for the optimal model (M17-s-s-XH-low; solid lines and circles) compared to the observations of SN 2020acat (dashed lines and crosses). From the bottom to top we show the \textit{UVM2} (magenta), u (cyan), B (blue), V (green), r (red), ugBVriz pseudo-bolometric (black), i (yellow), z (blue), J (red), H (green) and K (blue) light curves, which, for clarity, have been shifted by 5.0 4.3 2.0, 0.0, -2.3, -6.7, -9.0, -11.0, -13.0 and -15.0 mags, respectively.}
\label{f_acat_lightcurve_opt_nir}
\end{figure}

In Fig.~\ref{f_acat_matter_evo} we show the evolution of the temperature, electron fraction and radioactive energy deposition in the carbon-oxygen core, the inner and outer helium envelope and the hydrogen envelope (averaged over the spatial cells and compositional zones) as well as the evolution of the photosphere for the optimal model of SN 2020acat, whereas in Figs.~\ref{f_acat_spec_evo_opt_nir} and \ref{f_acat_lightcurve_opt_nir} we show the spectral evolution in the optical and NIR and the lightcurves in the UV, optical and NIR for the optimal model compared to the observed evolution of SN 2020acat. In addition, in Figs.~\ref{f_acat_spec_cell_evo}-\ref{f_acat_spec_trans_evo_ion_6} in Appendix~\ref{a_additional_figures} we show the contributions to the spectral evolution of the optimal model from the different spatial layers, compositional zones and radiative processes giving rise to the emission. Note, that there might be a slight shift in some quantities at 100 days when charge-transfer is turned on (see Appendix~\ref{a_configuration}).

Not surprisingly, the evolution of the model for SN 2020acat is qualitatively similar to that of the model for SN 2011dh \citepalias[see][figs.~2-4]{Erg22}, as they are both Type IIb SN models, although with different SN parameters. Initially ($\sim$5 days), the photosphere is near the inner boundary of the hydrogen envelope, which is relatively cool and mainly recombined, whereas the core is hot and highly ionised. The emission mainly originates from the hydrogen envelope, and the hydrogen signature is strong with lines from the Balmer and Paschen series, mainly seen in emission. After $\sim$10 days the photosphere begins to recede into the helium envelope and emission from therein increases. At the same time, the radioactive energy deposition increases outside the photosphere, and due to both these effects the helium lines begin to rise, and at $\sim$40 days they dominate the spectrum. The hydrogen line emission fades away on a similar time-scale (although H$\alpha$ and H$\beta$ remain in absorption), and completes the transition from a hydrogen- to a helium-dominated spectrum. 

Between $\sim$40 days and $\sim$60 days, the photosphere recedes through the inner parts of the helium envelope, and thereafter through the carbon-oxygen core until it disappears at $\sim$120 days when the SN becomes nebular. During this period emission from the carbon-oxygen core becomes increasingly important and at $\sim$120 days it dominates redwards the $B$-band. As a consequence, emission from heavier elements abundant in the core increases, in particular after $\sim$120 days, when the characteristic [\ion{O}{i}] 6300,6364 \AA~and [\ion{Ca}{ii}] 7291,7323 \AA~lines appear. During the nebular phase this trend continues while the temperature, electron fraction and energy deposition slowly decrease in the core. At 400 days emission from the carbon-oxygen core dominates the entire optical and NIR spectrum and the [\ion{O}{i}] 6300,6364 \AA~and [\ion{Ca}{ii}] 7291,7323 \AA~lines alone contribute about a quarter of the total luminosity.

Overall the agreement between the model and the observations of SN 2020cat is reasonable, and the main differences between SNe 2020acat and 2011dh discussed in Sect.~\ref{s_comparison_to_11dh} are reflected in our models. The luminosity is higher, the diffusion peak occurs earlier and is bluer, the line velocities are higher, the tail declines more slowly and the [\ion{O}{i}] 6300,6364 \AA~lines are stronger in the SN 2020acat model. However, there are also notable differences between our model and the observations of SN 2020acat. During the diffusion phase the peak luminosity is not entirely reproduced by the model, and is too low in all bands. In our models, the peak-to-tail ratio is sensitive to the mixing of the Ni/He material, and a better fit might be achieved by tweaking this parameter. The difference is more pronounced in the NIR than in the optical and even more so in the UV, where the \textit{UVM2} lightcurves is almost 2 mags too faint. As shown in Fig.~\ref{f_acat_lightcurve_UV_comp_z}, the \textit{UVM2} lightcurve is very sensitive to the metallicity, and more during the diffusion peak than on the tail, so the discrepancy in the \textit{UVM2} lightcurve could indicate a subsolar metallicity. However, the \textit{UVM2} lightcurve is also quite sensitive to the mass of the hydrogen envelope and the extinction (which was assumed to be zero in the host galaxy), so those factors may contribute as well.

\begin{figure}[tbp!]
\includegraphics[width=0.49\textwidth,angle=0]{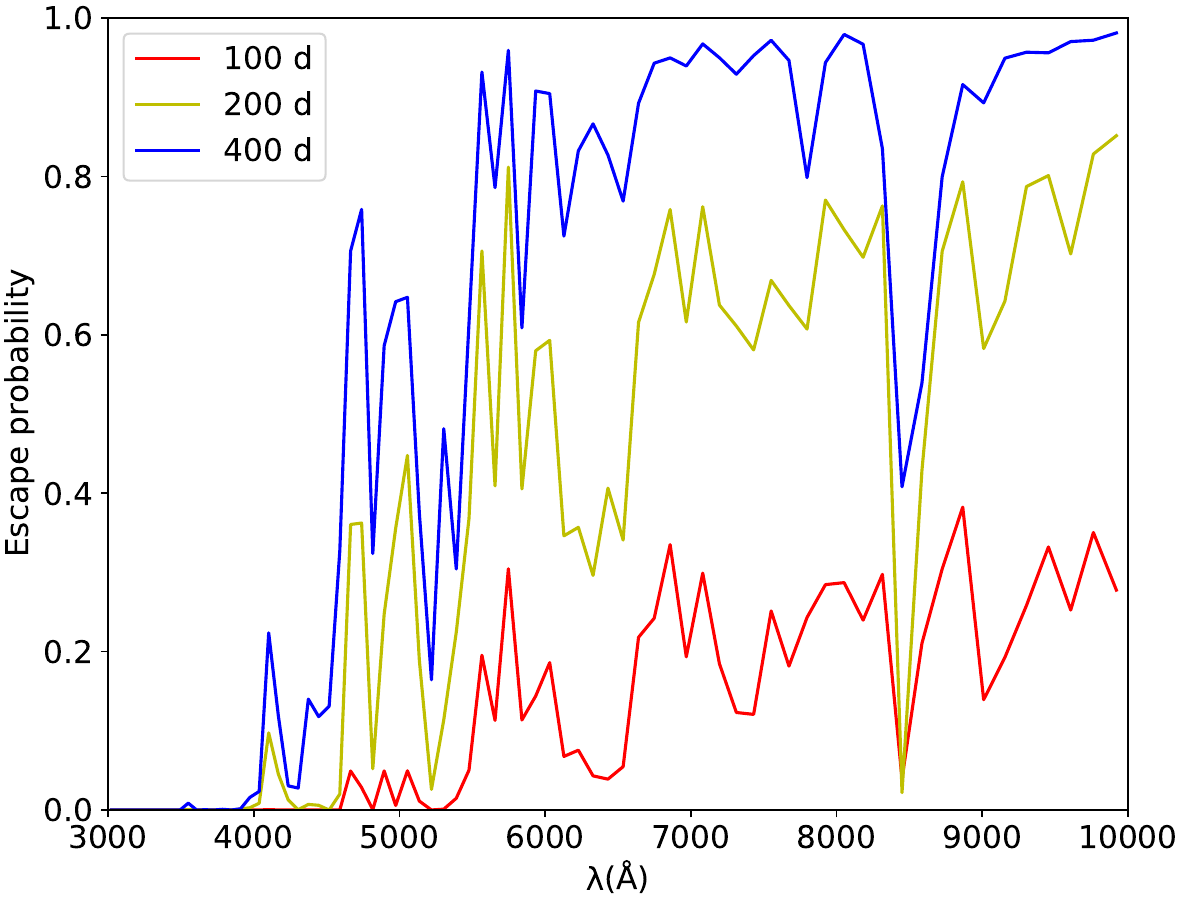}
\caption{Escape probability for photons emitted from the center of the SN as a function of wavelength at 100 (red), 200 (yellow) and 400 (blue) days for the optimal model, For clarity we have binned down the resolution ($\lambda/\Delta\lambda$) to 62.}
\label{f_acat_P_esc_lambda_100d_400d}
\end{figure}

On the tail, we see a growing excess in the NIR, in particular in the $K$ band, which is even more evident in the spectral comparison. This excess is reminiscent of SN 2011dh, where the excess was attributed to dust. However, in the case of SN 2011dh, a strong excess was also seen in the MIR, which underpinned this explanation. After $\sim$100 days a quite strong discrepancy also develops in the $B$ and $V$ bands, which are too bright in the model. We have not found any satisfying explanation for this by varying the parameters in our models, which indicates that the discrepancy originates from some process absent in our models. One such process is the formation of dust in the ejecta, which might absorb more strongly at bluer wavelengths. Another possible explanation is large-scale asymmetries in the ejecta, as the SN is still optically thick in this wavelength region at $\sim$200 days. This explanation is consistent with the fact that the agreement improves again towards $\sim$400 days. In general, it is important to note that the optical depths due to line scattering and fluorescence are quite high in the early nebular phase, so there is considerable reprocessing of the radiation. In particular in the blue, but also at longer wave-lengths. This is illustrated by Fig.~\ref{f_acat_P_esc_lambda_100d_400d}, which shows the escape probability from the center of the SN as a function of wavelength at 100, 200 and 400 days. Another aspect is that a large fraction of the emission from the oxygen-rich clumps is reprocessed in the Ni/He clumps, even at 200 days, so the arrangement of the clumps may also play a role.

\begin{figure}[tbp!]
\includegraphics[width=0.49\textwidth,angle=0]{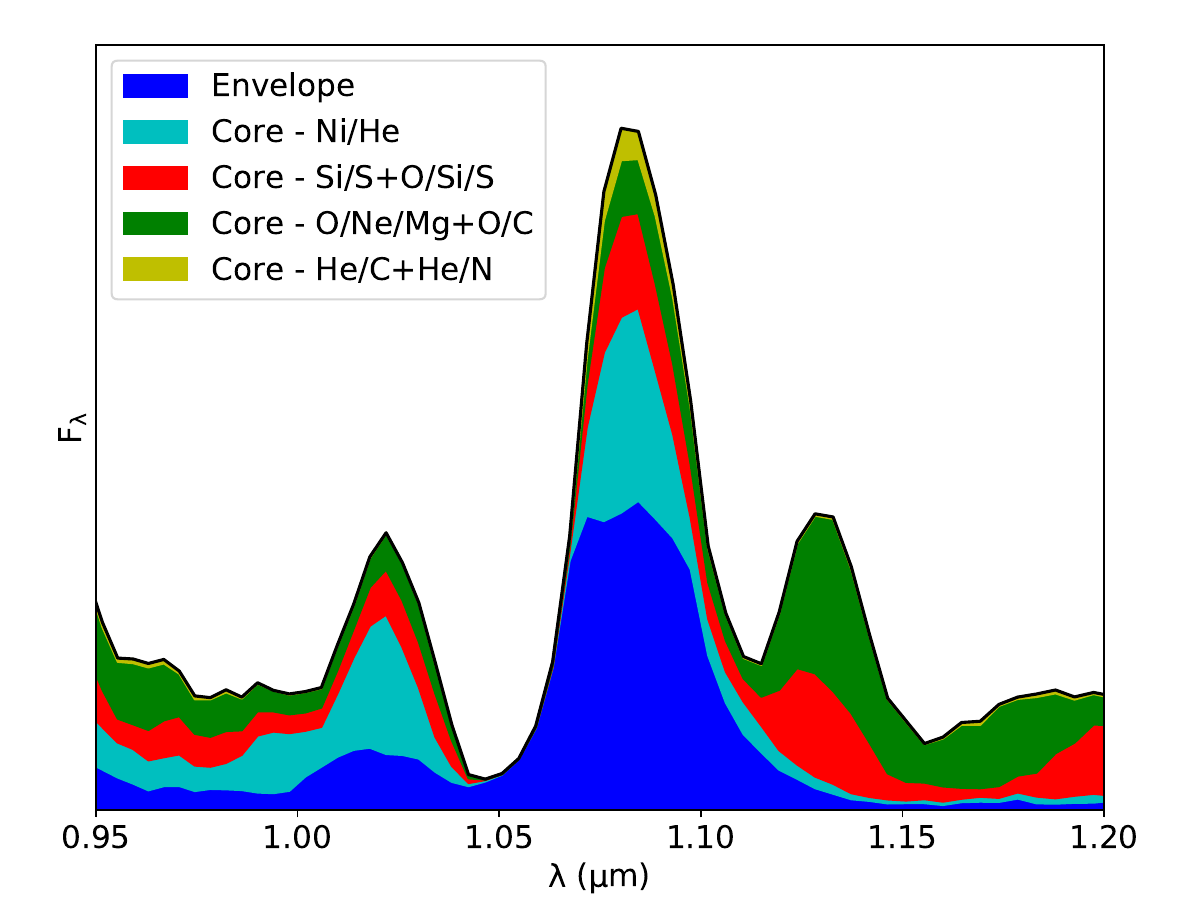}
\caption{Contributions (last emission or scattering event, excluding electron scattering) to the emission in the 1.1 $\mu m$ feature from the envelope (blue) and the Ni/He (cyan), Si/S+O/Si/S (red), O/Ne/Mg+O/C (green) and He/C+He/N (yellow) zones in the core for the optimal model of SN 2020acat at 150 days.}
\label{f_acat_spec_evo_He_1_mu}
\end{figure}

With respect to individual lines, the hydrogen and helium lines are relatively well reproduced (see Figs.~\ref{f_acat_spec_evo_H} and \ref{f_acat_spec_evo_He}) throughout the evolution. Actually, the agreement with observations is better than for the model of SN 2011dh \citepalias[compare][figs.~6 and 7]{Erg18}, likely thanks to the adjustments of the parameters of the hydrogen envelope done in Sect.~\ref{s_constrain}. However, the model fails to reproduce the flat-topped shape of the \ion{He}{i} 1.083 $\mu m$ and \ion{He}{i} 2.058 $\mu m$ lines discussed in \citetalias{Med23}. Focusing on the former, this turns out to be rather tricky as helium, silicon an sulphur in the Ni/He, Si/S and O/Si/S clumps in the core contributes quite strongly to the 1.1 $\mu m$ feature at later times. This is illustrated in Fig.~\ref{f_acat_spec_evo_He_1_mu}, which shows the contributions from the envelope and the different compositional zones in the core to the 1.1 $\mu m$ feature in our optimal model at 150 days. The small amount of helium envelope material mixed into the core in our model does not contribute significantly to the emission, and although it is true that the flat-topped line-profiles suggest weak such mixing \citepalias[see][]{Med23}, this condition is not sufficient to explain the shape of the line-profiles. Instead, it seems like we need to get rid of emission from the explosive nuclear burning material (i.e. the Ni/He, Si/S and O/Si/S zones) in the core. One possible way to achieve this would be to mix most of this material outside the carbon-oxygen core. Such extreme mixing seems a little odd in a spherically symmetric scenario, and might indicate large-scale asymmetries in the ejecta. An alternative explanation is that the explosive nuclear burning occurred in conditions quantitatively different from the original models in \citet{Woo07}; the He content under the NSE can be sensitive to, .e.g., the explosion energy. Note, however, that as the line has a clear P-Cygni profile there is also a scattering component, so even without emission from the core, we would still not expect a fully flat-topped profile as for a pure emission line.

\begin{figure}[tbp!]
\includegraphics[width=0.49\textwidth,angle=0]{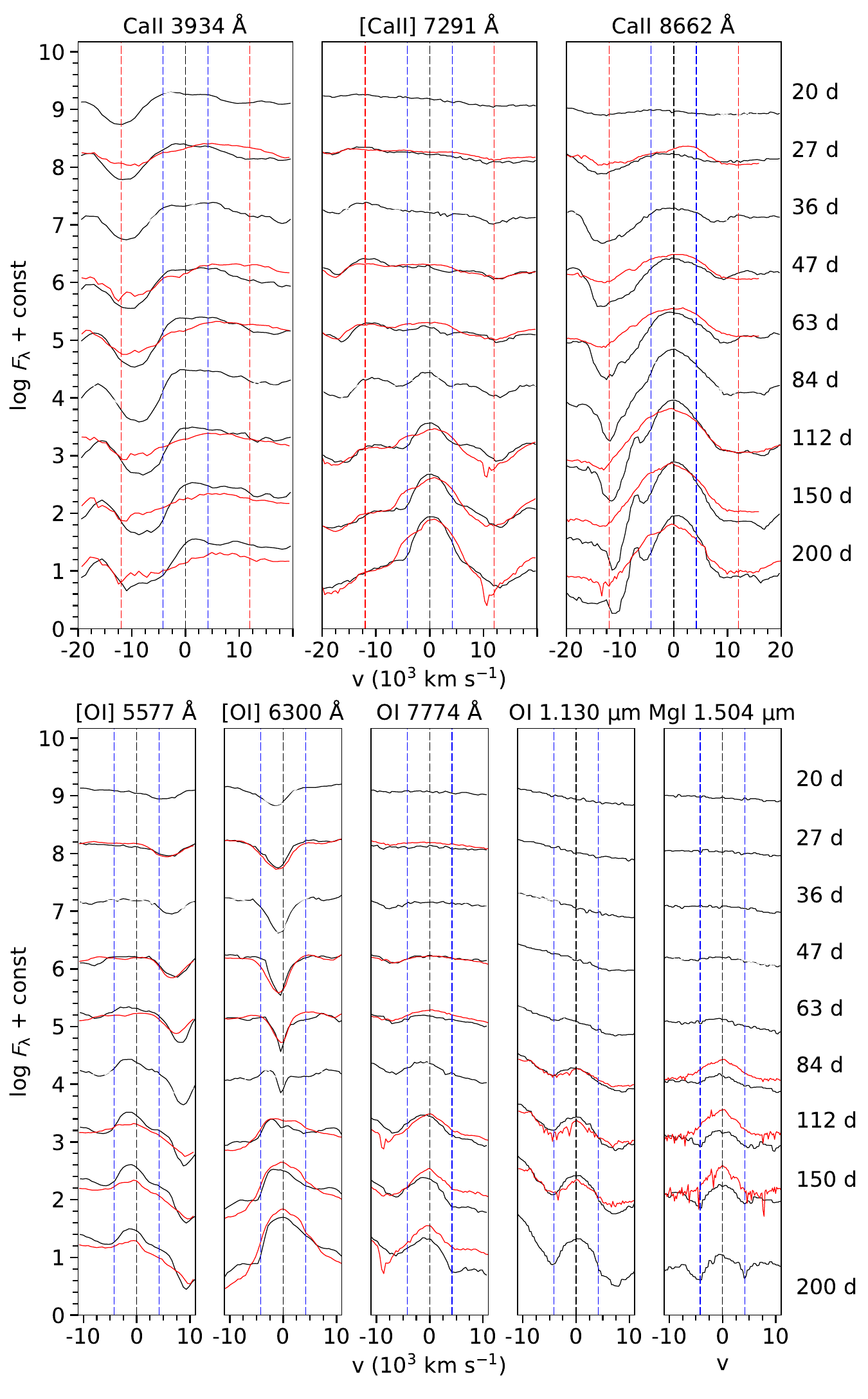}
\caption{Evolution of calcium, oxygen, and magnesium lines for the optimal model (M17-s-s-XH-low; black) compared to the observations of SN 2020acat (red). Spectra from  nine logarithmically spaced epochs between 20 and 200 days are shown. Otherwise the same as in Fig.~\ref{f_acat_spec_evo_H}.}
\label{f_acat_spec_evo_Ca_O_Mg}
\end{figure}

In Fig.~\ref{f_acat_spec_evo_Ca_O_Mg} we show a closeup of the calcium, oxygen and magnesium lines \citepalias[compare][fig.~9]{Erg18}. Like for SN 2011dh, the calcium and oxygen lines are reasonably well reproduced throughout the evolution by our optimal model. However, the \ion{Ca}{ii} NIR triplet and HK lines are overproduced by the model in absorption, a discrepancy not seen in the modelling of SN 2011dh, which has distinctly stronger absorption than SN 2020acat in these lines. The reproduction of the magnesium lines is not so good, where the \ion{Mg}{i} 1.504 $\mu m$ line is too weak in the model, in particular at early times, and the \ion{Mg}{i}] 4571 \AA~line is still absent at 400 days contrary to the observations. A similar discrepancy was seen for SN 2011dh \citep{Erg15,Jer15}, and as discussed in \citet{Jer15}, a possible explanation is the sub-solar magnesium abundance in the \citet{Woo07} models.

\begin{figure}[tbp!]
\includegraphics[width=0.49\textwidth,angle=0]{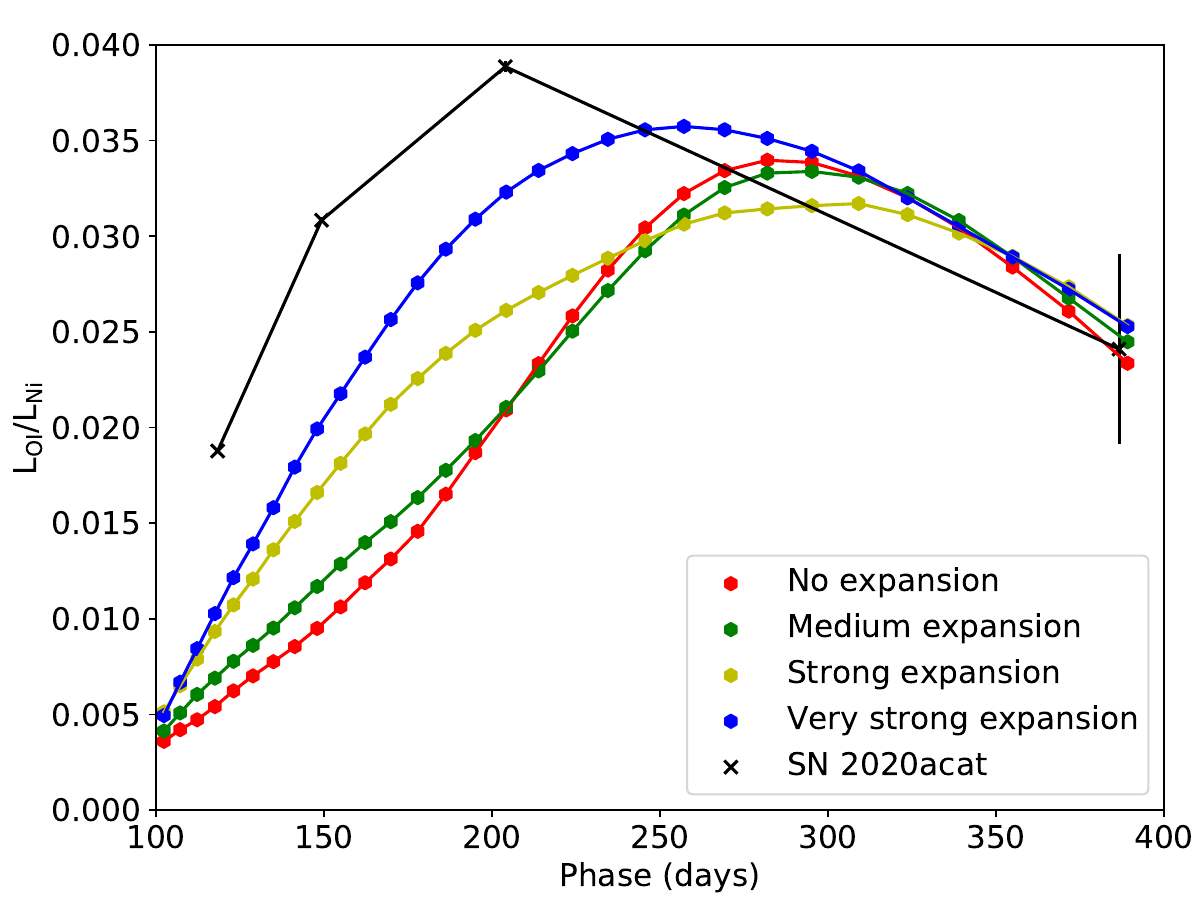}
\caption{Evolution of the luminosity in the [\ion{O}{i}] 6300,6364 \AA~lines normalized with the $^{56}$Ni decay luminosity for the JEKYLL models differing in the expansion of the radioactive material compared to the observations of SN 2020acat (black crosses). In the figure we also show a model with very strong (a contrast factor of 210) expansion of the radioactive material in the core.}
\label{f_acat_spec_line_flux_comp_OI_exp}
\end{figure}

\begin{figure}[tbp!]
\includegraphics[width=0.49\textwidth,angle=0]{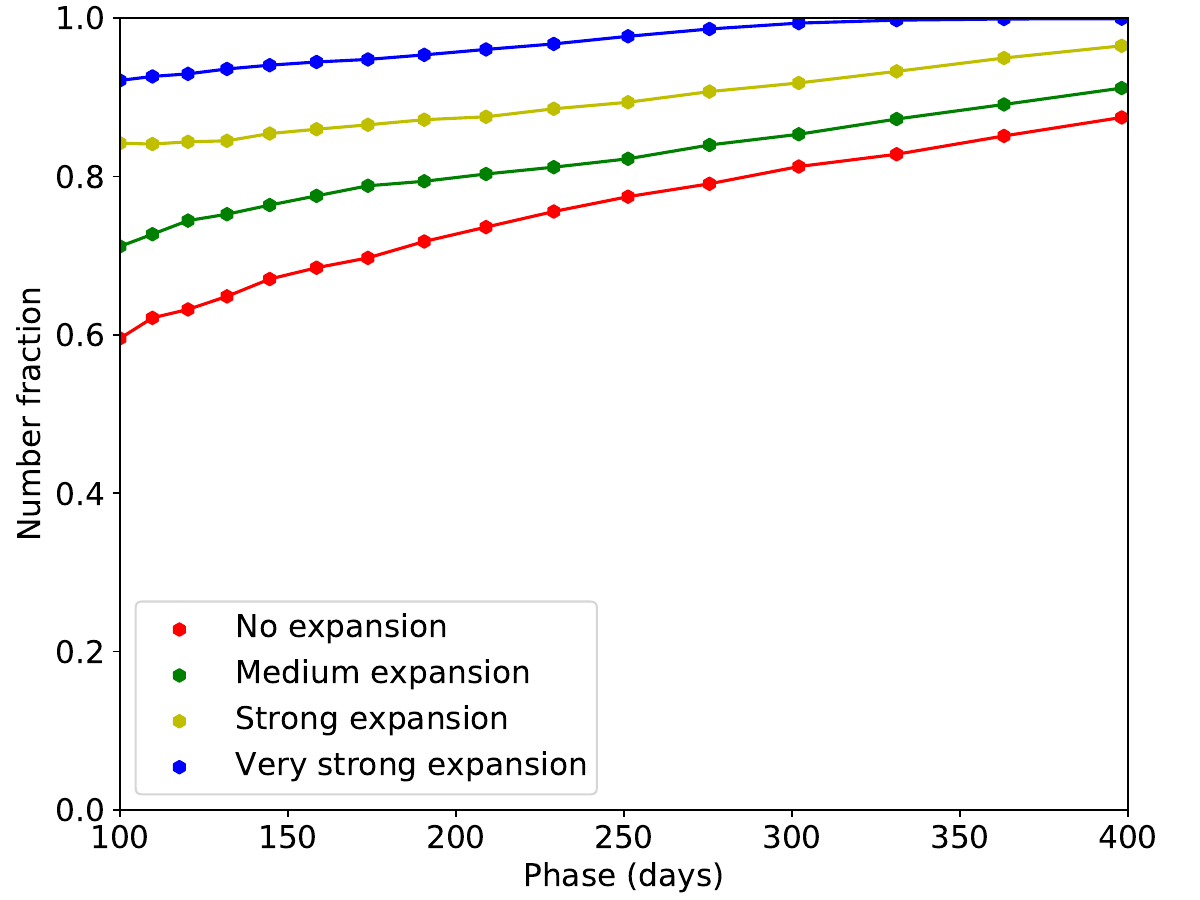}
\caption{Evolution of the fraction of \ion{O}{i} in the oxygen-rich clumps for the JEKYLL models differing in the expansion of the radioactive material. In the figure we also show a model with very strong (a contrast factor of 210) expansion of the radioactive material in the core.}
\label{f_acat_X_OI_evo}
\end{figure}

\begin{figure}[tbp!]
\includegraphics[width=0.49\textwidth,angle=0]{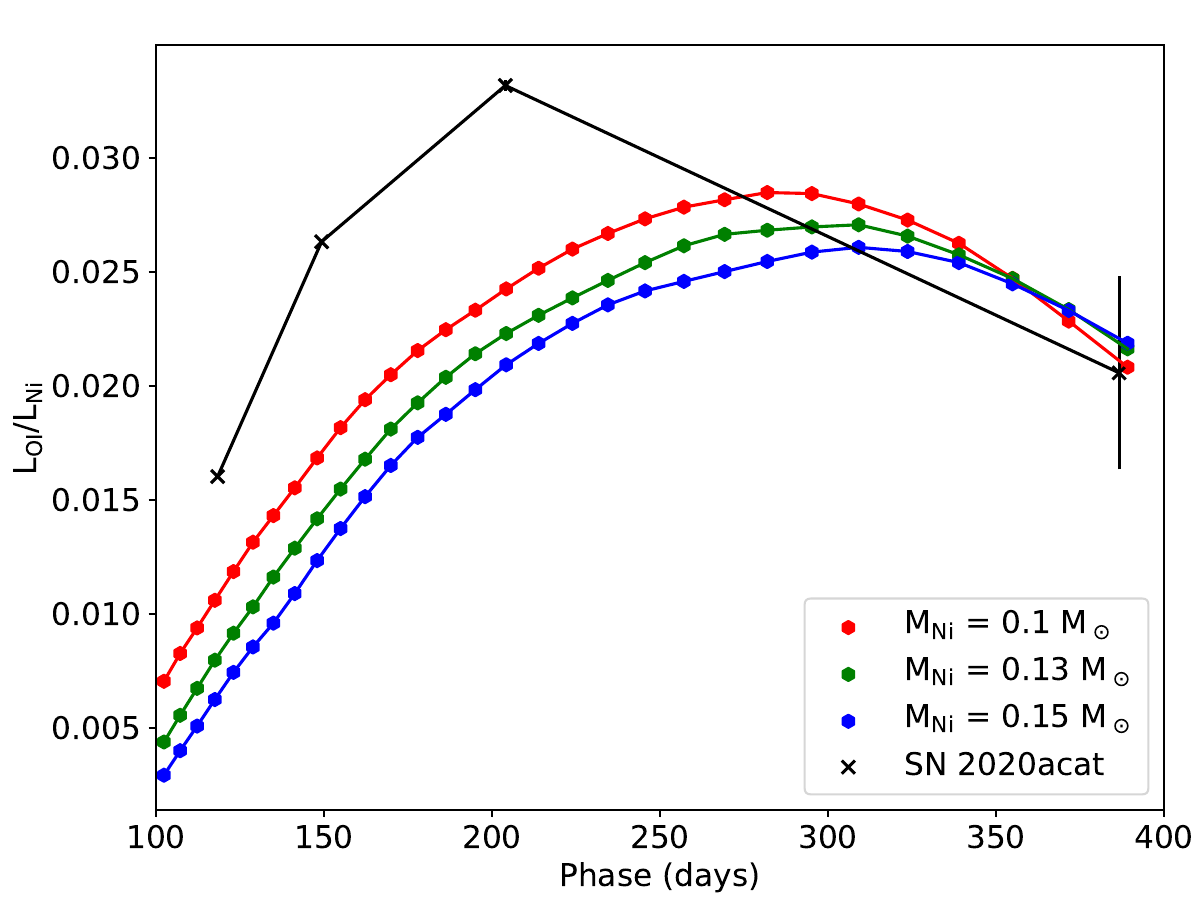}
\caption{Evolution of the luminosity in the [\ion{O}{i}] 6300,6364 \AA~lines normalized with the $^{56}$Ni decay luminosity for JEKYLL models with $^{56}$Ni masses of 0.1, 0.13 and 0.15 M$_\odot$ compared to the observations of SN 2020acat (black crosses).}
\label{f_acat_spec_line_flux_comp_OI_m_ni}
\end{figure}

\begin{figure}[tbp!]
\includegraphics[width=0.49\textwidth,angle=0]{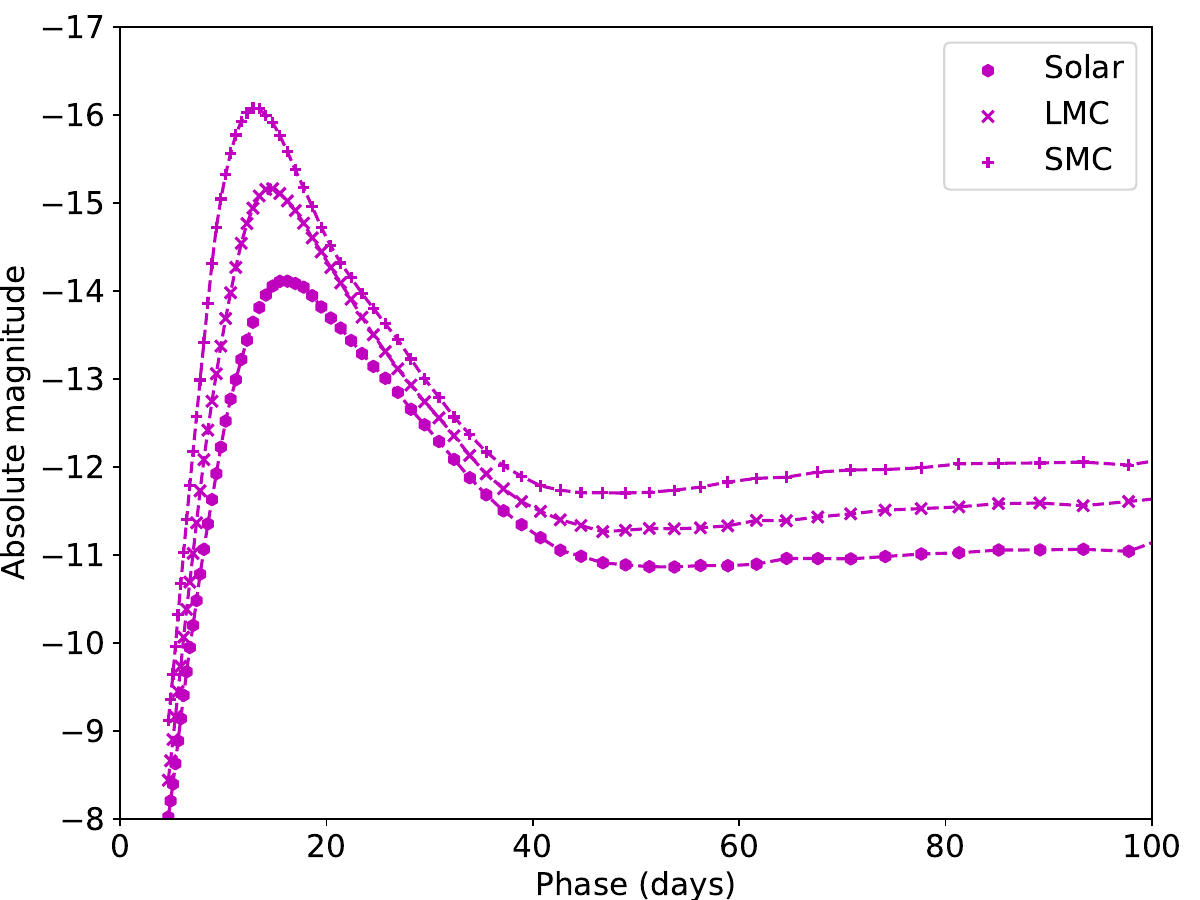}
\caption{\textit{UVM2} lightcurves for JEKYLL models with solar (circles), LMC (crosses) and SMC (pluses) metallicity.}
\label{f_acat_lightcurve_UV_comp_z}
\end{figure}

Also, as pointed out in Sect.~\ref{s_constrain} (and which is more clearly seen in Figs.~\ref{f_acat_spec_line_flux_comp_OI} and \ref{f_acat_spec_line_flux_comp_CaII}), the evolution of the [\ion{O}{i}] 6300,6364 \AA~lines differs somewhat from our models, and is faster. The reason for this is not entirely clear, and we have not been able to tweak our models to fully reproduce the evolution. A stronger expansion of the radioactive material in the core improves the agreement though. This is illustrated by Fig.~\ref{f_acat_spec_line_flux_comp_OI_exp}, where we show the evolution of the luminosity in the [\ion{O}{i}] 6300,6364 \AA~lines for models differing in the expansion of the radioactive material compared to the observations of SN 2020acat. In the figure we also show an additional model with very strong expansion of the radioactive material in the core (a contrast factor of 210). This model better reproduces the evolution in the early nebular phase, while the models with medium or none expansion give a considerably worse match. As shown in Fig.~\ref{f_acat_X_OI_evo}, this is partly explained by a higher fraction of \ion{O}{i} in the models with stronger expansion of the radioactive material, which have a higher density in the compressed oxygen clumps. However, other factors as for example the cooling rates also play a role. A lower mass of $^{56}$Ni, as might be inferred from the uncertainty in the distance, also improves the evolution in the early nebular phase. Likely due to decreased absorption of the [\ion{O}{i}] 6300,6364 \AA~emission in the Ni/He clumps. This is illustrated by Fig.~\ref{f_acat_spec_line_flux_comp_OI_m_ni}, where we show models with $^{56}$Ni masses of 0.1, 0.13 and 0.15 M$_\odot$, which are all consistent with the uncertainty in the distance (see Sect.~\ref{s_dist_ext}). 

Both the models differing in expansion of the radioactive material and the models differing in $^{56}$Ni mass tend to converge towards $\sim$400 days, which speaks in favour of using later epochs when trying to estimate the initial mass. This is likely due to a combination of increasing \ion{O}{i} fraction, increasing importance of the [\ion{O}{i}] 6300,6364 \AA~cooling, and decreasing absorption of the [\ion{O}{i}] 6300,6364 \AA~emission. Note, however, that at later epochs molecule cooling (which is not accounted for by JEKYLL) might decrease the [\ion{O}{i}] 6300,6300 \AA~emission from the O/Si/S and O/C clumps \citep[see e.g.][]{Jer15}. This is mainly a problem for models with relatively low initial mass, as the O/Ne/Mg zone dominates in more massive models like our optimal 17 M$_\odot$ model (see Appendix~\ref{a_ejecta_models}). Note also, that as previously discussed the optical depths are still relatively high in the early nebular phase. Only about 40-45 percent of the [\ion{O}{i}] 6300,6300 \AA~emission escapes at 150-200 days, whereas 90 percent escapes at 400 days. This means that like the emission in the $B$ and $V$ bands, the [\ion{O}{i}] 6300,6364 \AA~emission in the early nebular phase can be affected by large-scale asymmetries in the ejecta, which might provide an alternative explanation to the evolution of the [\ion{O}{i}] 6300,6364 \AA~lines.

\begin{figure}[tbp!]
\includegraphics[width=0.49\textwidth,angle=0]{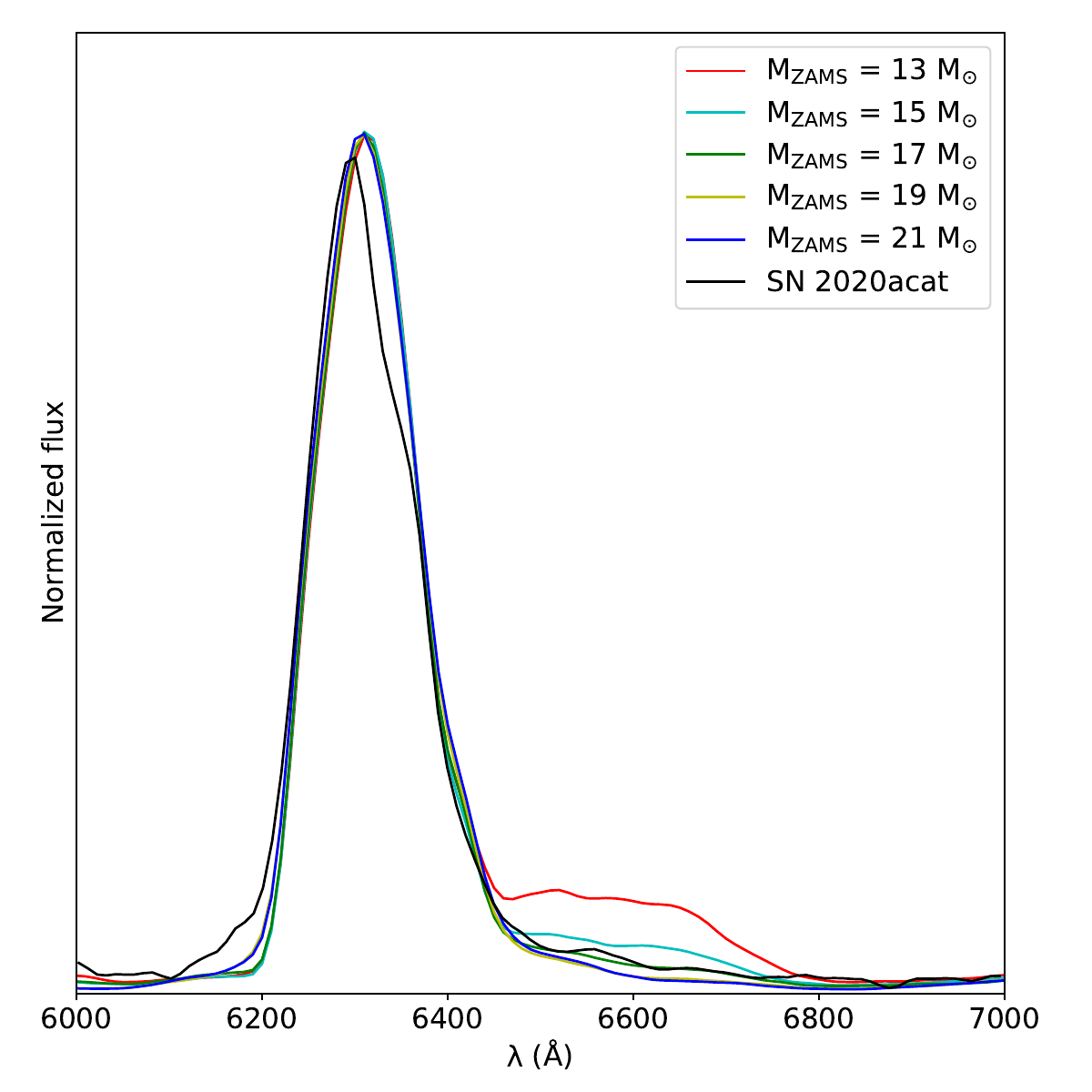}
\caption{[\ion{O}{i}] 6300, 6364 \AA~and [\ion{N}{ii}] 6548,6583 lines at 400 days normalized to the peak flux of the former for the JEKYLL models differing in initial mass compared to the observations of SN 2020acat.}
\label{f_acat_spec_comp_OI_NII}
\end{figure}

It is interesting to note that the quite strong [\ion{N}{ii}] lines at 6548,6583 \AA~emerging on the red shoulder of the [\ion{O}{i}] 6300,6364 \AA~lines~towards $\sim$300 days in SN 2011dh \citep[see][]{Jer15} seem to be much weaker for SN 2020acat. This difference is well reproduced by our optimal models for SNe 2011dh and 2020acat, and as the [\ion{N}{ii}] 6548,6583 \AA~lines originate from the He/N zone, it is explained by the much lower fraction of such material in models with higher initial mass (see \citealt{Jer15} for further discussion of this). The observed ratio of the [\ion{O}{i}] 6300,6364 \AA~and [\ion{N}{ii}] 6548,6583 \AA~lines gives further support for our conclusion that SN 2020acat originates from a progenitor with a considerably higher initial mass than SN 2011dh. This is illustrated by Fig.~\ref{f_acat_spec_comp_OI_NII}, where we show the [\ion{O}{i}] 6300,6364 \AA~and [\ion{N}{ii}] 6548,6583 \AA~lines at 400 days normalized by the peak flux of the former for our models differing in initial mass compared to the observations of SN 2020acat. As seen, the 17 M$_\odot$ model agrees best with the observations of SN 2020acat, whereas the 13 M$_\odot$ model seems to be excluded. 

Finally, we reiterate that some of the discrepancies between our optimal model and the observations of SN 2020acat might be related to large-scale asymmetries in the ejecta. Actually, a jet-disk-like geometry as the one proposed for SN 1998w by \citet{Mae03}, where most of the Ni/He material is in the jet-like component, and most of the oxygen material is in the disk-like component can not be excluded. Such a geometry also could provide an alternative explanation for the tension between the diffusion and the tail phases. However, as JEKYLL currently assumes spherical asymmetry on average, this hypothesis, as well as the effect of any other possible large-scale asymmetry can not be tested, and we leave such an investigation for future work.

\section{Conclusions}
\label{s_conclusions}

We present a set of Type IIb SN models calculated with the NLTE lightcurve and spectral synthesis code JEKYLL and compare those to observations of the Type IIb SN 2020acat. The bulk of the observations were presented in \citetalias{Med22} and \citetalias{Med23}, but we also present new late-time optical observations, and refine the photometry by applying S-corrections. To constrain the SN parameters for SN 2020acat, we have explored a parameter space in initial mass, mixing and expansion of the radioactive material, and the mass of the hydrogen envelope and the mass-fraction of hydrogen therein. In our phenomenological models, which are based on results from hydrodynamical models, the kinetic energy is fixed by the observed velocities.

The comparisons shows that a model with an initial mass of 17 M$_\odot$, strong mixing and expansion of the radioactive material, and an 0.1 M$_\odot$ hydrogen envelope with a low hydrogen mass-fraction ($X_\mathrm{H}$=0.027) gives the best overall agreement to the observations of SN 2020acat. Models with initial masses below 15 M$_\odot$ and above 19 M$_\odot$ seem to be excluded, as well as models which do not have strong expansion and at least medium mixing of the radioactive material. To be more precise, in our model grid the strong expansion scenario corresponds to a contrast factor of 60 in the core and 30 in the envelope, and the medium mixing scenario to 50 percent of the radioactive material mixed into the inner half of the helium envelope. Note, however, that the degree of expansion required depends on the assumed clump size, and with larger clumps less expansion would be needed.

Nevertheless, the strong expansion of the clumps containing radioactive material is a particularly interesting result. Without strong expansion of the "Ni bubbles", there is a tension between the diffusion phase and the subsequent evolution, and models that fit the nebular phase give rise to diffusion peaks that are too broad. This is in line with the results for SN 2011dh in \citetalias{Erg22}, where models without such expansion resulted in diffusion peaks broader than observed. As discussed in detail in \citetalias{Erg22}, the expansion of the "Ni bubbles" decrease the effective opacity and the diffusion time, and the width of the diffusion peak is therefore sensitive to this. The effect of the expansion of the "Ni bubbles" on the diffusion phase lightcurves has not been taken into account in previous lightcurve modelling of Type IIb and other SE SNe, and their ejecta masses might therefore have been systematically underestimated. It should be cautioned though, that the magnitude of the effect is uncertain, depends on weakly constrained properties of the 3-D ejecta structure, and might vary in different SNe, so further work is needed.

A tension between the diffusion phase and the tail phase, like the one we find for SN 2020acat, has been reported for other SE SNe \citep[e.g.][]{Whe15,Nag22}, and in particular for Type Ic-BL SNe \citep[e.g.][]{Mae03,Des17}, and typically large-scale asymmetries have been proposed to explain this. As JEKYLL currently assumes a geometry that is spherically symmetric on average, we can not rule out that large-scale asymmetries play a role in the case of SN 2020acat, and some evidence may also point in that direction. However, what we do show, is that small-scale asymmetries caused by expansion of the radioactive material may naturally resolve the tension, and provides an alternative explanation. To fully understand which role small- and large-scale asymmetries in the ejecta play, and to disentangle the effects they have on the observed lightcurves and spectra of SE SNe, full-fledged 3-D NLTE simulations, preferably based on 3-D explosion models, are needed. 

A detailed comparison of our optimal model with the observations of SN 2020acat is presented and the overall agreement is reasonably good, although distinct differences also exist. For example, our models do not fully reproduce the evolution of the flux in the [\ion{O}{i}] 6300,6364 \AA~lines, which is faster in SN 2020acat. A quite strong discrepancy between the optimal model and the observations of SN 2020acat also emerges in the $B$- and $V$-band lightcurves towards $\sim$200 days, although the agreement improves again towards $\sim$400 days. In addition, a growing excess emerges in the $K$ band after $\sim$100 days. This excess is reminiscent of SN 2011dh, where an IR excess attributed to dust developed at a similar timescale \citep[see][]{Erg15,Jer15}. Finally, our models are unable to reproduce the flat-topped line-profile of \ion{He}{i} 1.083 $\mu m$ emerging after $\sim$100 days discussed in \citetalias{Med23}. This might be interpreted as evidence of extreme mixing of the explosive nuclear burning material out of the carbon-oxygen core, but explosive nuclear burning in conditions different from those in the \citet{Woo07} models could also help to reduce helium emission from the core.

The relatively high initial mass of $\sim$17 M$_\odot$ estimated for the progenitor of SN 2020acat places it in the upper end of the mass distribution of Type IIb SN progenitors. \citet{Jer15} estimated initial masses well below 17 M$_\odot$ for the progenitors of SNe 2008ax, 2011dh and 1993J using modelling of their nebular spectra, which for the latter two is supported by stellar-evolutionary analysis of pre-explosion imaging of the progenitors. At the relatively high initial mass estimated for the progenitor of SN 2020acat a single star origin can not be excluded. Note, however, that the low estimated mass-fraction of hydrogen in the hydrogen envelope may be more in line with a binary origin. The modelling presented in this paper further demonstrates the capabilities of the JEKYLL code to self-consistently model the evolution of SNe from early to late times, and how this can be used to constrain the properties of SNe and their progenitor stars.

\begin{acknowledgements} 
This work has been supported by grants from the Swedish Research Council and the Swedish National Space Board, and the computations were performed with resources provided by the Swedish National Infrastructure for Computing (SNIC) at Parallelldator-centrum (PDC). H.K was funded by the Academy of Finland projects 324504 and 328898. K.M. acknowledges support from the Japan Society for the Promotion of Science (JSPS) KAKENHI grant JP18H05223, JP20H00174, and JP20H04737. K.M. and H.K. acknowledge support by the JSPS Open Partnership Bilateral Joint Research Projects between Japan and Finland (JPJSBP120229923). A.P. acknowledges support from PRIN-INAF 2022 project "Exploring new frontiers of the transient universe in the era of synoptic surveys". M. S. is funded by the Independent Research Fund Denmark (IRFD), grant number 10.46540/2032-00022B). We also thank Luc Dessart for fruitful and always interesting discussions on the effect of clumping on the lightcurves and spectra of SNe.
\end{acknowledgements}

\bibliographystyle{aa}
\bibliography{SN2020acat-paper}

\begin{appendix}

\section{Configuration}
\label{a_configuration}

JEKYLL was configured to run in time-dependent mode (with respect to the radiative transfer), and to use a full NLTE solution including the following; radiative bound-bound, bound-free and free-free processes, collisional bound-bound and bound-free processes, non-thermal excitation, ionisation and heating, as well as two-photon processes and charge-transfer. Note, however, that before 100 days charge-transfer was not included and non-thermal excitation was only included for He. The diffusion solver was used above an optical depth of 50, and a recombination correction based on the total recombination rates was used while still enforcing detailed balance. In addition, packet control \citepalias{Erg18} was turned on to assure good sampling of the radiation field in all frequency regions. The number of $\Lambda$-iterations per time-step was set to 4. As discussed in \citetalias{Erg18}, this gives a well converged solution, which has also been verified for the models used in this paper.

\section{Atomic data}
\label{a_atomic_data}

The used atomic dataset is the default choice described in \citetalias{Erg18} but has been extended with more levels and a full NLTE solution for ionisation stages V and VI. This makes only a small difference for the observed light curves and spectra, and tests show that the simulations are not sensitive to further changes in the number of levels and ionisation stages. Using online data provided by NIST\footnote{\url{www.nist.gov}} (National Institute of Standards and Technology) and R. Kurucz\footnote{\url{www.cfa.harvard.edu/amp/ampdata/kurucz23/sekur.html}}, these ions were updated to include 100 levels (or as many as were available) for elements lighter than Scandium and 300 levels (or as many as were available) for heavier elements. Total recombination rates for these ions were adopted from the online table provided by S. Nahar\footnote{\url{www.astronomy.ohio-state.edu/~nahar/_naharradiativeatomicdata/}} whenever available, and otherwise from \citet{Shu82}.

\section{Ejecta models}
\label{a_ejecta_models}

In Table \ref{t_M} we list the mass and in Tables \ref{t_X_M13}-\ref{t_X_M21} the composition for each zone in our models with initial masses of 13, 15, 17, 19, and 21 M$_\odot$. The ejecta mass and (kinetic) energy of these models are also listed in Table \ref{t_M}.

\begin{table*}[tb]
\caption{Ejecta mass (M$_\odot$), (kinetic) ejecta energy (B) and zone masses (M$_\odot$) for models with initial masses 13, 15, 17, 19, and 21 M$_\odot$.}
\begin{center}
\begin{tabular}{lllllllllll}
\toprule
M$_\mathrm{ZAMS}$ & M$_\mathrm{ej}$ & E$_\mathrm{ej}$ & Ni/He & Si/S & O/Si/S & O/Ne/Mg & O/C & He/C & He/N & H\\
\midrule
13 & 2.1e+00 & 8.2e-01 & 1.5e-01 & 6.8e-02 & 1.8e-01 & 3.1e-01 & 2.5e-01 & 2.4e-01 & 8.4e-01 & 5.0e-02\\
15 & 2.6e+00 & 9.2e-01 & 1.8e-01 & 8.0e-02 & 2.4e-01 & 4.5e-01 & 4.3e-01 & 7.5e-01 & 4.2e-01 & 5.0e-02\\
17 & 3.5e+00 & 1.0e+00 & 1.7e-01 & 1.1e-01 & 2.7e-01 & 1.2e+00 & 5.8e-01 & 9.3e-01 & 2.2e-01 & 5.0e-02\\
19 & 4.5e+00 & 1.3e+00 & 3.8e-01 & 1.2e-01 & 1.9e-01 & 1.9e+00 & 2.6e-01 & 1.2e+00 & 3.2e-01 & 5.0e-02\\
21 & 5.4e+00 & 1.4e+00 & 3.7e-01 & 1.1e-01 & 1.0e-01 & 2.8e+00 & 4.0e-01 & 1.3e+00 & 2.6e-01 & 5.0e-02\\
\bottomrule
\end{tabular}
\end{center}
\label{t_M}
\end{table*}

\begin{table*}[tb]
\caption{Zone composition for models with an initial mass of 13 M$_\odot$.}
\begin{center}
\begin{tabular}{lllllllll}
\toprule
Element & Ni/He & Si/S & O/Si/S & O/Ne/Mg & O/C & He/C & He/N & H\\
\midrule
H & 5.5e-06 & 8.9e-07 & 4.5e-08 & 3.7e-09 & 1.5e-09 & 8.0e-10 & 1.3e-07 & 5.4e-01\\
He & 1.5e-01 & 9.1e-06 & 5.0e-06 & 3.6e-06 & 4.2e-02 & 8.2e-01 & 9.9e-01 & 4.4e-01\\
C & 3.3e-07 & 2.0e-05 & 1.3e-03 & 6.6e-03 & 2.5e-01 & 1.5e-01 & 4.2e-04 & 1.2e-04\\
N & 2.0e-06 & 5.1e-07 & 2.9e-05 & 3.5e-05 & 1.3e-05 & 4.1e-05 & 8.4e-03 & 1.0e-02\\
O & 9.1e-06 & 1.1e-02 & 7.5e-01 & 7.2e-01 & 6.4e-01 & 1.3e-02 & 7.8e-04 & 4.7e-03\\
Ne & 1.1e-05 & 1.8e-05 & 2.4e-03 & 1.4e-01 & 5.6e-02 & 1.4e-02 & 1.4e-03 & 3.0e-03\\
Na & 7.0e-07 & 9.0e-07 & 3.7e-05 & 9.6e-04 & 1.9e-04 & 1.9e-04 & 1.7e-04 & 7.3e-05\\
Mg & 2.0e-05 & 1.4e-04 & 4.8e-02 & 9.8e-02 & 1.5e-02 & 1.9e-03 & 7.2e-04 & 7.2e-04\\
Al & 1.4e-05 & 2.2e-04 & 4.7e-03 & 8.0e-03 & 1.1e-04 & 6.5e-05 & 7.6e-05 & 6.9e-05\\
Si & 2.9e-03 & 3.9e-01 & 1.5e-01 & 2.3e-02 & 9.9e-04 & 8.6e-04 & 8.2e-04 & 8.2e-04\\
S & 5.5e-03 & 3.8e-01 & 3.4e-02 & 7.1e-04 & 2.4e-04 & 3.8e-04 & 4.2e-04 & 4.2e-04\\
Ar & 1.7e-03 & 5.8e-02 & 3.8e-03 & 8.2e-05 & 7.9e-05 & 9.7e-05 & 1.1e-04 & 1.1e-04\\
Ca & 3.5e-03 & 4.0e-02 & 1.0e-03 & 3.4e-05 & 2.7e-05 & 6.1e-05 & 7.4e-05 & 7.4e-05\\
Sc & 2.2e-07 & 4.9e-07 & 4.3e-07 & 1.5e-06 & 1.3e-06 & 3.9e-07 & 6.1e-08 & 4.5e-08\\
Ti & 8.4e-04 & 5.2e-04 & 2.3e-05 & 5.6e-06 & 5.1e-06 & 3.4e-06 & 3.4e-06 & 3.4e-06\\
V & 3.2e-05 & 1.3e-04 & 4.2e-06 & 6.0e-07 & 7.1e-07 & 5.2e-07 & 4.5e-07 & 4.3e-07\\
Cr & 2.4e-03 & 7.0e-03 & 7.6e-05 & 1.5e-05 & 1.2e-05 & 1.9e-05 & 2.0e-05 & 2.0e-05\\
Mn & 1.7e-05 & 2.1e-04 & 1.2e-05 & 5.7e-06 & 4.2e-06 & 1.0e-05 & 1.6e-05 & 1.5e-05\\
Fe & 2.8e-03 & 4.1e-02 & 9.3e-04 & 8.8e-04 & 8.0e-04 & 1.3e-03 & 1.4e-03 & 1.4e-03\\
Co & 3.1e-08 & 1.8e-08 & 1.3e-04 & 1.3e-04 & 1.8e-04 & 6.7e-05 & 4.4e-06 & 4.0e-06\\
Ni & 3.2e-02 & 2.4e-03 & 5.9e-04 & 4.5e-04 & 4.5e-04 & 9.3e-05 & 8.2e-05 & 8.2e-05\\
$^{56}$Ni & 7.7e-01 & 7.2e-02 & 4.8e-06 & 3.0e-05 & 1.3e-05 & 1.3e-06 & 2.5e-08 & 5.6e-11\\
$^{57}$Ni & 3.3e-02 & 1.5e-03 & 9.6e-06 & 1.4e-06 & 3.0e-08 & 7.4e-09 & 3.0e-09 & 1.7e-11\\
$^{44}$Ti & 2.7e-04 & 2.0e-05 & 3.1e-07 & 2.6e-10 & 7.4e-12 & 2.0e-13 & 1.4e-13 & 6.1e-16\\
\bottomrule
\end{tabular}
\end{center}
\label{t_X_M13}
\end{table*}

\begin{table*}[tb]
\caption{Zone composition for models with an initial mass of 15 M$_\odot$.}
\begin{center}
\begin{tabular}{lllllllll}
\toprule
H & 2.6e-06 & 4.0e-07 & 7.4e-09 & 1.6e-09 & 1.0e-15 & 1.0e-15 & 4.5e-08 & 5.4e-01\\
He & 2.4e-01 & 9.5e-06 & 4.0e-06 & 3.7e-06 & 1.6e-04 & 9.4e-01 & 9.9e-01 & 4.4e-01\\
C & 1.5e-06 & 1.3e-06 & 3.9e-04 & 8.5e-03 & 2.0e-01 & 3.9e-02 & 2.4e-04 & 1.2e-04\\
N & 2.4e-06 & 1.0e-15 & 2.9e-05 & 6.9e-05 & 1.3e-05 & 2.7e-03 & 9.1e-03 & 1.0e-02\\
O & 1.8e-05 & 1.0e-05 & 8.1e-01 & 6.8e-01 & 7.3e-01 & 5.6e-03 & 1.8e-04 & 3.2e-03\\
Ne & 2.0e-05 & 7.3e-06 & 1.4e-04 & 2.3e-01 & 5.0e-02 & 6.8e-03 & 1.1e-03 & 3.0e-03\\
Na & 9.0e-07 & 7.9e-07 & 2.1e-05 & 5.2e-03 & 1.9e-04 & 1.8e-04 & 1.8e-04 & 7.9e-05\\
Mg & 4.2e-05 & 1.4e-04 & 4.5e-02 & 6.2e-02 & 1.6e-02 & 7.3e-04 & 7.0e-04 & 7.2e-04\\
Al & 1.0e-05 & 2.0e-04 & 4.4e-03 & 4.0e-03 & 1.2e-04 & 7.2e-05 & 9.0e-05 & 6.9e-05\\
Si & 2.5e-04 & 4.1e-01 & 1.2e-01 & 4.8e-03 & 9.4e-04 & 8.2e-04 & 8.2e-04 & 8.2e-04\\
S & 2.3e-04 & 3.9e-01 & 1.9e-02 & 2.9e-04 & 2.2e-04 & 4.2e-04 & 4.2e-04 & 4.2e-04\\
Ar & 2.4e-04 & 5.5e-02 & 5.9e-04 & 8.4e-05 & 8.6e-05 & 1.1e-04 & 1.1e-04 & 1.1e-04\\
Ca & 2.8e-03 & 3.5e-02 & 2.8e-05 & 3.6e-05 & 2.6e-05 & 7.3e-05 & 7.4e-05 & 7.4e-05\\
Sc & 2.3e-07 & 2.2e-07 & 1.2e-07 & 1.2e-06 & 1.6e-06 & 7.3e-08 & 4.5e-08 & 4.5e-08\\
Ti & 1.7e-03 & 5.6e-04 & 8.1e-06 & 5.6e-06 & 7.0e-06 & 3.4e-06 & 3.4e-06 & 3.4e-06\\
V & 3.4e-05 & 1.5e-04 & 3.0e-06 & 4.1e-07 & 4.5e-07 & 4.7e-07 & 4.3e-07 & 4.3e-07\\
Cr & 2.4e-03 & 6.6e-03 & 8.2e-06 & 1.5e-05 & 1.2e-05 & 2.0e-05 & 2.0e-05 & 2.0e-05\\
Mn & 1.8e-06 & 2.8e-04 & 1.7e-06 & 6.7e-06 & 2.2e-06 & 1.7e-05 & 1.5e-05 & 1.5e-05\\
Fe & 7.9e-04 & 4.2e-02 & 3.4e-04 & 8.5e-04 & 5.5e-04 & 1.4e-03 & 1.4e-03 & 1.4e-03\\
Co & 2.4e-08 & 2.0e-09 & 1.9e-04 & 1.8e-04 & 2.0e-04 & 4.8e-06 & 4.0e-06 & 4.0e-06\\
Ni & 2.9e-02 & 2.3e-03 & 8.6e-04 & 4.4e-04 & 6.7e-04 & 8.2e-05 & 8.2e-05 & 8.2e-05\\
$^{56}$Ni & 6.9e-01 & 5.7e-02 & 1.9e-07 & 3.5e-07 & 1.1e-08 & 7.5e-08 & 2.0e-08 & 1.2e-10\\
$^{57}$Ni & 3.4e-02 & 1.4e-03 & 3.1e-06 & 2.6e-07 & 1.3e-08 & 6.6e-09 & 1.0e-15 & 2.2e-11\\
$^{44}$Ti & 5.3e-04 & 1.5e-05 & 1.1e-09 & 1.0e-15 & 1.0e-15 & 1.0e-15 & 1.0e-15 & 1.0e-15\\
\bottomrule
\end{tabular}
\end{center}
\label{t_X_M15}
\end{table*}

\begin{table*}[tb]
\caption{Zone composition for models with an initial mass of 17 M$_\odot$.}
\begin{center}
\begin{tabular}{lllllllll}
\toprule
Element & Ni/He & Si/S & O/Si/S & O/Ne/Mg & O/C & He/C & He/N & H\\
\midrule
H & 2.5e-06 & 1.2e-07 & 6.1e-08 & 1.7e-09 & 3.4e-10 & 2.5e-11 & 3.8e-08 & 5.4e-01\\
He & 1.3e-01 & 7.7e-06 & 3.3e-06 & 2.9e-06 & 4.5e-02 & 9.3e-01 & 9.9e-01 & 4.4e-01\\
C & 3.5e-07 & 2.0e-05 & 6.9e-05 & 1.5e-02 & 2.4e-01 & 4.5e-02 & 2.5e-04 & 1.2e-04\\
N & 1.5e-06 & 8.0e-07 & 1.3e-05 & 3.8e-05 & 1.1e-05 & 1.1e-03 & 9.1e-03 & 1.0e-02\\
O & 8.1e-06 & 1.6e-02 & 2.6e-01 & 6.9e-01 & 6.8e-01 & 1.1e-02 & 1.7e-04 & 3.2e-03\\
Ne & 9.3e-06 & 2.5e-05 & 1.1e-04 & 2.1e-01 & 2.2e-02 & 9.2e-03 & 1.1e-03 & 3.0e-03\\
Na & 9.0e-07 & 1.1e-06 & 1.3e-06 & 5.1e-03 & 2.0e-04 & 1.8e-04 & 1.8e-04 & 7.9e-05\\
Mg & 1.9e-05 & 1.9e-04 & 5.5e-04 & 5.8e-02 & 6.7e-03 & 7.4e-04 & 7.0e-04 & 7.2e-04\\
Al & 2.7e-05 & 2.8e-04 & 2.5e-04 & 4.5e-03 & 7.4e-05 & 7.3e-05 & 9.5e-05 & 6.9e-05\\
Si & 1.5e-02 & 4.3e-01 & 3.5e-01 & 1.3e-02 & 9.0e-04 & 8.3e-04 & 8.2e-04 & 8.2e-04\\
S & 2.7e-02 & 3.8e-01 & 3.2e-01 & 2.8e-03 & 3.0e-04 & 4.1e-04 & 4.2e-04 & 4.2e-04\\
Ar & 7.6e-03 & 5.3e-02 & 5.5e-02 & 4.1e-04 & 8.6e-05 & 1.1e-04 & 1.1e-04 & 1.1e-04\\
Ca & 1.1e-02 & 3.2e-02 & 2.2e-02 & 1.5e-04 & 4.4e-05 & 7.3e-05 & 7.4e-05 & 7.4e-05\\
Sc & 3.2e-07 & 6.2e-07 & 1.3e-06 & 1.4e-06 & 7.1e-07 & 8.8e-08 & 4.5e-08 & 4.5e-08\\
Ti & 1.1e-03 & 3.2e-04 & 1.6e-04 & 6.7e-06 & 4.9e-06 & 3.4e-06 & 3.4e-06 & 3.4e-06\\
V & 7.1e-05 & 1.2e-04 & 1.2e-05 & 6.5e-07 & 3.2e-07 & 4.9e-07 & 4.3e-07 & 4.3e-07\\
Cr & 7.4e-03 & 4.2e-03 & 2.0e-04 & 1.4e-05 & 1.6e-05 & 2.0e-05 & 2.0e-05 & 2.0e-05\\
Mn & 1.5e-04 & 2.9e-04 & 1.7e-05 & 5.4e-06 & 7.8e-06 & 1.6e-05 & 1.5e-05 & 1.5e-05\\
Fe & 1.2e-02 & 4.8e-02 & 1.4e-03 & 7.9e-04 & 1.0e-03 & 1.4e-03 & 1.4e-03 & 1.4e-03\\
Co & 3.4e-08 & 4.7e-08 & 8.8e-07 & 1.6e-04 & 1.2e-04 & 4.8e-06 & 4.0e-06 & 4.0e-06\\
Ni & 3.4e-02 & 2.3e-03 & 8.1e-04 & 4.8e-04 & 3.1e-04 & 8.2e-05 & 8.2e-05 & 8.2e-05\\
$^{56}$Ni & 7.3e-01 & 3.1e-02 & 2.9e-07 & 1.7e-05 & 1.8e-05 & 2.6e-07 & 2.5e-09 & 1.2e-10\\
$^{57}$Ni & 2.8e-02 & 9.2e-04 & 1.6e-05 & 8.8e-07 & 3.0e-08 & 6.2e-09 & 8.0e-10 & 2.2e-11\\
$^{44}$Ti & 2.6e-04 & 1.3e-05 & 5.9e-06 & 1.8e-08 & 5.1e-12 & 2.5e-13 & 2.9e-14 & 8.2e-16\\
\bottomrule
\end{tabular}
\end{center}
\label{t_X_M17}
\end{table*}

\begin{table*}[tb]
\caption{Zone composition for models with an initial mass of 19 M$_\odot$.}
\begin{center}
\begin{tabular}{lllllllll}
\toprule
Element & Ni/He & Si/S & O/Si/S & O/Ne/Mg & O/C & He/C & He/N & H\\
\midrule
H & 5.5e-05 & 1.2e-06 & 2.1e-07 & 6.8e-09 & 2.3e-09 & 6.2e-10 & 5.0e-02 & 5.4e-01\\
He & 5.5e-01 & 5.7e-05 & 2.3e-05 & 1.1e-05 & 1.0e-02 & 8.0e-01 & 9.5e-01 & 4.4e-01\\
C & 5.0e-07 & 4.8e-05 & 1.6e-04 & 4.0e-02 & 2.6e-01 & 8.0e-02 & 8.6e-05 & 1.2e-04\\
N & 1.2e-06 & 1.8e-06 & 1.9e-05 & 3.4e-05 & 2.2e-05 & 1.3e-04 & 2.5e-03 & 1.0e-02\\
O & 9.7e-06 & 4.1e-02 & 5.8e-01 & 7.5e-01 & 7.1e-01 & 1.2e-01 & 4.2e-05 & 3.2e-03\\
Ne & 7.7e-06 & 3.6e-05 & 1.5e-04 & 1.7e-01 & 1.6e-02 & 4.3e-03 & 2.2e-04 & 3.0e-03\\
Na & 7.2e-08 & 8.3e-10 & 1.0e-09 & 1.8e-06 & 4.2e-08 & 3.3e-09 & 7.4e-11 & 7.9e-05\\
Mg & 1.5e-05 & 2.3e-04 & 6.5e-04 & 3.1e-02 & 6.3e-03 & 6.4e-04 & 1.1e-04 & 7.2e-04\\
Al & 2.7e-05 & 3.1e-04 & 1.8e-04 & 2.8e-03 & 4.5e-05 & 1.7e-05 & 1.5e-05 & 6.9e-05\\
Si & 2.0e-02 & 4.4e-01 & 2.3e-01 & 6.8e-03 & 4.8e-04 & 1.9e-04 & 1.1e-04 & 8.2e-04\\
S & 3.3e-02 & 3.6e-01 & 1.6e-01 & 1.2e-03 & 1.2e-04 & 7.2e-05 & 5.1e-05 & 4.2e-04\\
Ar & 8.0e-03 & 4.6e-02 & 2.7e-02 & 1.5e-04 & 3.5e-05 & 1.7e-05 & 1.2e-05 & 1.1e-04\\
Ca & 9.5e-03 & 2.7e-02 & 8.3e-03 & 3.5e-05 & 1.3e-05 & 9.7e-06 & 7.2e-06 & 7.4e-05\\
Sc & 2.7e-06 & 2.7e-07 & 7.7e-07 & 3.0e-07 & 2.2e-07 & 2.7e-08 & 4.0e-09 & 4.5e-08\\
Ti & 8.5e-07 & 2.0e-05 & 7.3e-05 & 2.3e-06 & 2.0e-06 & 4.9e-07 & 2.8e-07 & 3.4e-06\\
V & 1.5e-08 & 2.2e-07 & 8.8e-08 & 1.8e-07 & 8.2e-08 & 5.1e-08 & 3.4e-08 & 4.3e-07\\
Cr & 4.9e-04 & 7.3e-04 & 1.0e-04 & 4.3e-06 & 4.2e-06 & 2.2e-06 & 1.5e-06 & 2.0e-05\\
Mn & 1.7e-04 & 2.6e-04 & 1.0e-05 & 1.2e-06 & 9.8e-07 & 1.6e-06 & 1.1e-06 & 1.5e-05\\
Fe & 2.1e-02 & 5.0e-02 & 1.4e-03 & 4.3e-04 & 4.1e-04 & 2.8e-04 & 1.9e-04 & 1.4e-03\\
Co & 2.2e-03 & 1.5e-03 & 7.6e-06 & 4.0e-05 & 4.6e-05 & 5.7e-06 & 2.7e-07 & 4.0e-06\\
Ni & 1.2e-02 & 1.2e-03 & 3.2e-04 & 1.7e-04 & 1.3e-04 & 1.5e-05 & 5.5e-06 & 8.2e-05\\
$^{56}$Ni & 3.3e-01 & 2.5e-02 & 2.5e-08 & 1.6e-11 & 5.6e-13 & 1.3e-13 & 2.5e-15 & 1.2e-10\\
$^{57}$Ni & 1.1e-02 & 5.7e-04 & 9.3e-08 & 1.8e-10 & 4.6e-12 & 5.3e-11 & 1.9e-11 & 2.2e-11\\
$^{44}$Ti & 9.5e-05 & 1.1e-05 & 2.6e-06 & 4.8e-09 & 2.6e-12 & 1.6e-13 & 2.6e-15 & 8.2e-16\\
\bottomrule
\end{tabular}
\end{center}
\label{t_X_M19}
\end{table*}

\begin{table*}[tb]
\caption{Zone composition for models with an initial mass of 21 M$_\odot$.}
\begin{center}
\begin{tabular}{lllllllll}
\toprule
Element & Ni/He & Si/S & O/Si/S & O/Ne/Mg & O/C & He/C & He/N & H\\
\midrule
H & 4.8e-05 & 3.7e-07 & 1.8e-07 & 2.0e-09 & 1.2e-09 & 3.0e-10 & 1.7e-02 & 5.4e-01\\
He & 6.0e-01 & 6.4e-05 & 2.4e-05 & 1.1e-05 & 1.9e-02 & 8.0e-01 & 9.8e-01 & 4.4e-01\\
C & 7.4e-07 & 3.6e-05 & 1.6e-04 & 8.4e-03 & 2.6e-01 & 7.9e-02 & 2.2e-04 & 1.2e-04\\
N & 1.3e-06 & 7.3e-07 & 6.9e-06 & 3.0e-05 & 1.7e-05 & 5.5e-05 & 2.5e-03 & 1.0e-02\\
O & 1.2e-05 & 1.1e-02 & 1.7e-01 & 7.5e-01 & 7.0e-01 & 1.2e-01 & 5.3e-05 & 3.2e-03\\
Ne & 9.4e-06 & 2.4e-05 & 1.0e-04 & 1.9e-01 & 1.2e-02 & 4.1e-03 & 2.7e-04 & 3.0e-03\\
Na & 1.1e-07 & 7.3e-10 & 1.6e-09 & 1.2e-06 & 2.6e-08 & 3.7e-09 & 1.2e-10 & 7.9e-05\\
Mg & 9.3e-06 & 1.9e-04 & 2.4e-04 & 3.5e-02 & 5.8e-03 & 4.7e-04 & 1.2e-04 & 7.2e-04\\
Al & 2.1e-05 & 3.3e-04 & 3.1e-04 & 3.2e-03 & 4.3e-05 & 1.6e-05 & 1.5e-05 & 6.9e-05\\
Si & 7.7e-03 & 4.4e-01 & 3.9e-01 & 7.1e-03 & 4.8e-04 & 1.9e-04 & 1.2e-04 & 8.2e-04\\
S & 1.5e-02 & 3.7e-01 & 3.4e-01 & 1.2e-03 & 1.2e-04 & 7.3e-05 & 5.2e-05 & 4.2e-04\\
Ar & 4.4e-03 & 5.0e-02 & 6.2e-02 & 1.6e-04 & 3.4e-05 & 1.7e-05 & 1.2e-05 & 1.1e-04\\
Ca & 6.2e-03 & 3.1e-02 & 3.0e-02 & 6.0e-05 & 1.3e-05 & 1.0e-05 & 7.4e-06 & 7.4e-05\\
Sc & 2.6e-06 & 2.7e-07 & 2.2e-06 & 3.7e-07 & 2.2e-07 & 2.4e-08 & 4.1e-09 & 4.5e-08\\
Ti & 9.3e-07 & 1.7e-05 & 2.4e-04 & 2.9e-06 & 2.0e-06 & 4.8e-07 & 2.8e-07 & 3.4e-06\\
V & 1.4e-08 & 2.8e-07 & 3.1e-07 & 1.9e-07 & 7.9e-08 & 5.2e-08 & 3.4e-08 & 4.3e-07\\
Cr & 4.3e-04 & 1.0e-03 & 4.7e-04 & 4.6e-06 & 4.2e-06 & 2.3e-06 & 1.5e-06 & 2.0e-05\\
Mn & 1.5e-04 & 3.6e-04 & 5.6e-05 & 1.1e-06 & 1.0e-06 & 2.1e-06 & 1.1e-06 & 1.5e-05\\
Fe & 9.3e-03 & 3.6e-02 & 3.7e-03 & 1.9e-04 & 2.1e-04 & 1.4e-04 & 9.8e-05 & 1.4e-03\\
Co & 2.2e-03 & 2.3e-03 & 2.7e-05 & 4.4e-05 & 4.3e-05 & 4.3e-06 & 2.8e-07 & 4.0e-06\\
Ni & 1.6e-02 & 1.8e-03 & 2.5e-04 & 1.9e-04 & 1.3e-04 & 1.2e-05 & 5.7e-06 & 8.2e-05\\
$^{56}$Ni & 3.3e-01 & 5.3e-02 & 3.5e-07 & 1.2e-10 & 2.7e-13 & 8.7e-14 & 2.3e-15 & 1.2e-10\\
$^{57}$Ni & 1.2e-02 & 9.5e-04 & 7.7e-07 & 5.5e-10 & 2.9e-12 & 5.8e-11 & 1.9e-11 & 2.2e-11\\
$^{44}$Ti & 9.2e-05 & 1.4e-05 & 1.4e-05 & 2.5e-08 & 1.8e-12 & 1.2e-13 & 2.7e-15 & 8.2e-16\\
\bottomrule
\end{tabular}
\end{center}
\label{t_X_M21}
\end{table*}

\section{Additional spectral figures}
\label{a_additional_figures}

\begin{figure*}[tbp!]
\includegraphics[width=1.0\textwidth,angle=0]{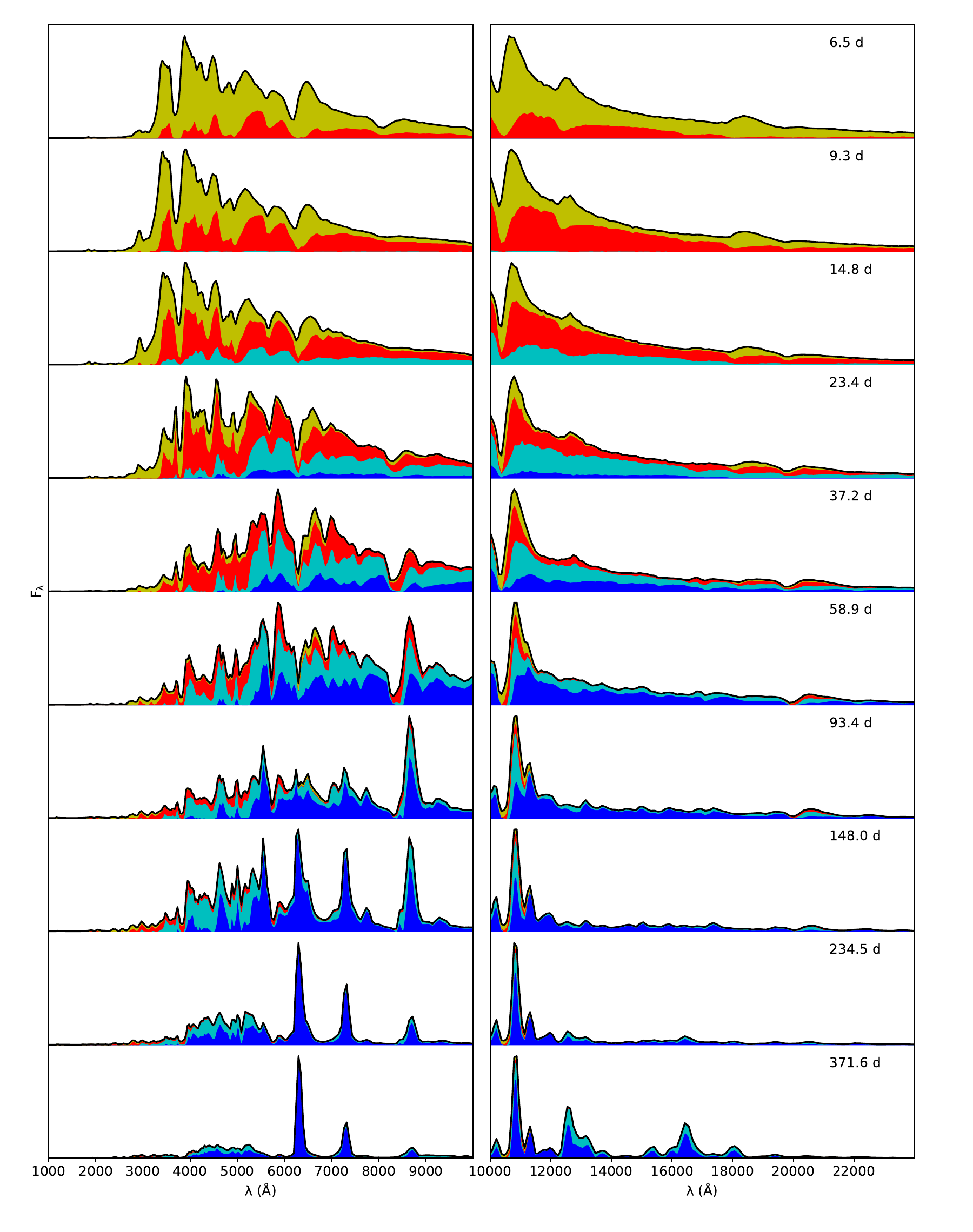}
\caption{Spectral evolution in the optical (left panel) and NIR (right panel) for the optimal model, where the NIR flux has been scaled as indicated in blue. In the spectra we show the contributions (last scattering or emission event, excluding electron scattering) to the flux from the carbon-oxygen core (blue), the inner (cyan) and outer (red) helium envelope and the hydrogen (yellow) envelopes.}
\label{f_acat_spec_cell_evo}
\end{figure*}

For convenience, we provide a set of additional spectral figures for the optimal model of SN 2020acat. First, in \ref{f_acat_spec_cell_evo} we show the contribution (last scattering or emission event excluding electron scattering) from the carbon-oxygen core, the inner and outer helium envelope and the hydrogen envelope to the spectral evolution.

\begin{figure*}[tbp!]
\includegraphics[width=1.0\textwidth,angle=0]{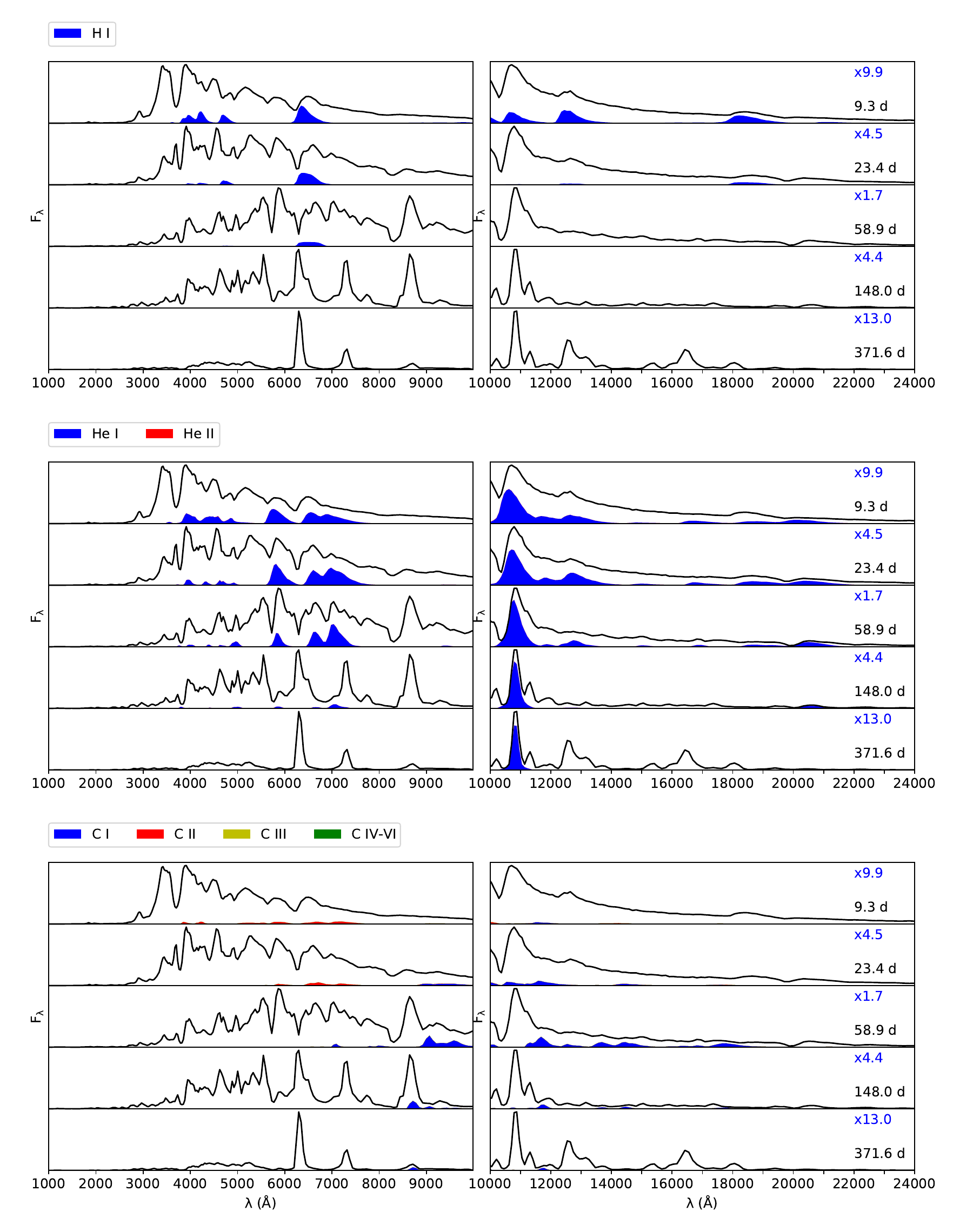}
\caption{Bound-bound contribution from ionisation stages I (blue), II (red), III (yellow), and higher (green) of hydrogen (upper panel), helium (middle panel), and carbon (lower panel) to the spectral evolution of the optimal model.}
\label{f_acat_spec_trans_evo_ion_1}
\end{figure*}

\begin{figure*}[tbp!]
\includegraphics[width=1.0\textwidth,angle=0]{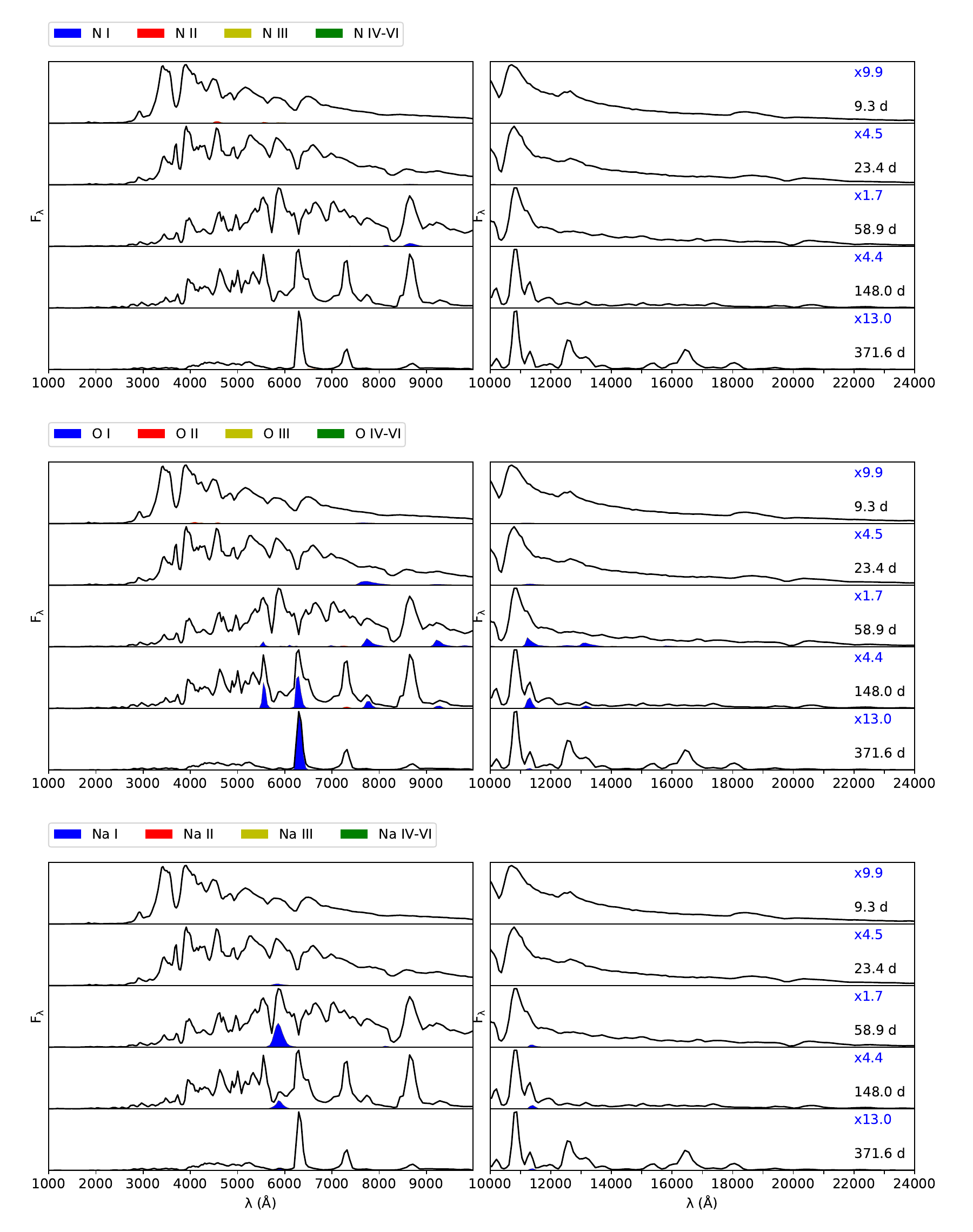}
\caption{Bound-bound contribution from ionisation stages I (blue), II (red), III (yellow), and higher (green) of nitrogen (upper panel), oxygen (middle panel), and sodium (lower panel) to the spectral evolution of the optimal model.}
\label{f_acat_spec_trans_evo_ion_2}
\end{figure*}

\begin{figure*}[tbp!]
\includegraphics[width=1.0\textwidth,angle=0]{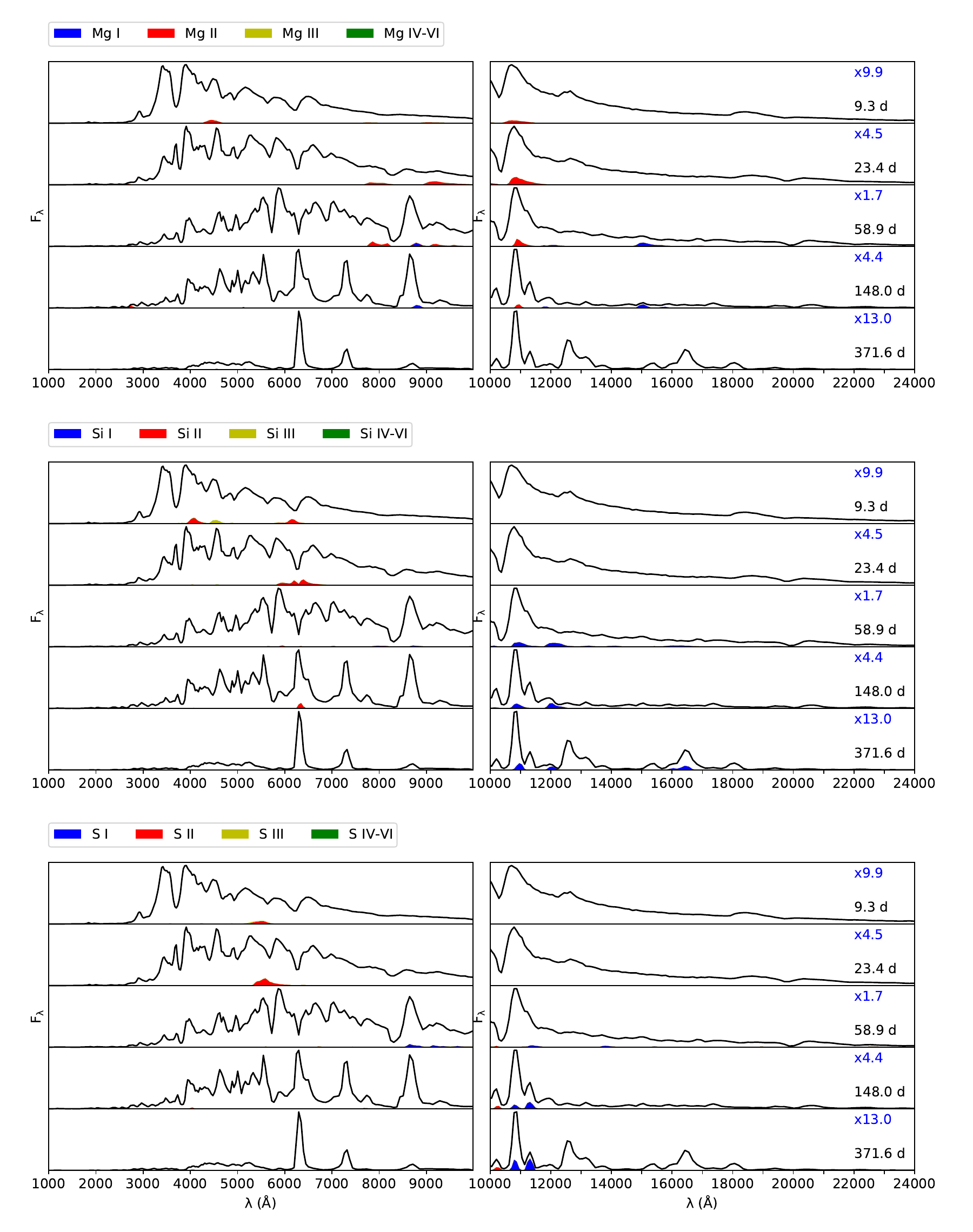}
\caption{Bound-bound contribution from ionisation stages I (blue), II (red), III (yellow), and higher (green) of magnesium (upper panel), silicon (middle panel), and sulphur (lower panel) to the spectral evolution of the optimal model.}
\label{f_acat_spec_trans_evo_ion_3}
\end{figure*}

\begin{figure*}[tbp!]
\includegraphics[width=1.0\textwidth,angle=0]{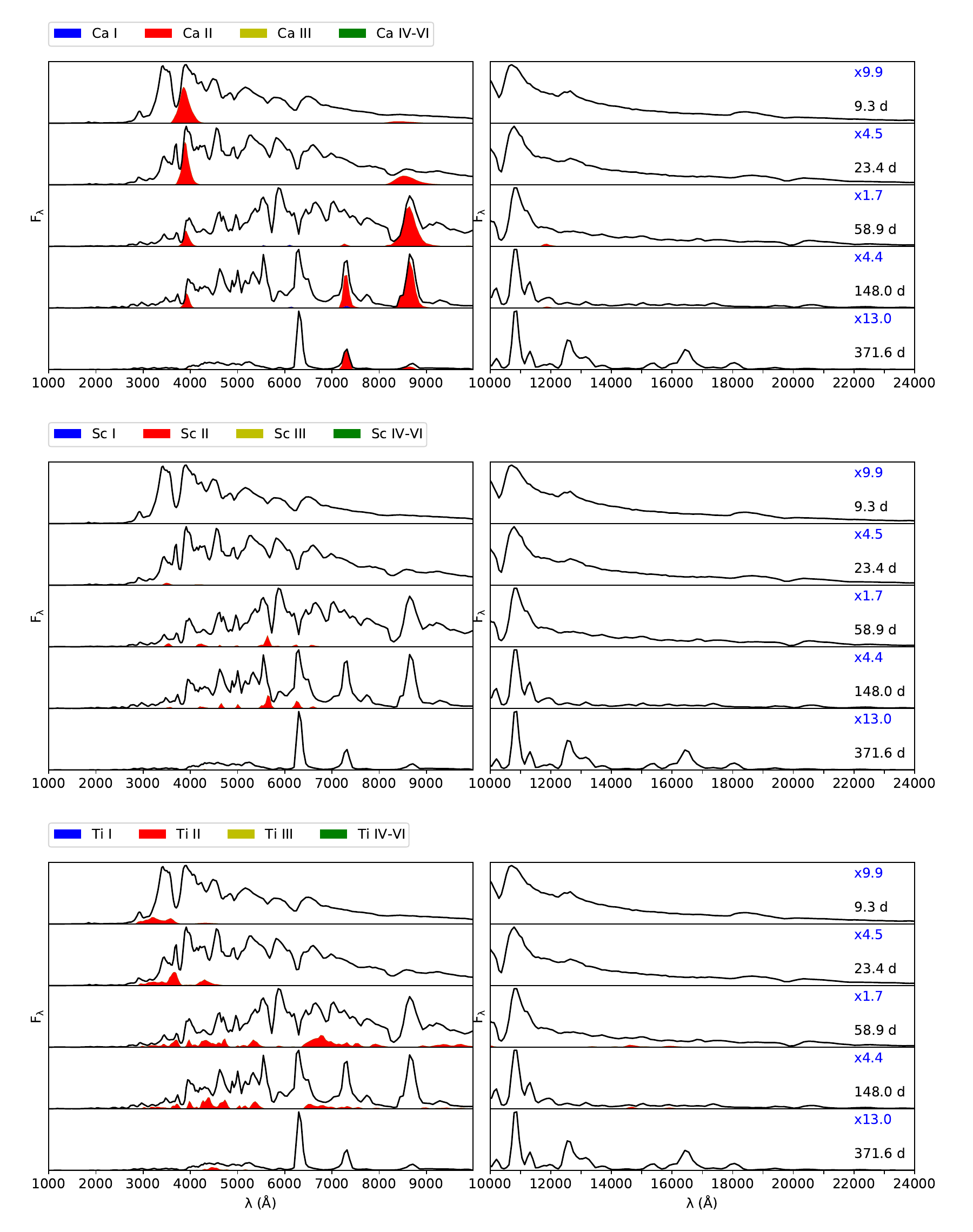}
\caption{Bound-bound contribution from ionisation stages I (blue), II (red), III (yellow), and higher (green) of calcium (upper panel), scandium (middle panel), and titanium (lower panel) to the spectral evolution of the optimal model.}
\label{f_acat_spec_trans_evo_ion_4}
\end{figure*}

\begin{figure*}[tbp!]
\includegraphics[width=1.0\textwidth,angle=0]{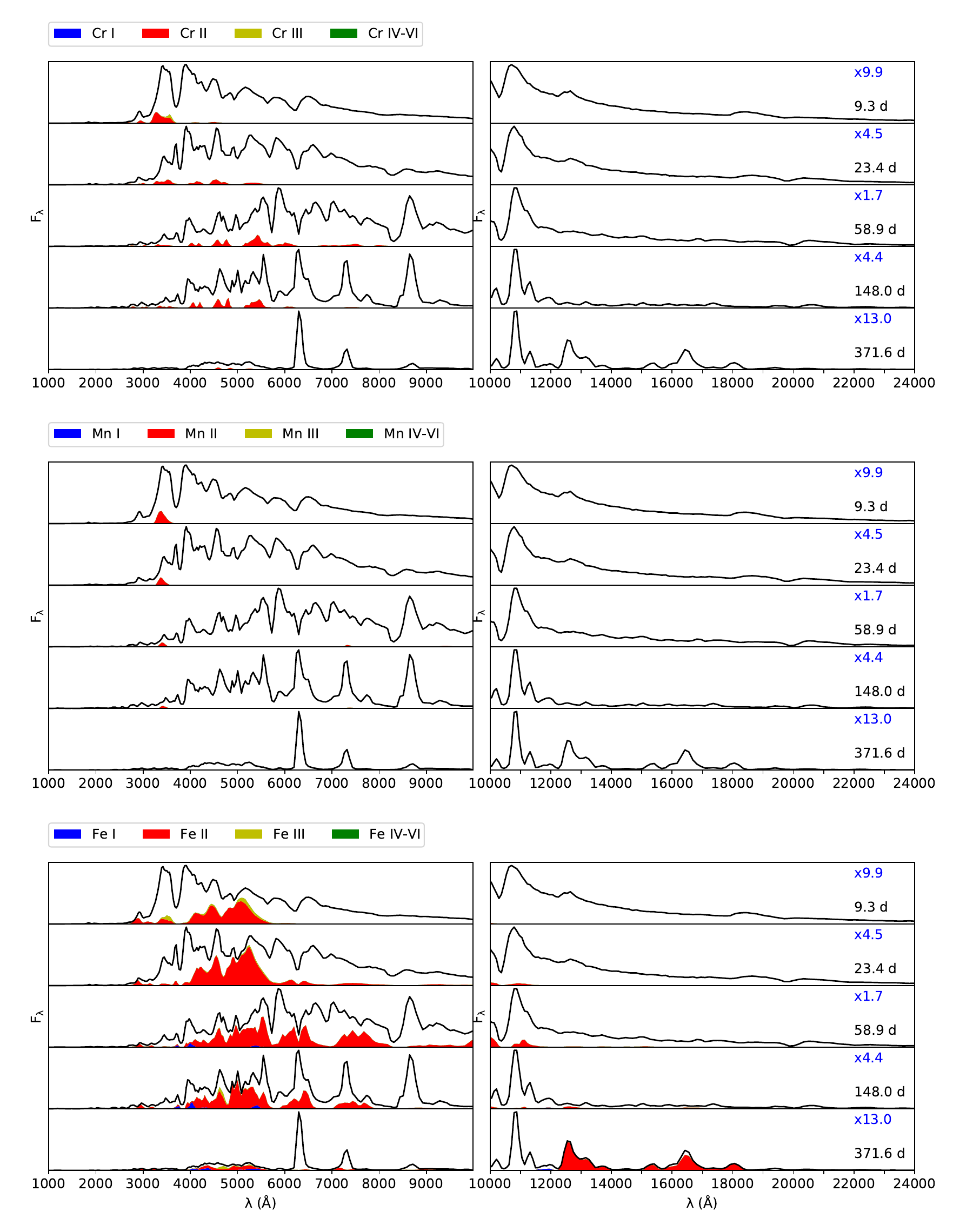}
\caption{Bound-bound contribution from ionisation stages I (blue), II (red), III (yellow), and higher (green) of chromium (upper panel), manganese (middle panel), and iron (lower panel) to the spectral evolution of the optimal model.}
\label{f_acat_spec_trans_evo_ion_5}
\end{figure*}

\begin{figure*}[tbp!]
\includegraphics[width=1.0\textwidth,angle=0]{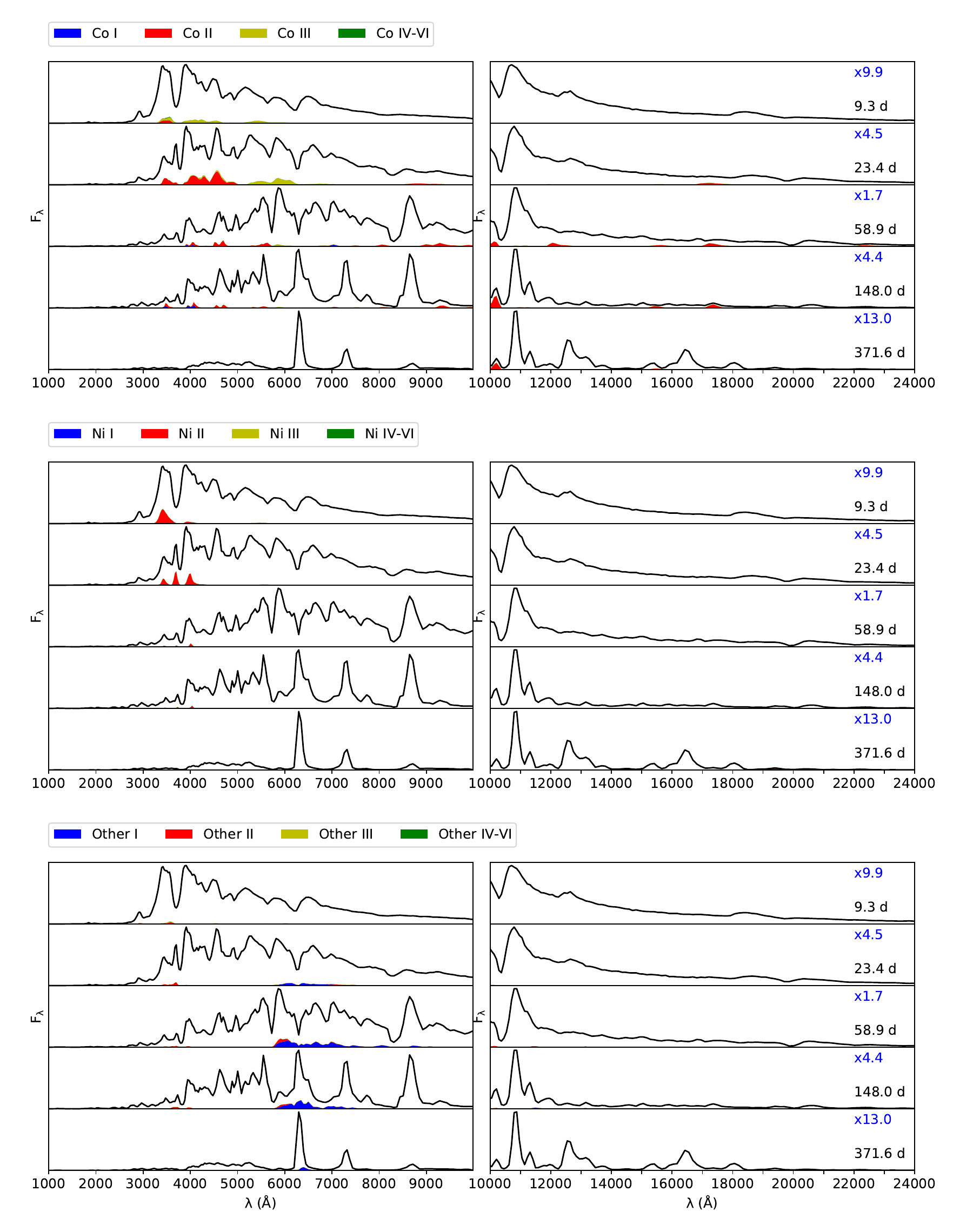}
\caption{Bound-bound contribution from ionisation stages I (blue), II (red), III (yellow), and higher (green) of cobalt (upper panel), nickel (middle panel), and other elements (lower panel) to the spectral evolution of the optimal model.}
\label{f_acat_spec_trans_evo_ion_6}
\end{figure*}

Second, in Fig.~\ref{f_acat_spec_trans_evo_ion_1}-\ref{f_acat_spec_trans_evo_ion_6} we show the bound-bound contribution (last scattering or emission event excluding electron scattering) from ionisation stages I, II, III, and higher of hydrogen, helium, carbon, nitrogen, oxygen, sodium, magnesium, silicon, sulphur, calcium, scandium, titanium, chromium, manganese, iron, cobalt, nickel, and other elements to the spectral evolution.

\begin{figure*}[tbp!]
\includegraphics[width=1.0\textwidth,angle=0]{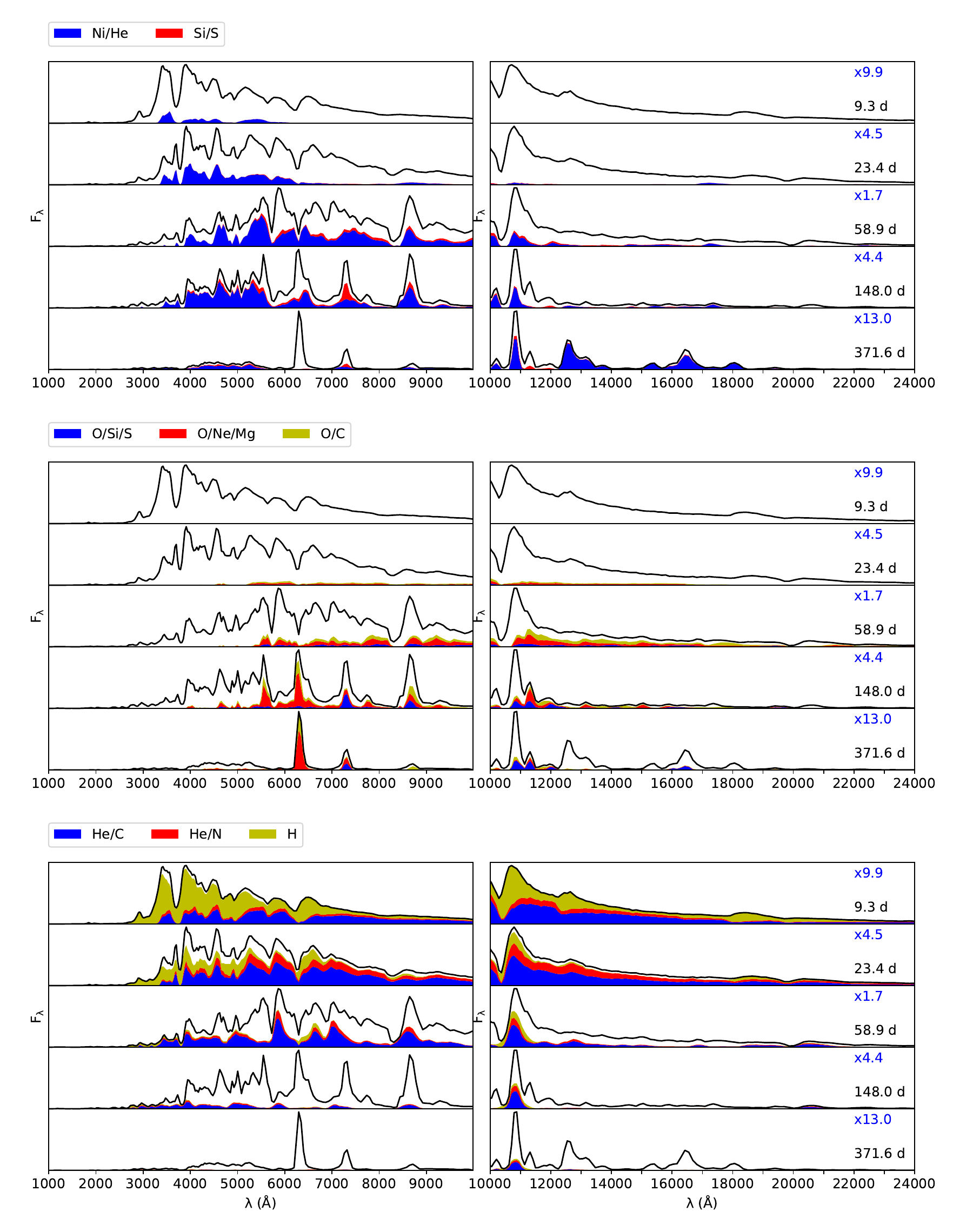}
\caption{Bound-bound contribution from the nickel-rich zones (Ni/He: blue, Si/S: red), the oxygen-rich zones (O/Si/S: blue, O/Ne/Mg: red, O/C: yellow), and the hydrogen- and helium-rich zones (H/C: blue, He/N: red, H: yellow) to the spectral evolution of the optimal model.}
\label{f_acat_spec_cell_evo_zone}
\end{figure*}

Finally, in Fig.~\ref{f_acat_spec_cell_evo_zone} we show the bound-bound contribution (last scattering or emission event excluding electron scattering) from the nickel-rich (Ni/He, Si/S), oxygen-rich (O/Si/S, O/Ne/Mg, O/C) and hydrogen- and helium-rich (He/C, He/N, H) compositional zones to the spectral evolution.

\end{appendix}

\end{document}